\documentclass[review,hidelinks]{elsarticle}

\usepackage[displaymath,mathlines,pagewise]{lineno}
\usepackage{hyperref}
\usepackage{amsmath}
\usepackage{amssymb}
\usepackage{booktabs}
\usepackage{multirow}
\usepackage{color}
\modulolinenumbers[1]

\DeclareSymbolFont{TXlettersA}{U}{txmia}{m}{it}
\SetSymbolFont{TXlettersA}{bold}{U}{txmia}{bx}{it}
\DeclareFontSubstitution{U}{txmia}{m}{it}
\DeclareMathSymbol{\deltaup}{\mathord}{TXlettersA}{14}

\newcommand{\pd}[2]{\frac{\partial #1}{\partial #2}}
\newcommand{\pdF}[2]{\frac{\deltaup #1}{\deltaup #2}}

\newcommand{\pdl}[2]{\partial #1 / \partial #2}

\newcommand{\Sec}[1]{Sec.\ \ref{#1}}

\newcommand{\Eq}[1]{Eq.\ (\ref{#1})}
\newcommand{\Eqs}[2]{Eqs.\ (\ref{#1}) and (\ref{#2})}
\newcommand{\Fig}[1]{Fig.~\ref{#1}}
\newcommand{\Figs}[2]{Figs.\ \ref{#1} -- \ref{#2}}

\newcommand{\lr}{\left(}
\newcommand{\rr}{\right)}

\newcommand{\ls}{\left[}
\newcommand{\rs}{\right]}
\newcommand{\lc}{\left\{}
\newcommand{\rc}{\right\}}

\newcommand{\av}[1]{\langle #1 \rangle}

\newcommand{\be}{\begin{equation}}
\newcommand{\blnm}{\begin{linenomath*}}
\newcommand{\ee}{\end{equation}}
\newcommand{\elnm}{\end{linenomath*}}

\newcommand{\bspl}{\begin{split}}
\newcommand{\espl}{\end{split}}
\newcommand{\bea}{\begin{eqnarray}}
\newcommand{\eea}{\end{eqnarray}}
\newcommand{\bagn}{\begin{align}}
\newcommand{\eagn}{\end{align}}
\newcommand{\bs}{\begin{split}}
\newcommand{\es}{\end{split}}
\newcommand{\bc}{\begin{center}}
\newcommand{\ec}{\end{center}}
\newcommand{\bp}{\begin{picture}(0,0)}
\newcommand{\ep}{\end{picture}}
\newcommand{\bfl}{\begin{flushleft}}
\newcommand{\efl}{\end{flushleft}}

\newcommand{\bx}{\mathbf{x}}

\newcommand{\bW}{\mathbf{W}}

\newcommand{\bu}{\mathbf{u}}

\newcommand{\bw}{\mathbf{w}}

\newcommand{\bn}{\mathbf{n}}
\newcommand{\bS}{\mathbf{S}}

\newcommand{\bng}{\mathbf{n}_\Gamma}

\newcommand{\eph}{\epsilon_h}
\newcommand{\alpx}{\alpha \lr  \bx,t  \rr}
\newcommand{\alpp}{\alpha \lr  \psi \lr \bx,t \rr \rr}
\newcommand{\alpps}{\alpha \lr  \psi \rr}
\newcommand{\deltaa}{\tilde{\delta} \lr  \alpha \rr}
\newcommand{\deltap}{\delta \lr  \Psi \rr}

\newcommand{\cO}{{\mathcal{O}}}

\newcommand{\psio}{\psi  \left( \alpha \right)}
\newcommand{\psix}{\psi  \left( \bx,t \right)}
\newcommand{\Psix}{\Psi  \left( \bx,t \right)}

\journal{International Journal of Multiphase Flow}

\biboptions{authoryear}

\begin{document}

\begin{frontmatter}

  
  \title{On a relation between the volume of fluid,
         level-set\newline and phase field interface models}


  \author{Tomasz Wac{\l}awczyk\fnref{}\corref{mycorrespondingauthor}}
\address{Department of Aeronautics, Institute of Aeronautics and Applied Mechanics,\\ Warsaw University of Technology,\\ Nowowiejska 24, 00653 Warszawa, Poland}


\cortext[mycorrespondingauthor]{Tomasz Wac{\l}awczyk}
\ead{twacl@meil.pw.edu.pl}


\begin{abstract}
  This paper discusses
  a relation
  between the re-initialization
  equation of the level-set
  functions derived
  by Wac{\l}awczyk [\emph{J.Comp.Phys.}, {{299}}, (2015)]
  and the condition for the
  phase equilibrium provided
  by the stationary
  solution to the
  modified Allen-Cahn
  equation [\emph{Acta Metall.}, {{27}}, (1979)].
  As a consequence,
  the statistical model of
  the non-flat  interface in the state
  of phase equilibrium is postulated.   
  This new physical model
  of the non-flat interface
  is introduced based
  on the statistical picture
  of the sharp interface
  disturbed by the field
  of stochastic forces,
  it yields the relation
  between the sharp and
  diffusive interface models.
  Furthermore,
  the new  
  techniques required  
  for the accurate solution
  of the model equations 
  are proposed.
  First it is shown,
  the constrained interpolation
  improves re-initialization of
  the level-set functions as it
  avoids oscillatory numerical errors
  typical for the second-order
  accurate
  interpolation schemes.
  Next,
  the new  
  semi-analytical,
  second order accurate
  Lagrangian scheme is put forward
  to integrate the advection
  equation in time avoiding
  interface curvature
  oscillations 
  introduced 
  by the second-order
  accurate flux
  limiters.
  These techniques
  provide means
  to obtain complete,
  second-order
  convergence 
  during
  advection
  and re-initialization 
  of the interface
  in the state of
  phase equilibrium.
\end{abstract}


\end{frontmatter}

\section{Introduction}
\label{sec1}
%
Experiments 
reveal 
the macroscopic 
interface is a region
of a finite thickness
$\eph \!\sim\! \sqrt{k_B T/\sigma}\,[m]$,
were $k_B\,[J/K]$ is 
the Boltzman constant,
$T\,[K]$ is absolute 
temperature and
$\sigma\,[J/m^2]$ is 
the surface tension 
coefficient
(\cite{vrij1973,aarts2004}).
In this region,
the liquid phase
and its vapor
co-exist
in the state
of phase equilibrium
(\cite{waals1979,smol1908}).
Similarly,
the ensemble 
averaged description
of interfaces 
interacting
with turbulence
introduces
the non-zero 
width $\eph$ 
of the
``surface layer''
or ``intermittency region''
(\cite{hong00,brocchini01a,brocchini01b,waclawczyk2015}).
Therein,
$\eph \!\sim\! D/C \,[m]$
where
$D\,[m^2/s]$ is
the diffusivity 
and $C\,[m/s]$
is characteristic 
velocity  
related to
local 
properties of
the ensemble averaged
turbulent velocity field.
Because
the macroscopic 
interface 
thickness 
$\eph \sim 0.5 \,[nm]$
is usually negligible
when compared with
the characteristic
flow scale,
the sharp 
interface 
model 
is most
often
used.
This is also 
the case in two phase
turbulent flows
as modeling of
$\eph \lr \bx,t \rr$  
is complex.
In the sharp
interface model
the interface
is approximated
using the three dimensional 
Heaviside function
$H \lr \bx,t \rr$ 
that
indicates  
presence
of the liquid 
phase.

The sharp interface model
is the cornerstone
of the volume of fluid
(VOF) family of numerical methods,
see  \cite{trygg11}.
The key 
problem 
there is numerical 
approximation 
of the transport equation
%
\blnm
  \be
  \pd{H}{t} + \bW \nabla H = 0,
  \label{eq1}
  \ee
\elnm
where $\bW\,[m/s]$ 
denotes velocity
of the sharp interface.
The position
of the sharp
interface
defined 
by the 
level-set 
$H \lr \bx,t \rr \!=\! 1/2$ 
is found
in the
geometrical
reconstruction
procedure.
The VOF methods
guarantee  exact 
satisfaction of
the law of mass 
conservation
providing
$\bW \!=\! \bu$,
where
$\bu\,[m/s]$ is
velocity of
incompressible
gas/liquid
phases continuous
at the interface.
In such case,  
the transport equation
for the phase indicator
function $H \lr \bx,t \rr$
can be derived directly
from the mass conservation
equation.
However, 
$H \lr \bx,t \rr$ 
is discontinuous
at the interface,
for this reason  
the VOF methods
require 
auxiliary numerical
techniques to
approximate the spatial
interface orientation
and curvature,
the exhaustive list
of these techniques
is provided by \cite{trygg11}.

Yet other way to
represent the sharp
interface is by the
zero level-set of
the function $\Psi \lr \bx,t \rr \!=\! 0$,
where $\Psi \lr \bx,t \rr$
denotes 
the signed-distance
from the 
sharp interface.
This is the staple
of the standard
level-set (SLS)
method introduced
by  \cite{osher1988}
and further developed
by others, see 
\cite{sussman94,sussman98,osher03}
to mention only 
the first works 
on the level-set
method(s).
Unlike in 
the VOF interface model,
in the SLS model
the sharp interface 
is captured
by the zero level-set 
of the smooth signed-distance 
function $\Psi \lr \bx,t \rr$
with the property
$|\nabla \Psi| \!=\! 1$.
The standard level-set (SLS)
method does not implicitly
obey the law of mass conservation,
but allows computing 
the interface
orientation 
$\bng \!=\! \nabla \Psi/|\nabla \Psi|$
and curvature
$\kappa \!=\! -\nabla^2 \Psi$
in the straightforward
and accurate
manner.

Although both 
the VOF and SLS
interface models 
reconstruct
the same sharp interface,
the SLS model
additionally requires
re-initializaton
of the signed distance function
in order to preserve
the  property 
$|\nabla \Psi|\!=\!1$.
Namely, beside solving
the advection equation
\blnm
 \be
  \pd{\Psi}{t} + \bW \nabla \Psi = 0,
 \label{eq2}
 \ee
\elnm
the stationary solution
to the re-initialization
equation
\blnm
 \be
  \pd{\Psi}{\tau} = -sgn[\Psi_0] \lr |\nabla \Psi| - 1 \rr + F \lr H,\Psi \rr |\nabla \Psi|,
 \label{eq3}
 \ee
\elnm
where $F \lr H,\Psi \rr$
is a known function,
is needed after
each advection step 
(see  \cite{sussman94,sussman98,osher03}).
In \Eq{eq3},
$\tau\,[s]$ denotes
``artificial'' time 
and $\Psi_0$ is 
the signed-distance
function after 
precedent 
solution of \Eq{eq2}.
$\Psi_0$ 
must be used
in \Eq{eq3}
as consecutive 
numerical solutions
of this equation 
have tendency
to move 
the interface
from $\Psi_0 \lr \bx,t \rr\!=\!0$
increasing 
the loss of mass,
see work of \cite{osher03}
and references therein.
One notices,
re-initialization 
\Eq{eq3} has no physical
interpretation in
the SLS model,
it is perceived
as a geometrical
constraint
required 
to preserve 
$|\nabla \Psi|\!=\!1$
during advection
of the sharp interface
$\Psi \lr \bx,t \rr \!=\! 0$.
Additionally,
in spite of
discretization of
\Eqs{eq2}{eq3} 
with the higher-order 
schemes: 
5-th order 
WENO in space 
and 4-th
order TVD Runge-Kutta
in time, typically,
only the second-order
accuracy is achieved
when \Eqs{eq2}{eq3}
are used to advect
the interface on
the uniform,
orthogonal 
grids 
(see \cite{herrmans05}).

An alternative
description
of the interface
is introduced 
by the diffusive
and/or the phase 
field (PHF)
interface models. 
These 
phenomenological
models  are based
on the assumption
about abrupt
but continuous
variation
of the liquid
phase density 
across the interface
with the non-zero
thickness
(\cite{waals1979,cahn1958,allen1979,anderson1998}).
The thickness 
of the interface
is $\eph > 0$
if the liquid
phase and its vapor
are below
critical
conditions.
The first 
mathematical 
model of 
the flat interface
in the state
of the thermodynamical
equilibrium
has been 
introduced by
\cite{waals1979}.
Therein,
the density
based functional
is put forward
to represent
the balance of
the Helmholtz
free energy
in the vicinity
of the flat, 
regularized
interface.
The
interfacial
energy density 
equilibrium
is established
due to local,
continuous 
distribution
of the liquid 
phase
density.
Later on,
it was recognized
the van der Waals
density based functional
is related to
the Ginzburg-Landau functional
derived from the theory
of the first and/or second-order
phase transitions
(see \cite{cahn1958, allen1979}).
In this
latter PHF model,
the material properties
are changing across
the interface by
means of 
the order parameter $\alpx$
allowing a smooth
transition between
the liquid phase
and its vapor.
\cite{allen1979}
obtained
$\alpx$ 
by a solution
of time-dependent,
non-linear equation
\blnm
 \be
 \pd{\alpha}{\tau} = 2D \nabla^2{\alpha}
                   - \frac{C}{\eph} \pdF{f \lr \alpha \rr}{\alpha} 
                   = \frac{2C}{\eph} \ls
                     \eph^2 \nabla^2{\alpha}
                   - \lr 1-2\alpha \rr \alpha \lr 1-\alpha \rr  \rs,
 \label{eq4}
 \ee
\elnm
where $\deltaup$ 
denotes the functional derivative,
$D\!=\!C\eph\,[m^2/s]$ is the diffusivity
coefficient, $C/\eph\,[1/s]$ is the kinetic
parameter,  $\eph\,[m]$ 
is the interface
width and
$f \lr \alpha \rr \!=\! \alpha^2 \lr 1-\alpha \rr^2\,[-]$ 
denotes the double
well potential.
According 
to 
\cite{allen1979} and
references therein,
the order parameter $\alpx$
in \Eq{eq4}
is not a conserved quantity
and therefore, 
it does not have a clear
physical interpretation.
In spite of 
aforementioned
limitations,
\cite{allen1979}
use the PHF
interface model 
defined by \Eq{eq4}
to investigate
the second-order 
phase transitions
in binary-fluids.
The profile
of the order parameter
obtained from the stationary
solution to
equation (\ref{eq4})
is given by
the Lipschitz
continuous
function related 
to the hyperbolic 
tangent 
(\cite{waals1979,cahn1958, allen1979,anderson1998}).

Subsequently, 
\cite{olsson05}
 introduced
the conservative 
level-set (CLS) method
to some extent
coupling the advantages
of the sharp and regularized
interface models
(see \cite{chiu11,balcazar2014}).
In the CLS method,
the interface is represented
by the level-set of
the regularized 
Heaviside function
$\alpx \!=\! 1/2$.
As in the SLS method,
the CLS method
also requires
re-initialization
of the conserved
level-set function
$\alpx$ 
to reduce numerical
errors introduced
during the advection
step.
The direct numerical
solution of Olsson and Kreiss 
re-initialization equation 
 \blnm
 \be
   \pd{\alpha}{\tau} = \nabla \cdot \ls D |\nabla \alpha|\bng - C \alpha \lr 1-\alpha \rr \bng \rs
   \label{eq5}
 \ee
 \elnm 
where $\bng \!=\! \nabla \alpha / |\nabla \alpha|$,
suffers from similar problems
as re-initialization performed
using \Eq{eq3}.
In particular,
when the number
of re-initialization
steps $N_\tau \!\to\! \infty$
consecutive solutions of \Eq{eq5}
lead to artificial 
deformations of 
the regularized interface,
(see \cite{mccaslin2014,twacl15}
and references 
therein).
Interestingly,
the profile of
the conserved
level-set function
obtained from the 
analytical solution
to \Eq{eq5} 
in  the steady state,
is given by the same
function 
as the profile of
the order parameter
in the  \cite{allen1979} 
phase field
model given by \Eq{eq4}.
Unlike in \Eq{eq4},
the stationary solution to \Eq{eq5}
is obtained in the
direction normal 
to the interface 
$\bng=\nabla \alpha/|\nabla \alpha|$.

Recently, 
using this latter 
property of \Eq{eq5},
its consistent solution
was proposed by 
\cite{twacl15}.
The consistent
solution uses both:
the signed-distance $\psio$ 
and conserved $\alpps$
level-set functions,
as the analytical solution
to \Eq{eq5} in steady
state reads
\blnm
 \be
 \alpp = \frac{1}{1+\exp{\lr -\psix/\eph \rr}}=
 \frac{1}{2} \ls 1+\tanh{\lr \frac{\psix}{2\eph} \rr} \rs.
 \label{eq6}
\ee
\elnm
For each
$\eph > 0$
the mapping 
between 
the level-set function
$\alpps$ and level-set
function $\psio$ 
can be derived directly 
from \Eq{eq6},
resulting in 
\blnm
\be
 \psi \lr \alpha \rr = \eph \ln{\ls \frac{\alpha\lr \psi \rr}
                                 {1 - \alpha \lr  \psi \rr} \rs}.
\label{eq7}
\ee
\elnm
The mapping given 
by \Eqs{eq6}{eq7} 
will be further 
denoted as
$\alpps\!-\!\psio$ 
emphasizing
$\psio$ is the inverse function
of $\alpps$.

The key idea 
introduced by
\cite{twacl15}
is to use the mapping
between 
the conserved $\alpps$
and signed-distance 
$\psio$ level-set 
functions
to calculate
analytically 
the gradient of 
more abruptly changing 
and hence more difficult 
to approximate on 
discrete grids 
function 
$\alpps$.
This gradient reads
 \blnm
 \be
  \nabla \alpha = \frac{\deltaa}{\eph} \nabla \psi,
  \label{eq8}
 \ee
 \elnm 
where $\deltaa=\alpha \lr 1-\alpha \rr$.
Computing  $| \nabla \alpha |$
in \Eq{eq5}
with \Eq{eq8} 
allows
reduction of 
numerical errors as
$|\nabla \psio| \!=\! 1$ for all
$\eph \!>\! 0$, whereas
$| \nabla \alpps | \!\to\! \infty$
when $\eph \!\to\! 0$.
The present author 
has also shown, 
\Eq{eq8}
can be further used 
to obtain the second-order 
spatial derivative of $\alpps$
reducing the
approximation errors
of the stationary
interface curvature.

Noting 
$\pdl{\alpha}{t} = \deltaa/\eph \pdl{\psi}{t}$
and substituting \Eq{eq8}
into \Eq{eq5}
allows 
to rewrite 
the advection
and re-initialization 
equations
of the level-set functions 
$\alpps \!-\! \psio$ in the form
\blnm
\be
   \pd{\alpha}{t} + \bw \nabla \alpha = \frac{\deltaa}{\eph} \ls \pd{\psi}{t}+ \bw \nabla \psi \rs = 0,
   \label{eq9}
\ee
\elnm
\blnm
\be
   \pd{\alpha}{\tau} = \nabla \cdot \ls C \deltaa \lr |\nabla\psi|-1 \rr \bng \rs,
   \label{eq10}
\ee
\elnm
where $\bw\,[m/s]$ denotes
velocity of the regularized
interface and
$\bng \!=\! \nabla \alpha/|\nabla \alpha| \!=\! \nabla \psi/|\nabla \psi|$, see \Eq{eq8}.
Let notice,
the right hand 
side (RHS)
of \Eq{eq10} 
equals zero
when $|\nabla\psio|\!=\!1$ 
or $\tilde{\delta} \lr H \rr \!=\! H \lr 1-H \rr \!=\! 0$.
The former 
condition holds
when $\alpps$ 
is given by the
hyperbolic tangent 
profile as this allows 
to derive the mapping
between $\alpps-\psio$,
see \Eqs{eq6}{eq7}.
The latter condition, $\tilde{\delta} \lr H \rr \!=\! 0$,
is satisfied in the limit of $\eph \to 0$.
In this limit,
the advection 
equation (\ref{eq9})
reduces to 
\Eq{eq1} where
$\bw=\bu$,
and re-initialization
\Eq{eq10} is
reduced to $\pdl{H}{\tau}\!=\! \delta \lr \Psi \rr \pdl{\Psi}{\tau} \!=\!0$.

\cite{twacl15} 
has observed
yet other 
feature of 
\Eq{eq10}, for $\eph > 0$
this equation can be
rewritten in the form
\blnm
\begin{align}
 \begin{split}
  \pd{\psi}{\tau} 
  &= \lr 1-2\alpha \rr |\nabla \psi| \lr |\nabla \psi| -1 \rr\\
  &+ \bng \cdot \nabla \lr |\nabla \psi| -1 \rr \eph \\
  &- \eph \lr |\nabla \psi| -1 \rr \kappa,
 \label{eq11}
 \end{split}
\end{align}
\elnm
where
$C=1\,[m/s]$ 
and  $\kappa = - \nabla \cdot \bng$.
Since
$sgn [\psi] = - sgn \ls \lr 1-2\alpha \rr \rs$,
\Eq{eq11}
resembles re-initialization
\Eq{eq3} introduced
in the SLS method;
one notices,
the similarity
between
\Eqs{eq3}{eq11} 
occurs in
the limit $\eph \!\to\! 0$.
For above named reasons,
\Eqs{eq9}{eq10}
yield
the analytical relation
between
the sharp interface model
used in the VOF and SLS methods 
and regularized interface model
used in the PHF and CLS methods.

In the present 
paper,
the physical interpretation
of the model equations (\ref{eq9})-(\ref{eq10})
is postulated.
First, 
the picture of 
the sharp interface 
agitated
by the  
stochastic 
velocity field
is presented
and 
its description 
in terms of
mean and fluctuating
components
is introduced.
Next, 
it is argued
the correct
stationary solution to
the re-initialization equation (\ref{eq10})
can be interpreted as finding
the minimum of
the modified Ginzburg-Landau
functional representing 
the interfacial
density of 
the Helmholtz free energy
in the
vicinity of 
the non-flat 
interface.
This is achieved 
by introduction
of an additional term
into the original Ginzburg-Landau functional,
accounting for 
the interfacial energy density
required to deform
the flat interface.
The additional
contribution to
the interfacial energy density
is stored in the local shape
and/or size
of the deformed
interface as
to create
the interface 
of small droplet
with
large curvature
more energy must
be supplied to
the liquid/gas phase.
As a consequence 
of this relation, 
it is shown
the conservative
level-set (CLS) method is
in fact the phase field model
of  the non-flat interface
in the state of phase equilibrium,
where the order parameter
$\alpps$ is interpreted as
the probability of finding one
of the two phases sharing
the regularized interface;
the probability $\alpps$
is a conserved quantity.
This result
and the results 
presented by \cite{twacl15},
provide the analytical relation 
between 
the sharp and diffusive
interface models.

In the second part
of the present work,
two new  techniques for 
a numerical  
solution 
of the statistical
interface model
equations (\ref{eq9})-(\ref{eq10})
are introduced. 
First, 
the constrained interpolation
is used to approximate
the RHS fluxes in 
the re-initialization 
equation (\ref{eq10}).
We demonstrate,
it reduces 
interpolation
errors typical for 
the second-order
accurate discretization 
schemes.
Afterwards, 
to improve
accuracy of advection, 
the semi-analytical
Lagrangian scheme 
for solution 
of the
equation (\ref{eq9})
is put forward.
The new Lagrangian scheme
avoids errors
introduced by 
the second-order flux limiters
and reaches
the second-order
convergence rate
of the interface shape
and curvature.
The constrained interpolation
and new Lagrangian scheme
permit to construct 
re-initialization 
and advection
procedures
with
the convergence
rates
the same 
as the theoretical 
orders of
accuracy of 
the schemes used
to approximate
\Eqs{eq9}{eq10}.

The present 
paper 
is organized
as follows.
In \Sec{sec2}, 
the statistical
model of the non-flat interface 
in the state of phase equilibrium
is postulated and 
its relation to Allen-Cahn
phase field model
is discussed. 
In \Sec{sec3},
the numerical techniques
required to obtain complete
second-order convergence
during advection
and re-initialization
of the interface are
put forward.
Therein, 
performance 
of  the new numerical
schemes for solution
of the statistical 
interface model
equations (\ref{eq9})-(\ref{eq10}) 
is investigated
in several numerical
experiments.
Finally in \Sec{sec4},
conclusions based on
the results obtained 
in the present work
are given.

\section{A statistical model of the non-flat interface in the state of phase equilibrium}
\label{sec2}
%
%
In what follows, 
the derivation 
of the ensemble
averaged equations 
of the sharp interface
disturbed by 
the field of stochastic
forces is shortly revisited. 
Afterwards, 
the relation 
of \Eq{eq10} 
to the modified 
Allen-Cahn 
phase field model
is established.
It is shown,
finding
stationary solution 
of \Eq{eq10}
can be related to
finding the minimum
of the modified Ginzburg-Landau
functional representing
the interfacial  
Helmholtz 
free energy
density 
of the non-flat, 
regularized interface.
%

%
\subsection{Statistical model of the sharp interface disturbed by stochastic velocity field}
\label{sec21}
\begin{figure}[!ht] \nonumber
 \begin{minipage}{.5\textwidth}
  \centering\includegraphics[width=0.85\textwidth,height=0.65\textwidth,angle=0]{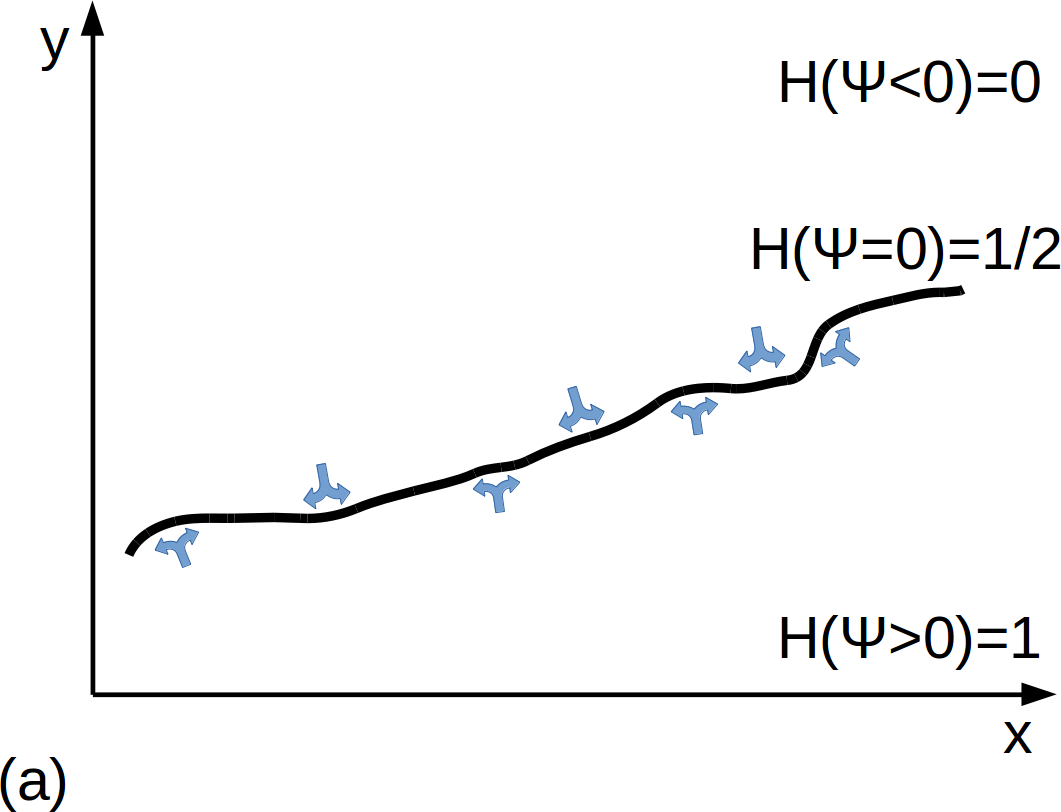}
 \end{minipage}
 \begin{minipage}{.5\textwidth}
  \centering\includegraphics[width=0.85\textwidth,height=0.65\textwidth,angle=0]{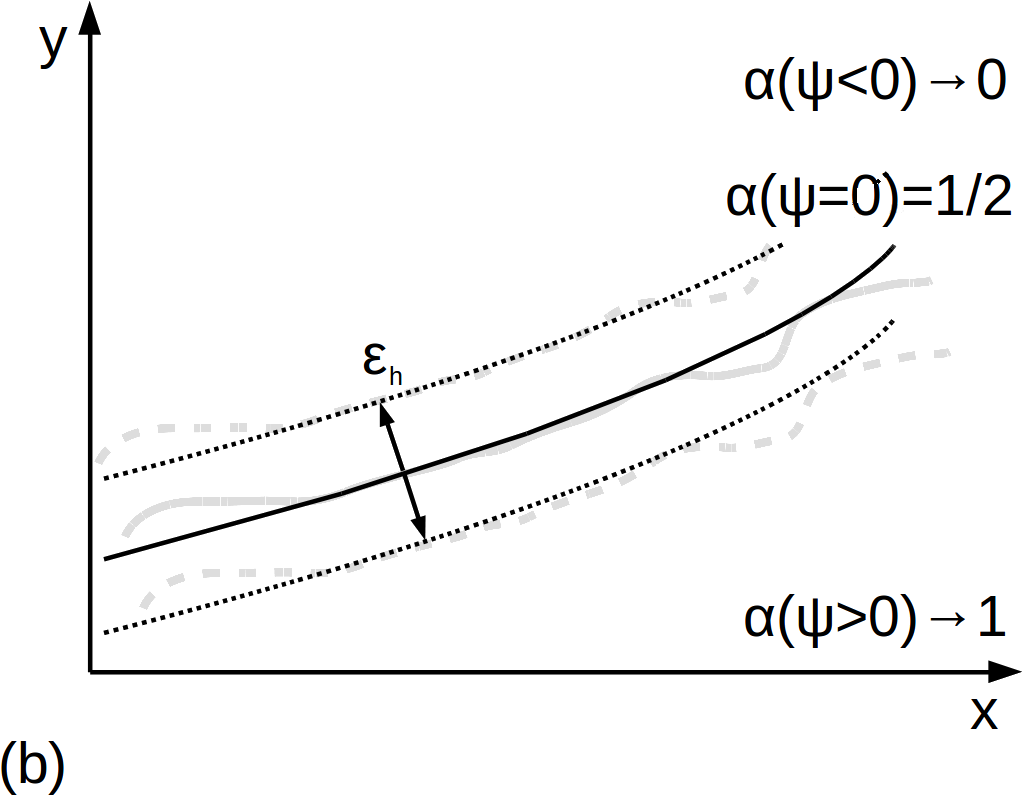}
 \end{minipage}
 \caption{\small{Schematic picture of an instantaneous 
                 sample of the sharp interface 
                 disturbed by the field of stochastic forces   (a)
                 and its ensemble averaged counterpart
                 (b). In the case (a) $\eph\!=\!0$, in the case (b) $\eph > 0$. }}
 \label{fig1}
 \end{figure}

%
At first, 
we consider 
the sharp 
interface 
between
two-phases
given by 
the level-set
of the phase 
indicator 
function 
$H \lr \Psi\!=\!0 \rr\!=\!1/2$,
where $\Psi \lr \bx,t \rr$
is the signed distance 
from the sharp interface.
Let now assume,
the sharp interface
is subjected to
the action 
of the field 
of stochastic 
forces
inducing
its instantaneous
velocity 
$\bW\,[m/s]$
what is schematically
presented in \Fig{fig1}(a).
Depending upon
character of the 
force field and 
chosen time/length scales,
\Eqs{eq9}{eq10}
can be interpreted
as the mesoscopic
or macroscopic
statistical models
of the interface.
In the former case,
fine grained deformation 
of the sharp interface
is caused by 
the thermal fluctuations 
(\cite{vrij1973,aarts2004}). 
Therefore,
$\bw$
in \Eq{eq9} 
describes motion of 
the idealized fluid elements
as the true particles 
of which fluid 
is composed have additional random, 
thermal motion.
In this sense, 
the idealized (macroscopic)
interface 
represented 
by $\alpps$ 
where $\eph \!\sim\! \sqrt{k_B T/\sigma}$
is advected by 
the idealized 
(averaged) velocity 
of fluid elements.

In the 
macroscopic 
interpretation 
of \Eqs{eq9}{eq10}, 
the force field 
disturbing
the sharp
interface with
velocity $\bW$
may be related to 
the instantaneous
velocity of
turbulent 
eddies.
Such interpretation
is possible
because
the characteristic
length scale of 
turbulence is
typically 
much larger
then the 
thickness
of the 
interface
disturbed by
thermal fluctuations.
Here, velocity
$\bw$ in \Eq{eq9}
represents 
the ensemble
averaged velocity
of the turbulent 
fluid phase
and $\alpps$ where $\eph \sim D/C$,
defines 
the intermittency
region, i.e.,
domain where
the sharp interface 
can be found with
non-zero probability
(\cite{hong00,brocchini01a,brocchini01b,waclawczyk2015}).
As these two
pictures are
similar,
it is assumed
a similar
mathematical 
formalism 
describes
the evolution
of the sharp 
interface on
the mesoscopic and
macroscopic scales.

In the present
work, 
the statistical 
description 
of the sharp
interface 
evolving with
velocity $\bW$
in direction 
$\bng \!=\! \nabla \Psi/|\nabla \Psi|$
is introduced.
A sample 
space $\xi\,[m]$ 
of 
the considered  
stochastic 
process are
all allowable
by \Eq{eq1}
values of
the signed
distance function
$\Psi$
recorded at
the given point 
$\bx$ and time $t$.
The ensemble 
average
operator $\av{\cdot}$ 
is defined
as a mean 
over infinitely
many independent
realizations  
or the integral 
over all elements $\xi$
in the sample space
weighted with
their probabilities.
In particular, 
the mean
phase indicator
function
$\av{H \lr \Psi \rr}$
is defined as
\blnm
\be
 \av{H\lr \Psi \rr} = \int_{-\infty}^\infty H \lr \xi \rr f_\Psi \lr \xi, \bx,t  \rr d\xi,
\label{eq12}
\ee
\elnm
where $f_\Psi \lr \xi, \bx, t \rr d\xi$
denotes 
the probability
that $\xi \!<\!  \Psix  \!<\! \xi + d\xi$
and  p.d.f. 
$f_\Psi \lr \xi,\bx,t \rr$
can be obtained 
as $f_\Psi \lr \xi, \bx,t \rr \!=\! \av{\delta \lr  \Psix -\xi \rr}$
(see \cite{,pope88,mwaclawczyk11}).
Further,
in this and
next
sections 
we will argue
$\av{\bW \lr \Psi \rr} \!=\! \bw \lr \psi \rr$ and
$\av{H\lr \Psi \rr} \!=\! \alpps$,
where 
$\bw \lr \psi \rr$
denotes the velocity of the regularized interface,
$\alpps$ is the regularized  Heaviside function
and $\psix$ 
is the signed 
distance
from the regularized
interface  
$\alpha \lr \psi\!=\!0 \rr \!=\! 1/2$.
\blnm
\elnm 

If all
details of 
the sharp 
interface evolution
in \Fig{fig1}(a)
are accounted for,
then $\bw\!=\!\bW$, 
$\psi \!=\! \Psi$,
$\eph\!=\!0$.
Hence,
the Heaviside 
function 
$H \lr \Psi \rr$
is  interpreted 
as the cumulative 
distribution function
(c.d.f.) and
its derivative:
the exact Dirac
delta  function 
$\deltap$,
as the probability 
density function
(p.d.f.)
of finding the
instantaneous position
of the sharp
interface.
\Fig{fig1}(b) 
schematically
shows during
the averaging process
some information
about a fine structure
of the interface
is lost,
it must
be reconstructed 
by the appropriate
model.

Let now
apply averaging
defined by \Eq{eq12}
to \Eq{eq1}.
During 
the ensemble averaging,
$\bW$ 
is decomposed
into the sum 
of its mean $\av{\bW}$
and fluctuation $\bW'$,
hence,
the ensemble 
averaged
\Eq{eq1}
can be written 
in the general 
form as
\blnm
\be
 \pd{\alpha}{t} + \bw \nabla \alpha = - \av{\bW' \nabla H} = - \av{\bW' \! \cdot \! \bng \deltap},
\label{eq13}
\ee
\elnm
where 
$\nabla H \!=\! \deltap \bng$,
$\bng\!=\!\nabla \Psi/|\nabla \Psi|$
and
we  use notation:
$\av{\bW}\!=\!\bw$,
$\av{H} \!=\! \alpha$.
To derive \Eq{eq13}
it is assumed
the fluid phase 
and its vapor 
are incompressible,
leading to the condition
$\nabla \!\cdot\! \bW \!=\! 0$.

The RHS term 
in \Eq{eq13}
represents 
a non-zero 
correlation
between the velocity 
fluctuation
in the direction 
normal
to the sharp
interface $\bW' \! \cdot \! \bng$
and its instantaneous 
position 
indicated by the
Dirac's delta 
function $\deltap$.
This term is closed by
the approach introduced
by \cite{mwaclawczyk11}
for modeling of the interaction
between stratified flows
and turbulence.
Therein, application 
of the eddy diffusivity
model and ensemble 
averaging allows 
to model the RHS 
in \Eq{eq13} 
by the
sum of 
diffusion
and counter
gradient
diffusion,
where
the latter 
term is
closed using
the non-conservative
model.
This
leads to $\alpha \lr \bx,t \rr$
represented by the normal
distribution as it was 
suggested by \cite{brocchini01a,brocchini01b}
in the context of the interfaces
agitated by the turbulent eddies.
An alternative
approach used
by  \cite{waclawczyk2015},
is the conservative
closure of the
counter gradient
diffusion term:
$ - \nabla \cdot \ls C \alpha \lr 1-\alpha  \rr  \bng \rs$,
$\bng\!=\!\nabla \psi/|\nabla \psi|$;
after separation
of advection and 
re-initialization 
in \Eq{eq13}
this latter assumption
allows 
to derive 
\Eq{eq9} and \Eq{eq5},
respectively.

In works 
of \cite{twaclawczyketal14} 
and \cite{waclawczyk2015}, 
the coefficients
$D\lr \bx,t \rr$, 
$C \lr \bx,t \rr$  
in \Eq{eq5}
are related 
to local
properties 
of the ensemble averaged
turbulent velocity field.
In the present 
work, 
we assume 
$D\!=\!C \eph$ 
where $C\!=\!const.$, $\eph\!=\!const.$,
hence,
the discussed 
conservative
closure of 
the counter
gradient diffusion
leading to \Eq{eq5}
results in
the probability 
$\alpps$ defined
in terms
of the logistic
distribution,
where, the c.d.f. $\alpps$
is given by \Eq{eq6}
and p.d.f. 
by $\deltaa/\eph$ in \Eq{eq8}.
The scale parameter
$\eph > 0$ in these equations
is related
to the standard deviation
$d=\eph\pi/\sqrt{3}\,[m]$.
Moreover,
the mapping between
$\alpps-\psio$ given by \Eq{eq7}
is the quantile function
of the logistic distribution
with the expected value
equal to zero, see \cite{bala1992}.

For aforementioned 
reasons,
the zero level-set
of the signed-distance
function $\psi \lr \alpha\!=\!1/2 \rr \!=\! 0$
describes 
the expected 
position of the 
regularized interface and
$\alpha \lr \psi\!=\!0 \rr \!=\!1/2$
indicates the probability with which
one of the two phases sharing
the regularized interface  
can be found.
One notices,
the sum of
probabilities 
of finding 
the phase one
or finding 
the phase two
in every point 
of the considered 
domain:
$\alpha_1 + \alpha_2 = 1$
is a conserved quantity.
This 
picture of
the averaged
or regularized interface 
is schematically presented
in \Fig{fig1}(b).

In \Sec{sec1},
the relation
between 
the SLS and VOF
sharp interface
models
has been discussed,
see description
of Eqs.~(\ref{eq9})-(\ref{eq11}). 
Next,
we will
show that \Eq{eq10}
is the conservative
form of modified Allen-Cahn
phase field model
given by
\Eq{eq4}.
This observation
permits 
the physical
interpretation
of re-initialization 
in the level-set methods
and introduces
the phase field
model with
the order parameter
that is a conserved
quantity.

\subsection{Conservative phase field model of the non-flat, regularized interface}
\label{sec22}
%
In order to show
the model given by \Eqs{eq9}{eq10}
describes the evolution of 
non-flat regularized interface 
in the state of phase equilibrium,
in the present paper 
the relation between
\Eq{eq10} and modified \Eq{eq4}
is established.
To derive it,
we use \Eq{eq8},
and \Eq{eq5} in
the non-conservative
form to arrive at
\blnm
\be
\pd{\alpha}{\tau} = \frac{C}{\eph} \ls
                 \eph^2 \nabla^2{\alpha}
                 - \lr 1 - 2 \alpha \rr\alpha \lr 1-\alpha \rr |\nabla \psi|
                 + \eph \alpha \lr 1-\alpha \rr \kappa \rs,
\label{eq15}
\ee
\elnm
where $\kappa \!=\! - \nabla \cdot \bng$
and  $\bng \!=\! \nabla \psi /\nabla \psi$.
If $\alpps$ is 
given by \Eq{eq1}
then $|\nabla \psio| \!=\! 1$ 
in \Eq{eq15}
and hence, 
the first two terms 
on the RHS of \Eq{eq15}
are identical  
to the RHS terms 
in \Eq{eq4}.
%
%

The Allen-Cahn equation (\ref{eq4})
is  obtained by computation 
of the functional derivative 
of the Ginzburg-Landau functional
representing 
the interfacial density 
of the Helmholtz free energy
\blnm
\be
  F \! \ls \alpha \rs = \int_V \sigma \ls \eph^2 |\nabla \alpha|^2 + f\lr \alpha \rr   \rs dV,
\label{eq16}
\ee
\elnm
where $\sigma$ is a constant
with dimension $[J/m^2]$
(see \cite{allen1979,anderson1998,moelans08,kim14}).
The contribution
from the
term 
accounting for
the interface deformation,
to the best of
this author's knowledge,
is absent in the definitions
of the Ginzburg-Landau functional 
known in the literature
(see e.g. \cite{cahn1958,allen1979,anderson1998,yue2007,brassel11,kim14,pashos2015,fedeli2017}).
We recall 
after  \cite{allen1979}
the original form of
the Ginzburg-Landau functional
given by \Eq{eq16} 
is equivalent to the
\cite{waals1979}
density
based functional
derived  
only for the
flat interfaces.

As a consequence
of \Eqs{eq15}{eq16}
in the present paper
it is proposed to add
the new term 
to the RHS of \Eq{eq16}.
This term, further denoted as
$k \lr \alpha \rr$
has to satisfy
the relation
\blnm
\be
 \int_V \frac{\sigma}{2 \eph} \, \pd{k}{\alpha} dV \deltaup \alpha
= - \int_V  \sigma  \, \alpha \lr 1-\alpha \rr/\eph \kappa\, dV \deltaup \alpha.
\label{eq17}
\ee
\elnm
Next,
we show the presence
of the new term in
the interfacial energy density balance
is essential
to guarantee the state of
phase equilibrium of the
non-flat interface.
As the functional
derivative of $k \lr \alpha \rr$
given by \Eq{eq17},
resembles 
the capillary term:
$\sigma \deltaa/\eph |\nabla \psi|\bng \kappa  = \sigma \nabla \alpha \kappa$
added to the momentum balance
in the one-fluid formulation
exploiting the sharp interface
model,
the new term
may be interpreted as contribution
to the total interfacial energy density
from the energy required
to deform the flat interface.
%
%

The contribution
to interfacial energy density
due to geometrical deformation
of the system is absent
in \Eq{eq16},
for this reason,
the modified Ginzburg-Landau
functional reads
\blnm
\begin{align}
  F_k\!\ls \alpha \rs =
  \int_V \sigma \ls \eph^2 |\nabla \alpha|^2
  + f \lr \alpha \rr
  + \eph k \lr \alpha \rr \rs dV.
\label{eq18}
\end{align}
\elnm
If we assume 
that at the
boundaries 
of the domain
of interest
characterized
by the normal vector $\bn$ 
the condition
$\nabla \alpha \!\cdot\! \bn \!=\! 0$
is satisfied,
the variation
of $F_k\! \ls \alpha \rs$ is
obtained in the form
\blnm
\begin{align}
  \deltaup F_k\! \ls \alpha \rs = \int_V
  \sigma \ls -2 \, \eph^2 \nabla^2 \alpha 
  + 2 \, \alpha \lr 1-\alpha \rr \lr 1-2\alpha \rr
  + \eph \pd{k}{\alpha}  \rs dV \deltaup \alpha = 0,
\label{eq19}
\end{align}
\elnm
as we search
for the minimum of $F_k\!\ls \alpha \rs$
with the respect to $\alpps$.
Since 
the volume integral
in \Eq{eq18}
is calculated over
arbitrary $V$, 
the only way
\Eq{eq19} is equal
to zero, is that
\blnm
\begin{align}
    \sigma \ls \nabla^2 \alpha 
  - \alpha \lr 1-\alpha \rr \lr 1-2\alpha \rr/\eph^2
  - \frac{1}{2\eph} \pd{k}{\alpha} \rs = 0.
\label{eq20}
\end{align}
\elnm
As it was pointed out
above in \Eq{eq17},
\Eq{eq15} predicts 
\blnm
\begin{align}
  \frac{\sigma}{2\eph}\pd{k}{ \alpha} = 
        \sigma \alpha \lr 1-\alpha \rr/\eph \nabla \cdot \lr \frac{\nabla \alpha}{|\nabla \alpha|} \rr,
\label{eq21}
\end{align}
\elnm
where we used
$\kappa = -\nabla \cdot \lr \nabla \alpha/|\nabla \alpha| \rr$.
After rearrangement of terms
in \Eq{eq21} with the help of \Eq{eq8},
using the property of 
the signed distance
function $|\nabla \psio|=1$,
the following formula is obtained
\blnm
\begin{align}
  \frac{\sigma}{2\eph}\pd{k}{\alpha} = \sigma \ls \nabla^2 \alpha - \alpha \lr 1-\alpha \rr \lr 1-2\alpha \rr/\eph^2 \rs.
\label{eq22}
\end{align}
\elnm
Substitution of \Eq{eq22}
into \Eq{eq19} or \Eq{eq20}
leads to the condition of 
the phase equilibrium;
the functional derivative
of the modified Ginzburg-Landau
functional $F_k\!\ls \alpha \rs$
representing the chemical potential,
is equal to zero
\blnm
\begin{align}
  \frac{\deltaup F_k\! \ls \alpha \rs}{\deltaup \alpha} = 0.
\label{eq23}
\end{align}
\elnm
Thus, $F_k\!\ls \alpha \rs$ has
the  extremum when $\alpps$ 
is given by \Eq{eq1} or 
equivalently
$|\nabla \psio| \!=\! 1$.
Later in this paper,
we will argue  using
results of
numerical 
simulations 
\Eq{eq23} provides
the condition
required
for existence
of the
$F_k \! \ls \alpha \rs$
minimum,
up to this moment,
we  assume that
this case is met.
In the following
section it is  shown
how condition
given by \Eq{eq23} 
can be interpreted.
%
%
\subsection{Velocity of  the regularized interface}
\label{sec23}
%
Subsequently, 
it is demonstrated 
that physical 
interpretation
of functions 
$\alpps-\psio$  
and their  
re-initialization 
equation (\ref{eq10}) 
postulated 
in \Sec{sec21} and \Sec{sec22}
is plausible;
using again \Eq{eq8} to eliminate $\nabla \alpha$
from \Eq{eq15}, we arrive at  
\blnm
\be
\pd{\psi}{\tau} = C \lc
                    \lr 1-2\alpha \rr \ls \lr \nabla \psi \rr^2 - |\nabla \psi| \rs
                    + \eph \ls \nabla^2{\psi} - \nabla \!\cdot\! \lr \frac{\nabla \alpha}{|\nabla \alpha|} \rr \rs \rc,
\label{eq24}
\ee
\elnm
where the 
two terms on the RHS
are equal to zero
only when
the mapping
between $\alpps\!-\!\psio$
is possible.
\Eq{eq24} 
may be 
interpreted as
the formula 
for the
normal 
velocity
component
$\bw \lr \psi \rr \cdot  \bng$
of the expected interface position
$\psi \lr \alpha=1/2 \rr = 0$
as $\pdl{\psi}{\tau}\,[m/s]$.
If the interface
is in the state of 
phase equilibrium,
then, the RHS
of \Eq{eq24}
and thus the
normal velocity
component
of the expected
interface position
is equal to zero.
Let us notice,
the first RHS
term acts only
away from the interface,
therefore, when 
$\alpha \lr \psi\!=\!0 \rr\!=\!1/2$
only the second RHS term in \Eq{eq24}
affects the velocity
$\pdl{\psi}{\tau}\,[m/s]$.
This result
agrees with
the prediction
of  \cite{allen1979}
showing the 
normal component
of the  interface 
velocity
is proportional to the interface
curvature.

As in 
the phase field
interface models 
based on \Eq{eq16}
the term $k \lr \alpha \rr$ 
is absent in the
interfacial energy density 
functional,
spontaneous loss or gain of mass
due to non-zero velocity $\bw \lr \psi \rr \!\cdot\! \bng \!=\! \pdl{\psi}{\tau}$
may be the consequence.
Such artificial
phenomenon
is described in details
by \cite{yue2007}
and was confirmed
by  \cite{bao2012}
in the case
of simulations with
the Cahn-Hilliard equation
derived based on the original
form of the Ginzburg-Landau functional
given by \Eq{eq16}.
The main mechanism
of this spurious
phase transition
is the flow
of energy between the first
and the second
RHS terms in \Eq{eq16}.
As it is explained by \cite{yue2007}
``this flow is perfectly permissible within
Cahn-Hilliard framework but would violate
mass conservation for the drop''.
The modified
functional $F_k\!\ls \alpha \rs$
given by \Eq{eq18}
guarantees satisfaction
of the law of mass conservation
as its functional
derivative is always,
exactly equal to zero
if $\alpps$ is given
by \Eq{eq1} and hence
$|\nabla \psio|\!=\!1$.
%

%
\section{Numerical solution of the model equations}
\label{sec3}
%
As it was mentioned
during interpretation 
of \Eq{eq24},
incorrect numerical solution
of \Eq{eq10} can cause
the spurious phase transition
leading to
the artificial
decay or gain of 
mass.
In \Sec{sec3} 
we have shown 
\Eq{eq10} guarantees 
satisfaction of 
the condition 
given by \Eq{eq23}
required for 
the phase equilibrium,
therefore,
the main challenge
for the numerical schemes
is to keep
this balance
of interfacial 
energies
unchanged.

If the interfacial 
energy density is
away from its minimum
because $\alpps$ is not 
given by \Eq{eq1},
for instance due to
artificial deformation caused
by the numerical errors,
the robust numerical scheme
must be able to overcome this
departure from the equilibrium
state and,
after some
re-initialization steps,
assure  satisfaction
of the condition given
by \Eq{eq23}.
This is possible
only if \Eq{eq23}
provides the condition
for the minimum of
the modified energy
functional $F_k \ls \alpha \rs$ given
by \Eq{eq18}.
In the case
\Eq{eq23} provides
the condition for
existence of the maximum 
of $F_k \ls \alpha \rs$,
the divergence
of the numerical solution
would be the expected consequence
of any departure from 
the equilibrium
state.

In this section,
we  discuss
two numerical techniques 
allowing to minimize impact
of the discretization errors on
the numerical solution 
of \Eqs{eq9}{eq10}.
The main
prerequisites for
the remaining part
of this paper are:
the velocity $C\!=\!1\,[m/s]$
and the width of the interface
is $\eph\!=\!\sqrt{K}\Delta x/4\,[m]$, $K=1,2$
are kept constant.
The second-order
accurate finite volume
method is used
for spatial discretization
of \Eqs{eq9}{eq10}.
If it is not stated otherwise,
\Eq{eq9} is advanced
in time using the second
order accurate implicit
Euler scheme (TTL)
(see \cite{peric02,schaefer06}).
The stationary
solution of 
re-initialization
\Eq{eq10}
in time $\tau\,[s]$
is obtained
using the third-order
accurate
TVD Runge-Kutta
method introduced
by \cite{gottlieb98}.

At first,
the constrained
interpolation 
is introduced 
to improve 
accuracy of 
approximation of 
the RHS fluxes 
in \Eq{eq10}.
Afterwards,
in order to avoid
errors introduced by
the second-order
accurate flux limiters,
the new semi-analytical
Lagrangian scheme 
for discretization
of \Eq{eq9} 
is put forward.
The new
schemes for 
the numerical solution
of \Eqs{eq9}{eq10} 
provide means 
to obtain
the third-order
convergence rate
of advection
and 
re-initialization
in time,
and 
second-order
convergence rate
of the interface
shape (volume)
and curvature.
These temporal
and spatial
convergence
rates are 
the same as
theoretical
orders of accuracy
of the 
numerical
schemes
used to 
approximate
\Eqs{eq9}{eq10}.

\subsection{Constrained interpolation}
\label{sec31}
%
%
In what follows we show
how to exploit
relation between 
$\alpps-\psio$
during numerical solution
of \Eq{eq10}.
In particular,
the constrained 
interpolation scheme
(CIS) introduced 
in this section
is used to approximate 
the RHS fluxes in \Eq{eq10},
see also \ref{appB}.
This scheme
permits to use
the steep profile of
the hyperbolic tangent
with the disretization errors
typical for interpolation 
of linear functions.

The idea
of CIS 
is based on 
a simple 
observation.
Since the mapping
between $\alpps-\psio$
is possible,
then, instead
interpolating
$\alpha_f$ 
directly (subscript $f$ denotes
the value interpolated to the face
$f$ of the given control volume $P$),
we can interpolate
$\psi_f$ and 
afterwards
calculate
$\alpha_f \!=\! \alpha \lr \psi_f \rr$
using
the profile
given by 
\Eq{eq6}
as a constraint.
In the case 
of  simplest
linear interpolation
of $\psi_f$
the constrained
interpolation 
is summarized below
\blnm
\begin{align}
 \begin{split}
    \psi_f &\approx \frac{1}{2} \lr \psi_P + \psi_F \rr + \mathcal{O} \lr \Delta x^2\rr, \\
    \alpha_f &= \alpha \lr \psi_f \rr = \frac{1}{1+\exp{\lr -\psi_f/\eph \rr}},
  \label{eq25}
 \end{split}
\end{align}
\elnm
where subscripts $F,f,P$ 
denote the neighbor control volume $F$ 
and face $f$ of the given 
control volume $P$, 
respectively.
One notices, 
no approximation is needed
to compute  $\alpha_f$
in Eqs.~(\ref{eq25}),
the numerical 
error of the constrained
interpolation scheme 
is introduced
only during
the linear interpolation
used to obtain $\psi_f$.
It is almost 
immediately
clear from  Eqs.~(\ref{eq25})
the constrained computation
of $\deltaa \!=\! \alpha \lr 1-\alpha \rr$ 
in \Eq{eq10} should 
be less prone 
to the dispersive
errors introduced
when the linear
interpolation (LIS)
is used directly
to compute
$\alpha_f$. 
In what follows, 
we provide 
quantitative 
arguments
for the above 
statement.

To investigate
properties of the constrained
interpolation, 
re-initialization of the
one-dimensional
regularized Heaviside 
function is studied
in the computational domain 
$\Omega = <\!0,1\!>\,[m]$.
The interface $\Gamma$ 
is located at $x_\Gamma=0.6\,[m]$
to avoid symmetry between the
uniform grid nodes distribution
and re-initialized profile of
hyperbolic tangent;
$\Delta x=1/N_c$ where $N_c=128$ 
is the number of control volumes.
At all boundaries 
of the computational 
domain $\Omega$,
the Neumann 
boundary condition
for $\alpps$ is used.

In order
to compare 
the constrained
interpolation scheme (CIS) with
the second-order accurate
linear interpolation scheme (LIS)
two tests are performed.
In the first test, 
the initial support
of $\alpps$ profile 
is four times smaller 
than in the final profile 
where $\eph=\Delta x$;
in this test the diffusion 
causes widening of
the interface.
\blnm
 \begin{figure}[!ht] \nonumber
 \begin{minipage}{.5\textwidth}
  \includegraphics[width=.75\textwidth,height=1.\textwidth,angle=-90]{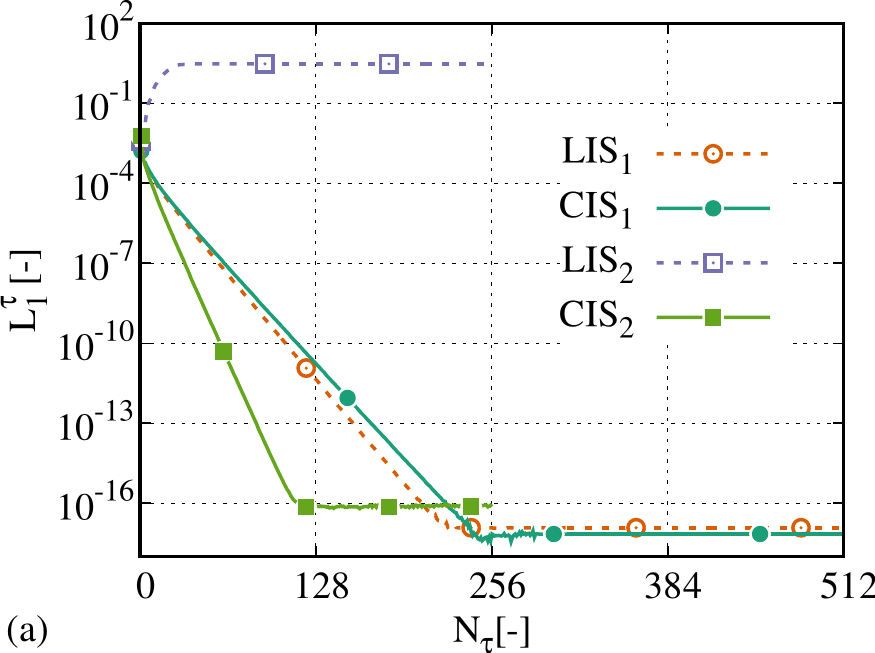}
 \end{minipage}
 \begin{minipage}{.5\textwidth}
  \includegraphics[width=.75\textwidth,height=1.\textwidth,angle=-90]{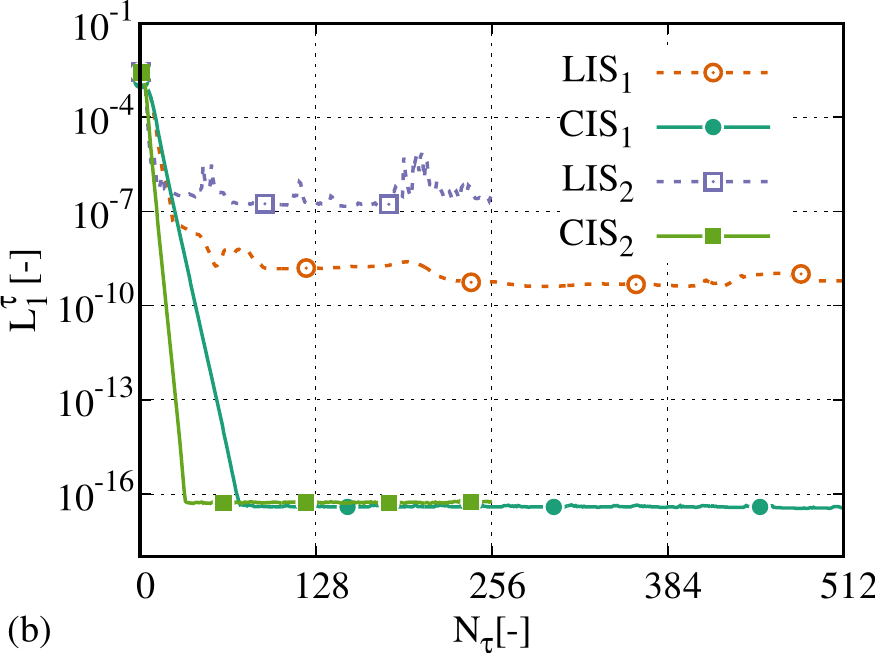}
 \end{minipage}
 \caption{\small{Convergence
                 of the solutions
                 to the diffusion (a),
                 and compression (b)
                 dominated test problems,
                 $L_1^\tau$ denotes norm defined by \Eq{Aeq1}.
                 In both tests
                 the linear interpolation (LIS) or
                 constrained interpolation (CIS)
                 are used,
                 subscripts $1,2$ denote simulation with time steps:
                 $\Delta \tau_1=\Delta x/4$, $\Delta \tau_2=2\Delta \tau_1$.
                 Results 
                 in \Figs{fig3}{fig5}
                 were evaluated
                 at the end of
                 the re-initialization
                 process after $N_\tau=256$ or $N_\tau=512$
                 re-initialization steps, respectively;
                 only the convergent
                 results  are
                 illustrated therein.}}
 \label{fig2}
 \end{figure}
 \elnm
In the second test case,
the initial support
of $\alpps$ profile 
is four times 
wider than 
the final one,
where 
$\eph=\Delta x/4$.
Here, the counter-gradient
diffusion leads to reduction
of the interface thickness. 
To closely inspect
sensitivity
of the solution on
the re-initialization
time step size,
solutions of \Eq{eq10} obtained 
with the two time step sizes
$\Delta \tau_1 = \Delta x/4$
and
$\Delta \tau_2 = 2\Delta \tau_1$
are compared.
In the case 
$\Delta \tau_1$ and $\Delta \tau_2$,
$N_\tau^1 = 512$ and $N_\tau^2 = 256$
re-initialization  steps
are carried out, respectively,
to assure the same total
re-initialization time.

The results
presented in 
\Fig{fig2},
illustrate
convergence of 
the re-initialization equation (\ref{eq10}) 
in time $\tau$
visualized
using the $L_1^\tau$ norm defined 
by \Eq{Aeq1}.
These results were
obtained with
the LIS or CIS interpolation
and two time steps
$\Delta \tau_1,\, \Delta \tau_2$
denoted using
subscripts $1,2$,
respectively.
We note, 
usage of the larger
time step $\Delta \tau_2 = 2\Delta \tau_1$
with LIS leads 
to divergence of
the simulation results 
in the case dominated 
by the diffusion,
see \Fig{fig2}(a).
In the compression 
dominated case, 
the accuracy 
of the solution
obtained with 
the time steps
$\Delta \tau_k$, $k=1,2$ 
and LIS is lower 
than this 
with CIS.
At the same time,
the convergence rates
and levels of
accuracy  
obtained with
CIS seem to be only
slightly 
affected by 
the selected time
step size $\Delta \tau_k$, $k=1,2$,
compare results
in \Fig{fig2}(a)(b).
When CIS is used
in both test cases,
the  machine accuracy
is achieved independent
from the time step size chosen;
compare results obtained with
CIS (solid lines, solid symbols)
and LIS (dashed lines, hollow symbols)
depicted in
\Fig{fig2}(a)-(b).
 \begin{figure}[!ht] \nonumber
 \begin{minipage}{.5\textwidth}
  \includegraphics[width=.75\textwidth,height=1.\textwidth,angle=-90]{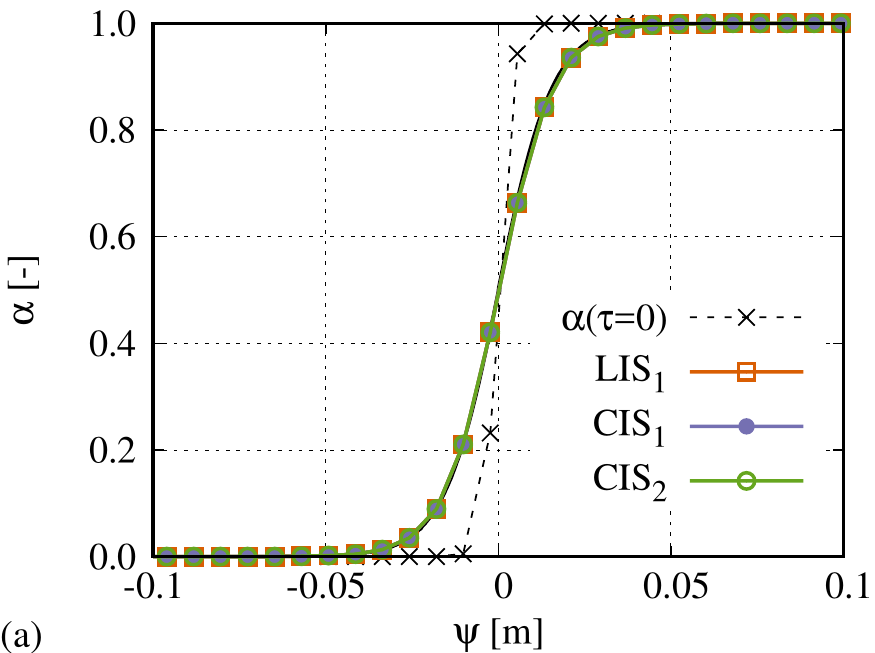}
 \end{minipage}
 \begin{minipage}{.5\textwidth}
  \includegraphics[width=.75\textwidth,height=1.\textwidth,angle=-90]{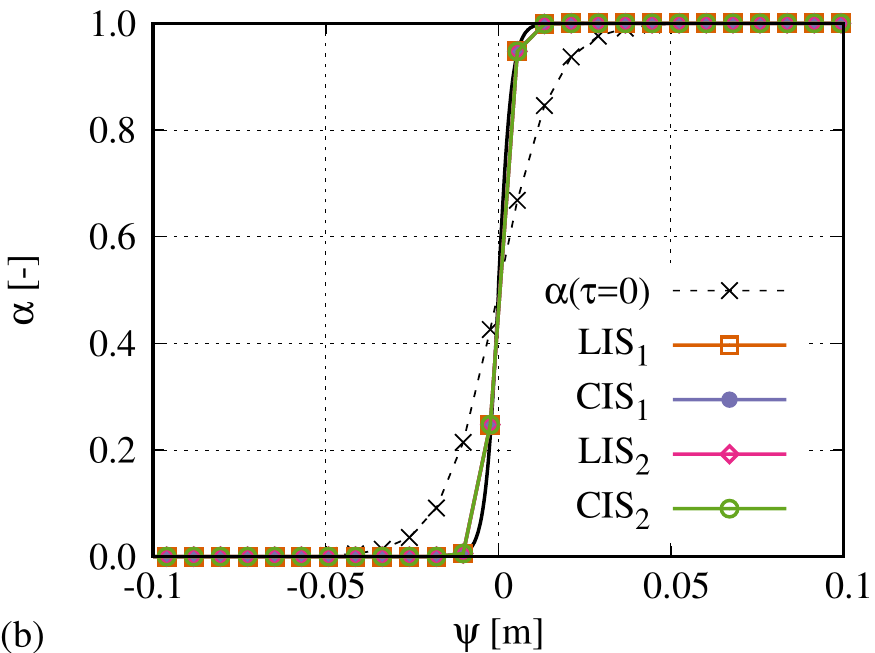}
 \end{minipage}
  \begin{minipage}{.5\textwidth}
  \includegraphics[width=.75\textwidth,height=1.\textwidth,angle=-90]{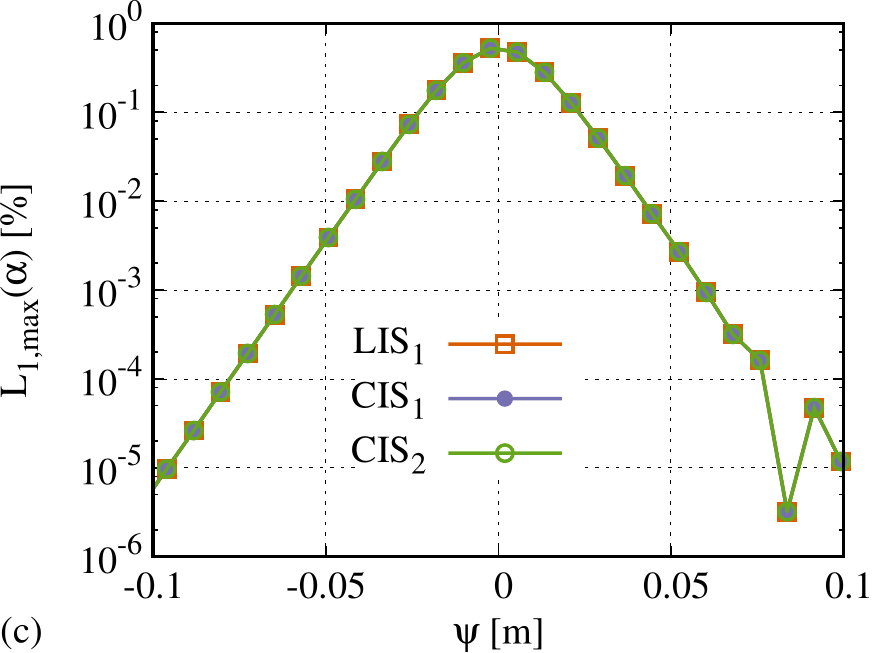}
 \end{minipage}
 \begin{minipage}{.5\textwidth}
  \includegraphics[width=.75\textwidth,height=1.\textwidth,angle=-90]{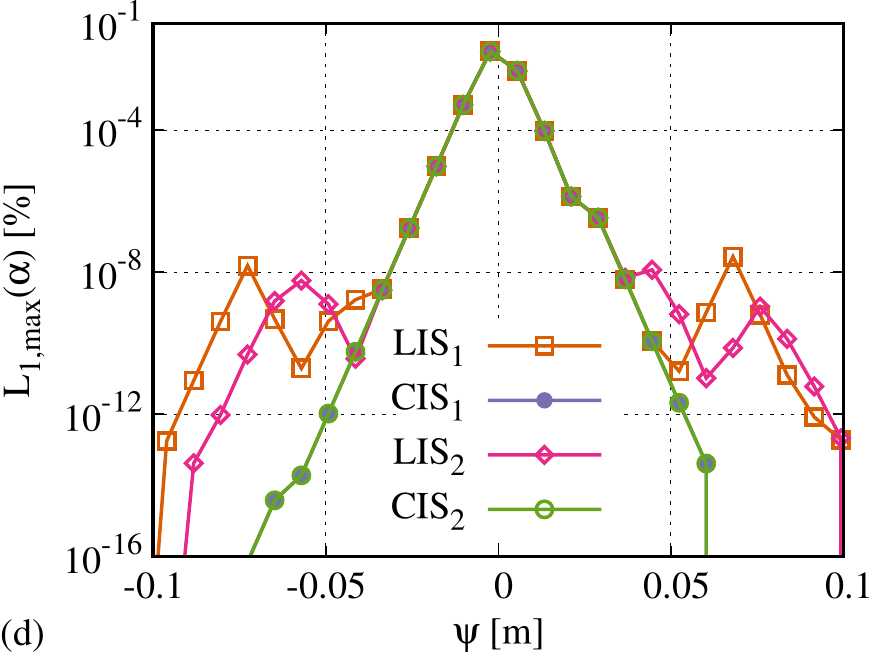}
 \end{minipage}
 \begin{minipage}{.5\textwidth}
  \includegraphics[width=.75\textwidth,height=1.\textwidth,angle=-90]{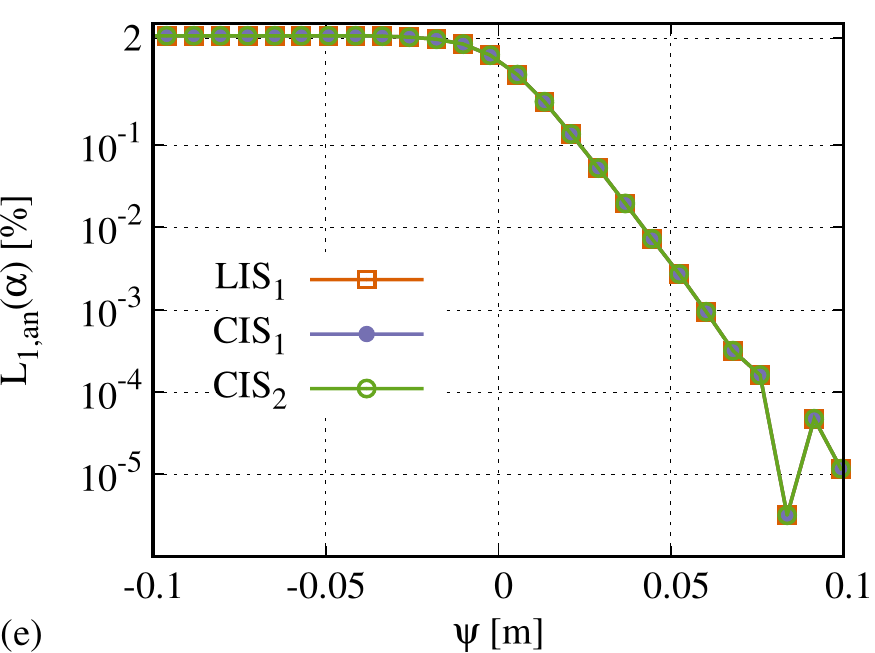}
 \end{minipage}
 \begin{minipage}{.5\textwidth}
  \includegraphics[width=.75\textwidth,height=1.\textwidth,angle=-90]{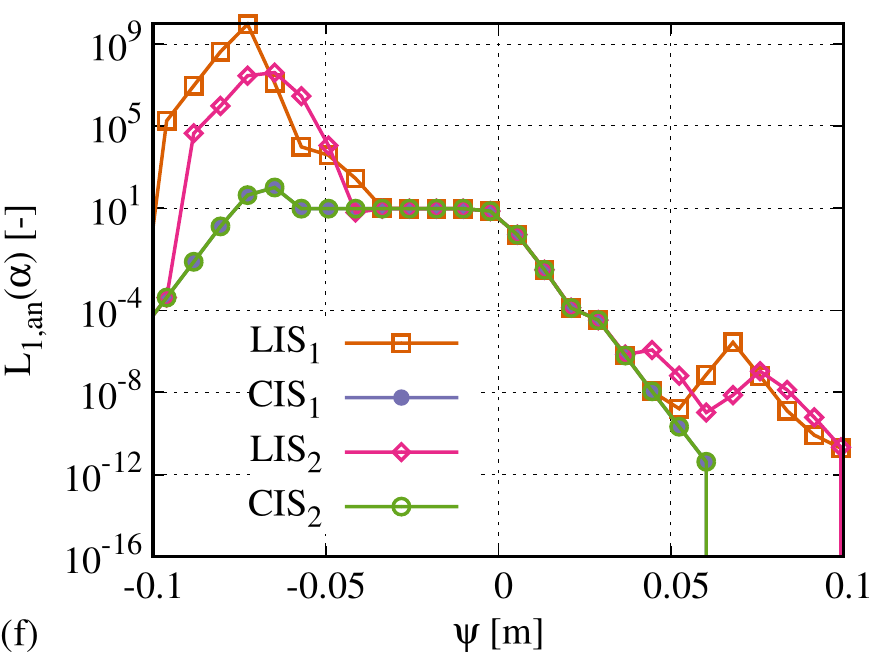}
 \end{minipage}
 \caption{\small{The comparison of
                 $\alpps$ profiles
                 reconstructed
                 with the LIS, CIS interpolation
                 during              
                 the diffusion (left)
                 and compression (right)
                 dominated test cases from \Fig{fig2}.
                 In diagrams (a),(b) 
                 numerical results 
                 are compared
                 with the analytic 
                 solution (black solid line),
                 diagrams (c)-(f) 
                 depict corresponding errors
                 defined by \Eqs{Aeq2}{Aeq3};                          
                 subscripts $1,2$ denote simulations with 
                 $\Delta \tau_1=\Delta x/4$, 
                 $\Delta \tau_2=2\Delta \tau_1$.}}  
 \label{fig3}
 \end{figure}

The convergence of 
the re-initialization process 
with two different interpolation
schemes LIS or CIS
is reflected in distribution 
of the numerical errors
after it is ceased.
These errors 
are obtained by 
the comparison
of the numerical solution
with the known analytical
profiles of $\alpps$ 
given by \Eq{eq6}
and the first components
of its first/second-order
spatial derivatives.
\Figs{fig3}{fig5} 
illustrate 
these profiles
as well as 
the errors of LIS and CIS 
interpolation schemes 
using the $L_{1,an}$ 
and $L_{1,max}$
norms defined by
 \Eqs{Aeq2}{Aeq3}.
All results depicted
in \Figs{fig3}{fig5}
were evaluated
at the end of
the re-initialization
process,
only the convergent
results from \Fig{fig2}
have been illustrated 
therein.
\begin{figure}[!ht] \nonumber
 \begin{minipage}{.5\textwidth}
  \includegraphics[width=.75\textwidth,height=1.\textwidth,angle=-90]{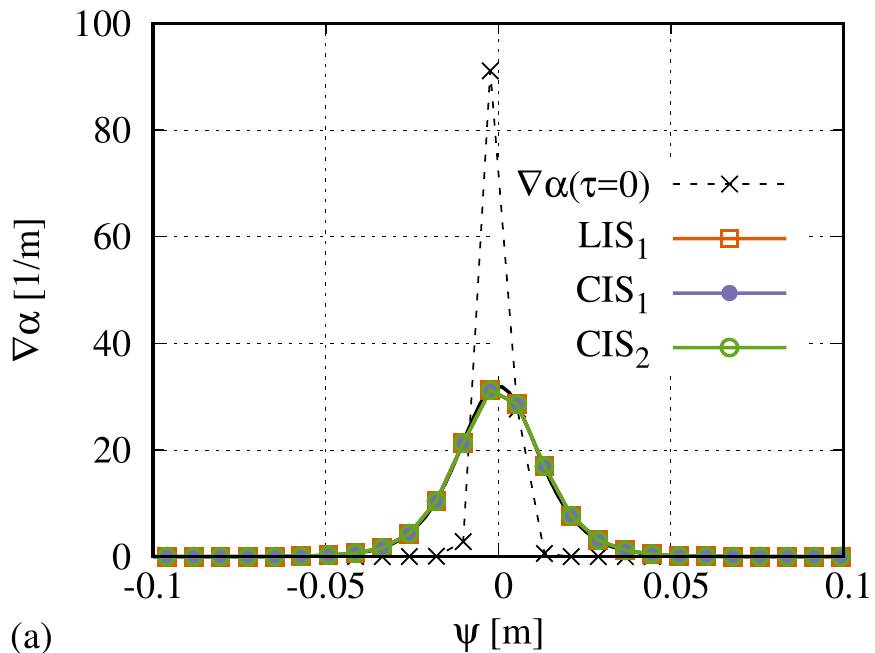}
 \end{minipage}
 \begin{minipage}{.5\textwidth}
  \includegraphics[width=.75\textwidth,height=1.\textwidth,angle=-90]{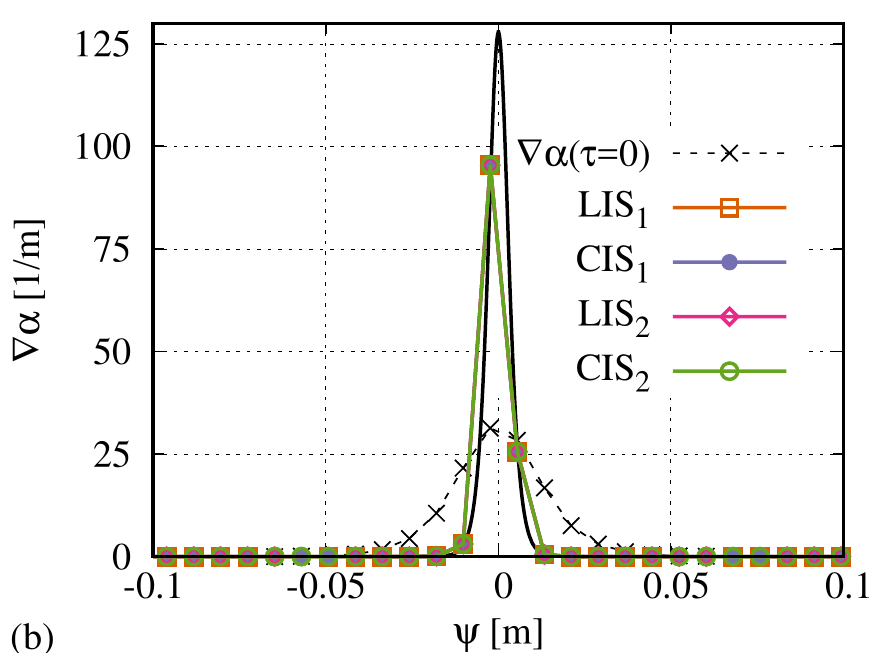}
 \end{minipage}
 \begin{minipage}{.5\textwidth}
  \includegraphics[width=.75\textwidth,height=1.\textwidth,angle=-90]{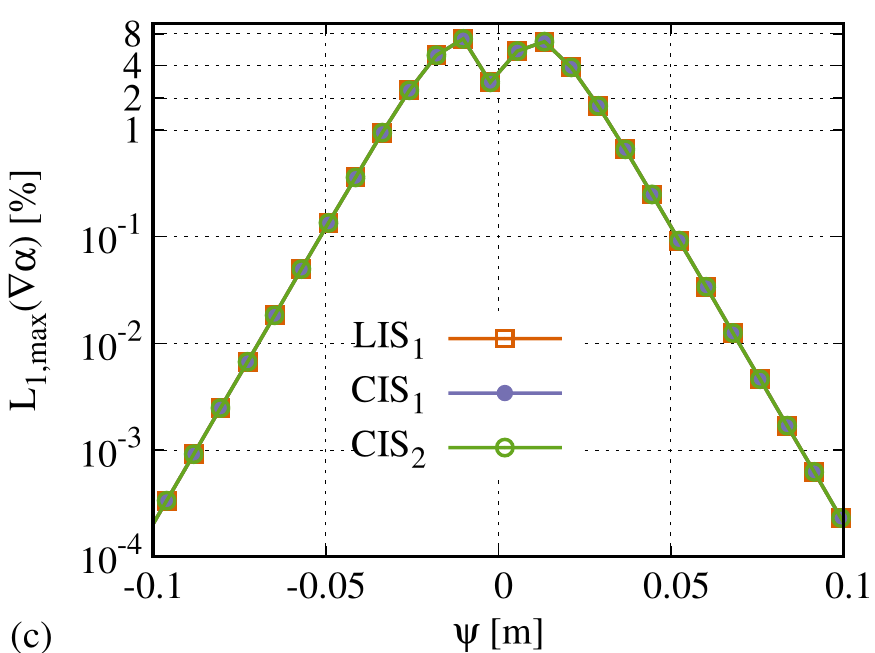}
 \end{minipage}
 \begin{minipage}{.5\textwidth}
  \includegraphics[width=.75\textwidth,height=1.\textwidth,angle=-90]{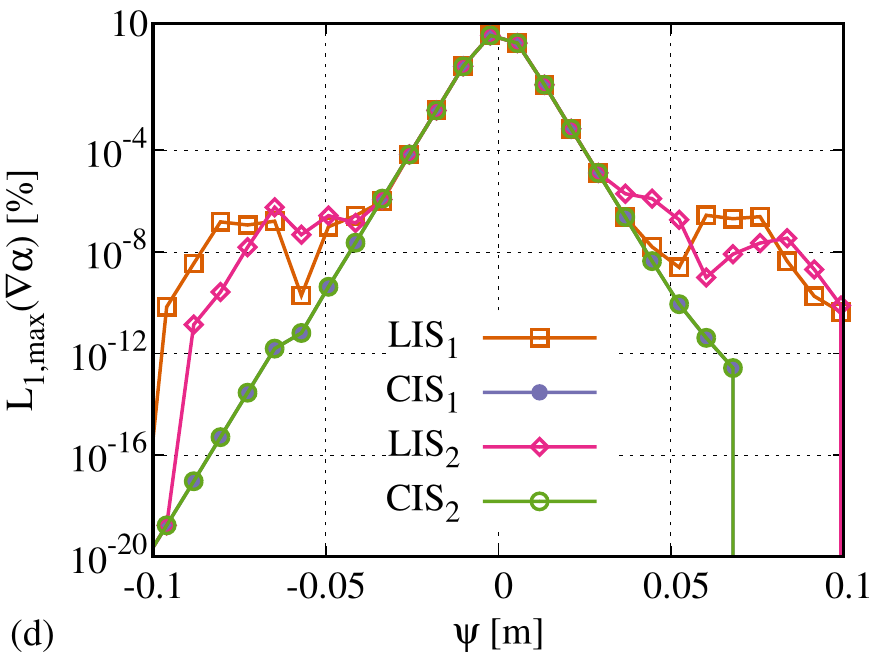}
 \end{minipage}
  \begin{minipage}{.5\textwidth}
  \includegraphics[width=.75\textwidth,height=1.\textwidth,angle=-90]{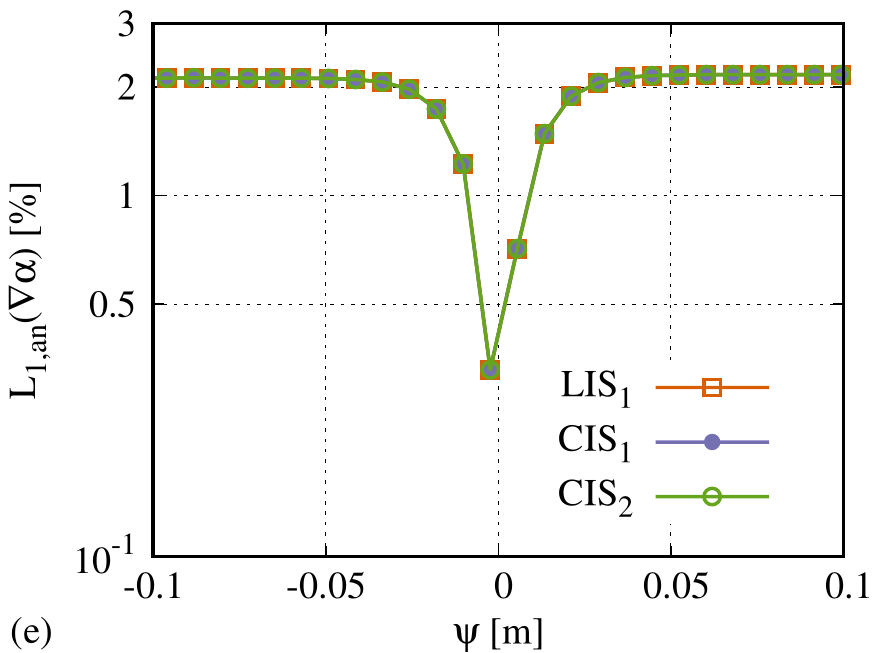}
 \end{minipage}
 \begin{minipage}{.5\textwidth}
  \includegraphics[width=.75\textwidth,height=1.\textwidth,angle=-90]{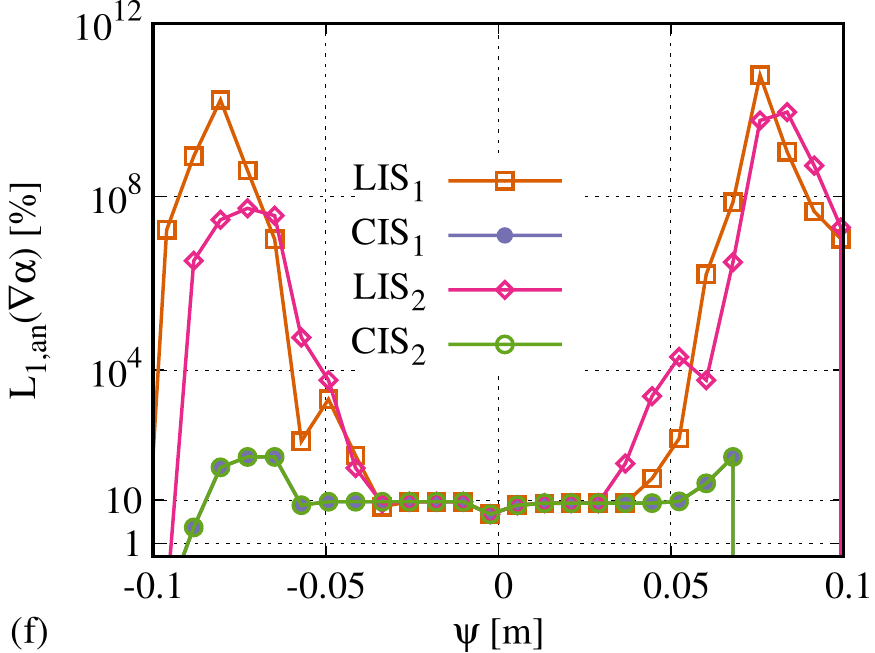}
 \end{minipage}
 \caption{\small{The comparison of the
                 first component of $\nabla \alpps$
                 reconstructed with 
                 the LIS, CIS interpolation
                 during
                 the diffusion (left)
                 and compression (right)
                 dominated test cases from \Fig{fig2}.
                 In diagrams (a),(b) 
                 numerical results 
                 are compared
                 with the analytic 
                 solution (black solid line),
                 diagrams (c)-(f) 
                 depict corresponding errors
                 defined by \Eqs{Aeq2}{Aeq3};                          
                 subscripts $1,2$ denote simulations 
                 with  $\Delta \tau_1=\Delta x/4$, 
                 $\Delta \tau_2=2\Delta \tau_1$.}}  
 \label{fig4}
\end{figure}
 \begin{figure}[!ht] \nonumber
 \begin{minipage}{.5\textwidth}
  \includegraphics[width=.75\textwidth,height=1.\textwidth,angle=-90]{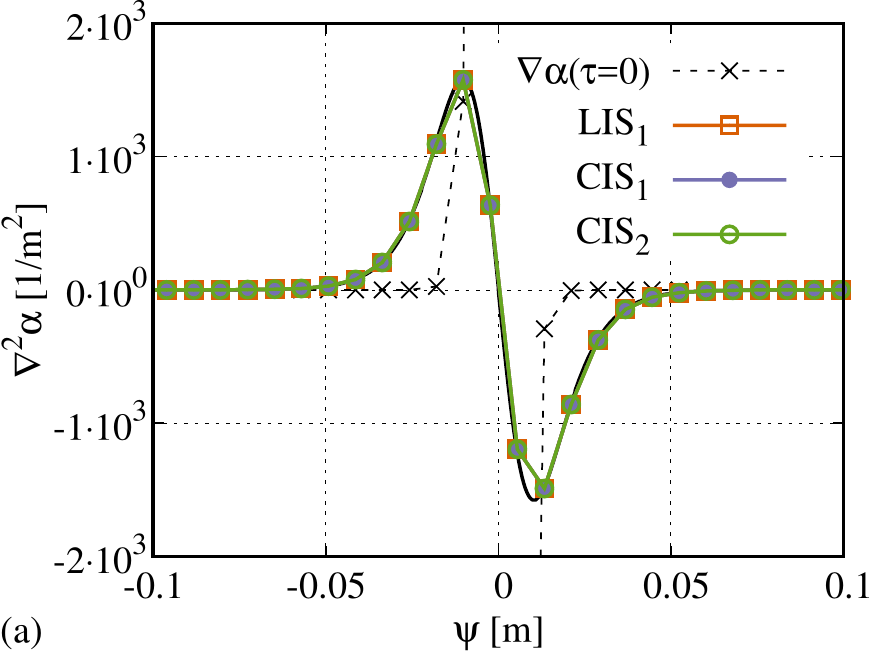}
 \end{minipage}
 \begin{minipage}{.5\textwidth}
  \includegraphics[width=.75\textwidth,height=1.\textwidth,angle=-90]{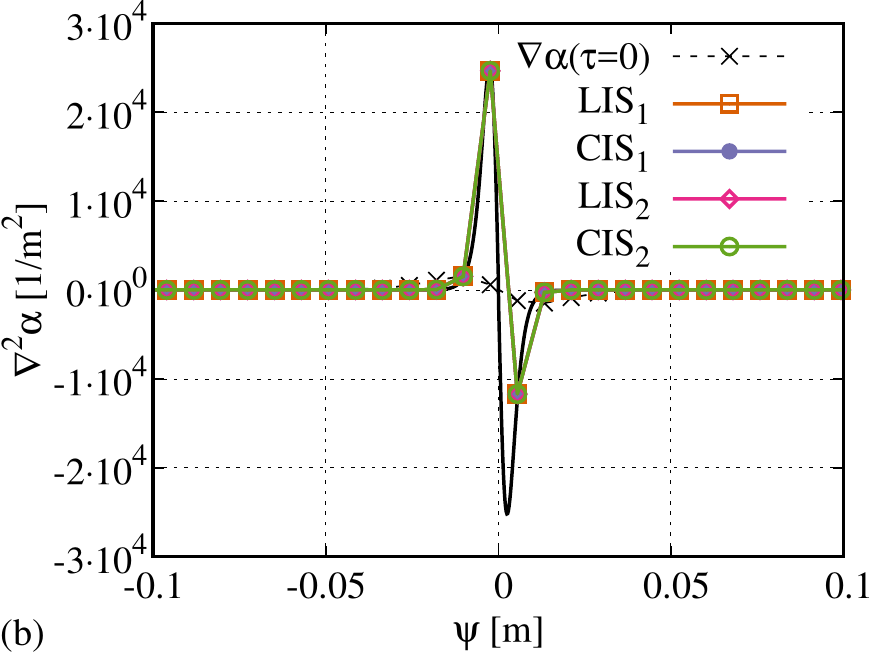}
 \end{minipage}
 \begin{minipage}{.5\textwidth}
  \includegraphics[width=.75\textwidth,height=1.\textwidth,angle=-90]{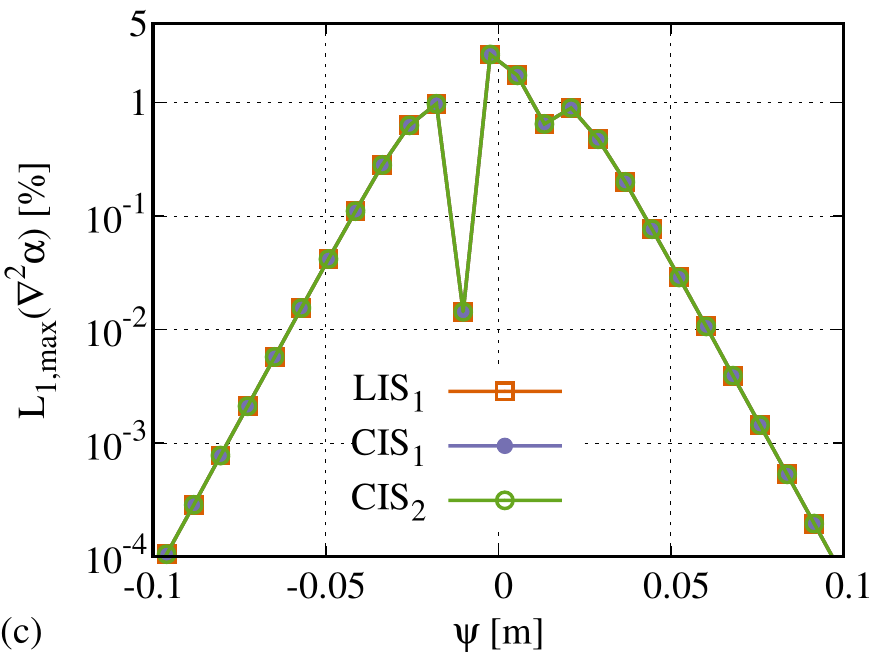}
 \end{minipage}
 \begin{minipage}{.5\textwidth}
  \includegraphics[width=.75\textwidth,height=1.\textwidth,angle=-90]{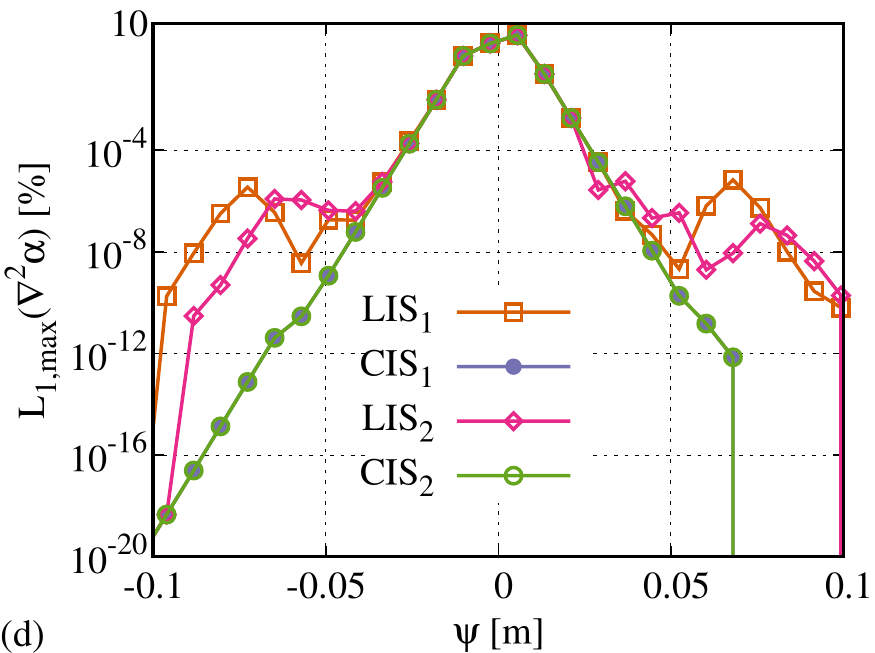}
 \end{minipage}
  \begin{minipage}{.5\textwidth}
  \includegraphics[width=.75\textwidth,height=1.\textwidth,angle=-90]{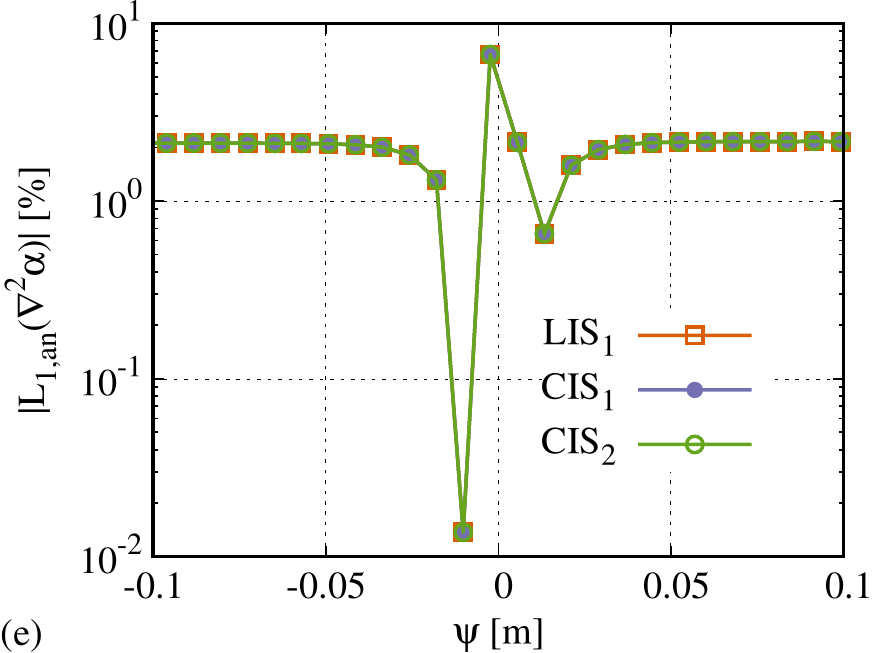}
 \end{minipage}
 \begin{minipage}{.5\textwidth}
  \includegraphics[width=.75\textwidth,height=1.\textwidth,angle=-90]{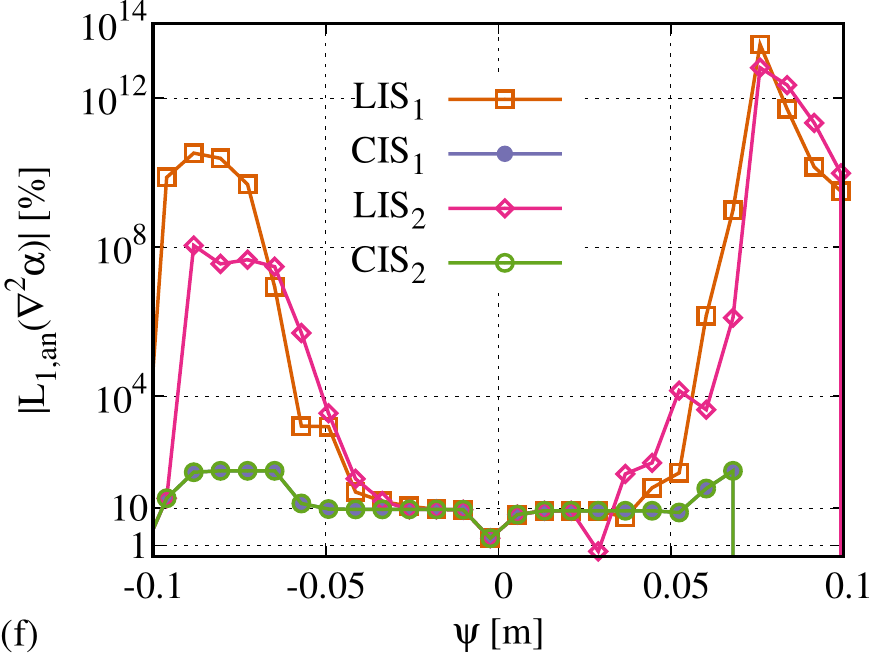}
 \end{minipage}
 \caption{\small{The comparison of
                 the first component of $\nabla^2 \alpps$
                 reconstructed with
                 the LIS, CIS interpolation
                 during
                 the diffusion (left)
                 and compression (right)
                 dominated test cases 
                 from \Fig{fig2}.
                 In diagrams (a),(b) 
                 numerical results 
                 are compared
                 with the analytic 
                 solution (black solid line),
                 diagrams (c)-(f) 
                 depict corresponding errors
                 defined by \Eqs{Aeq2}{Aeq3};                          
                 subscripts $1,2$ denote simulations 
                 with  $\Delta \tau_1=\Delta x/4$, 
                 $\Delta \tau_2=2\Delta \tau_1$.}} 
 \label{fig5}
\end{figure}

In the diffusion
dominated
test case depicted
in the left column 
of \Figs{fig3}{fig5}, 
almost no differences
can be observed
in the distribution
of the errors
when LIS or CIS
are used to reconstruct
$\alpps$,
regardless of
the selected time
step size,
see Figs.~\ref{fig3}(c)(e)--\ref{fig5}(c)(e).
This result is expected 
as the convergence
rates of $L^\tau_1$ norms 
obtained using $\text{LIS}_1$ and $\text{CIS}_1$
are almost identical, 
see \Fig{fig2}(a).

In the 
compression dominated
test case presented
in the right column
of \Figs{fig3}{fig5},
it can be seen 
the errors in 
results obtained
using CIS are
localized in 
the vicinity of
the interface $\psi \lr \alpha =1/2 \rr = 0$
and they do not
exhibit signs of
numerical dispersion
manifested in
the oscillations
of the reconstructed
solutions and
corresponding
errors.
The behavior
of CIS 
errors
is in contrast
with the oscillatory
results obtained
when LIS is used,
compare results 
in Figs.~\ref{fig3}(d)(f)--\ref{fig5}(d)(f).

As profiles in 
Figs.~\ref{fig3}(c)(d)--\ref{fig5}(c)(d) 
are normalized
by maximal 
value of $\alpha$, $\nabla \alpha$ 
and $\nabla^2\alpha$, respectively,
the distributions
of the $L_{1,max}$ norm
is symmetrical 
around the expected
position of the interface 
$\psi \lr \alpha \!=\! 1/2 \rr \!=\! 0$.
In the case 
of normalization
with the exact 
$\alpha$, $\nabla \alpha$ or $\nabla^2 \alpha$  
values
as it is illustrated 
in Figs.~\ref{fig3}(e)(f)--\ref{fig5}(e)(f),
the increment
of the  errors levels
defined by 
the $L_{1,an}$ norm 
away from the
interface $\psi \lr \alpha =1/2 \rr = 0$
is caused by 
the division
of small numbers
$\sim \cO \lr 10^{-12} \rr$
differing by the order 
of magnitude. 

The main difference 
between both interpolation
strategies is the lack 
of numerical solution 
and hence errors
oscillations 
when CIS is used.
Moreover, results 
obtained with CIS
are less sensitive 
to the selected
time step size 
$\Delta \tau_k$, $k=1,2$
as is shown by 
the $L_1^\tau$
recordings depicted
in \Fig{fig2}.
The difference
in the performance 
of both interpolation
schemes is explained 
by usage of  
$\alpps$ profile given
by \Eq{eq6} 
in  CIS  
constraining
values interpolated to
the faces of the given
control volume,
see Eqs.~(\ref{eq25})
and \ref{appB}.

The results presented
in \Fig{fig2} and 
\Figs{fig3}{fig5} clearly
demonstrate  advantages
of CIS over LIS,
therefore,
CIS will 
be preferred
for discretization
of the RHS fluxes
in \Eq{eq10}
in the remaining
part of the
present paper 
where
the results
of simulations 
with advection 
of $\alpps$ and $\psio$
level-set functions
are presented.

\subsection{Lagrangian advection scheme}
\label{sec32}

In this section,
a semi-analytical
Lagrangian
scheme for 
discretization 
of \Eq{eq9}
governing
advection
of $\alpps-\psio$
level-set functions
in the external
velocity field
is put forward.
The main motivation
for its introduction
are  problems
with obtaining
the theoretical
convergence rates
of the re-initialization process
and interface curvature
on gradually
refined grids
during advection
of the solid objects
in the divergence free
velocity fields.
In particular,
when the second-order
accurate 
spatial discretization
and third order
accurate TVD Runge-Kutta
method 
are used alongside
in solution
of the advection 
and re-initialization equations 
(\ref{eq9})-(\ref{eq10}),
respectively.
In the majority 
of works where \Eq{eq9}
is solved using 
the first/second-order
accurate spatial
discretization 
authors consider only
numerical accuracy
(or the convergence rate)
of the advected interface
shape;
plethora
of examples
is available
in the literature
see for instance
works of
\cite{waclawczyk06, waclawczyk08_2, waclawczyk08_3, osher03, olsson05, chiu11, 
      balcazar2014, mccaslin2014, twacl15}.
Moreover, 
it is hard 
to find works
where 
detailed 
information  
about 
the convergence rate 
of re-initialization 
during advection of 
the level-set functions 
$\alpps$ and/or $\psio$
is presented.

In order
to derive
a more accurate
advection scheme,
\Eq{eq8} is used 
again
this time
to obtain
the correct
numerical
solution
of \Eq{eq9}.
In this regard,
the new Lagrangian
advection scheme 
also uses
the profile of $\alpps$
as a constraint
because it is
assumed
\Eq{eq8}
holds after each
re-initialization
cycle.
This assumption
is reasonable
if stationary
solution to \Eq{eq10}
is obtained with the
smallest possible error,
see results
in \Sec{sec31}
and discussion 
in \cite{twacl15}.
After substitution of
\Eq{eq8} into \Eq{eq9}
we arrive at
\blnm
\be
\pd{\alpha}{t} + \frac{\deltaa}{\eph} |\nabla \psi| \bw \cdot \bng = 0.
\label{eq26}
\ee
\elnm
The rearrangement of terms 
in  equation (\ref{eq26}) 
leads to
\blnm
\be
\frac{1}{\alpha \lr 1-\alpha \rr}\pd{\alpha}{t} = -\frac{1}{\eph} |\nabla \psi| \bw \!\cdot\! \bng.
\label{eq27}
\ee
\elnm
The left hand side is now integrated
between $\alpha^n$ and $\alpha^{n+1}$,
whereas the right hand side between
$t^n$ and $t^{n+1}$ to obtain 
\blnm
\be
\ln{\lr \frac{ \alpha}{1-\alpha} \rr} \Biggr|_{\alpha^n}^{\alpha^{n+1}}
= - \frac{1}{\eph} \int_{t^n}^{t^{n+1}}  \!\!\!  |\nabla \psi| \bw \!\cdot\! \bng  dt,
\label{eq28}
\ee
\elnm
where $n,\,n\!+\!1$
denotes old and new time levels,
respectively.
Integration 
given by \Eq{eq28} 
allows 
to derive the
following formula for advancement
of $\alpps-\psio$ in time $t$ which
is given by the formula
\blnm
\be
  \alpha^{n+1} = \frac{\alpha^n \exp{\ls I\lr t^n \rr \rs}}{1-\alpha^n \lr 1- \exp{\ls I\lr t^n \rr \rs} \rr},
\label{eq29}
\ee
\elnm
where the RHS integral in \Eq{eq28}
is denoted as $I(t^n)$.
This integral 
must be approximated
by the appropriate quadrature;
in the present work we adopt the
second-order Adams-Bashforts method 
leading  to 
\blnm
\be
I \lr t^n \rr \approx -\frac{1}{\eph}\ls \frac{3}{2} f\lr t^{n},\psi^{n} \rr
- \frac{1}{2}  f \lr t^{n-1},\psi^{n-1} \rr \rs  \Delta t,
\label{eq30}
\ee
\elnm
where $f=|\nabla \psi| \bng \cdot \bw$.
The semi-analytical, explicit 
scheme given by \Eqs{eq29}{eq30}
is second-order accurate in time
and no spatial discretization
of $\alpps$ is needed as it
exploits Lagrangian 
form of \Eq{eq9}.
A lack of spatial 
disretization
in the Lagrangian scheme
may be an advantage when
compared with
the second-order
TVD MUSCL
controlling
only the slope of
the local solution.
However,
at the same time
the main disadvantage
of the Lagrangian scheme
is its non-conservative and
explicit formulation,
see \Eq{eq26} and
Eqs.~(\ref{eq29})-(\ref{eq30}),
respectively.

Subsequently, 
the properties
of the new Lagrangian scheme
are compared with 
the standard
second-order TVD MUSCL 
used to approximate
convective term
in \Eq{eq9}.
The comparison
is carried out
during
advection of 
the circular interface 
in the divergence
free velocity field
$\bu\!=\!(u_1,u_2)\!=\!V_0/L(y-0.5,0.5-x)$
where $V_0\!=\!1\,[m/s]$ and $L\!=\!1\,[m]$.
In what follows, 
advection
and re-initialization
equations (\ref{eq9})-(\ref{eq10})
are solved alongside to advance
the circular interface
without deformation.
In such case 
we set $\bw=\bu$
in \Eq{eq9} and hence in
\Eqs{eq29}{eq30}.
The present investigations 
are performed in quadratic
domain 
$\Omega = <\!0,1\!>\!\times\!<\!0,1\!>\,[m^2]$  
on four gradually refined grids
$m_k=2^{4+k}\times 2^{4+k},\,k=2,\ldots,5$ 
with the uniform grid nodes 
distribution;
the Neumann boundary condition
is used at all boundaries of 
the computational
domain $\Omega$.
Initially at $t=0$,
the center of the circular
interface
with the radius $R=0.15\,[m]$ 
is located 
at the point
$(x_0,y_0)=(0.65,0.5)\,[m]$.
The time step
size $\Delta t_l$ 
during 
solution of \Eq{eq5}
is chosen to
satisfy three
CFL conditions:
$C\!u_0 \!\approx\! 0.35$,
$C\!u_1 \!\approx\! 2  C\!u_0$,
$C\!u_2 \!\approx\! 4  C\!u_0$,
where
\blnm
\be
 C\!u_l = \sum_{f=1}^{n_b} max \lc \frac{\bu_f \bS_f \Delta t_l}{V_P}, 0 \rc \quad \text{and} \quad l\!=\!0,1,2,
 \label{eq31}
\ee
\elnm
$n_b$ denotes
the number of
neighbor control volumes,
$\bS_f$ is the surface 
of the control volume's $P$ 
face $f$ and $V_P$ is the volume
of the control volume $P$. 
The interface 
width is set to 
$\eph\!=\!\sqrt{2}\Delta x/4$
and
$\Delta \tau \!=\! D/C^2\!=\!\eph$
similarly 
to the
advection tests 
performed by
\cite{twacl15}.
\blnm
 \begin{figure}[!ht] \nonumber
  \begin{minipage}{.425\textwidth}
  \includegraphics[width=1.0\textwidth,height=.75\textwidth,angle=0]{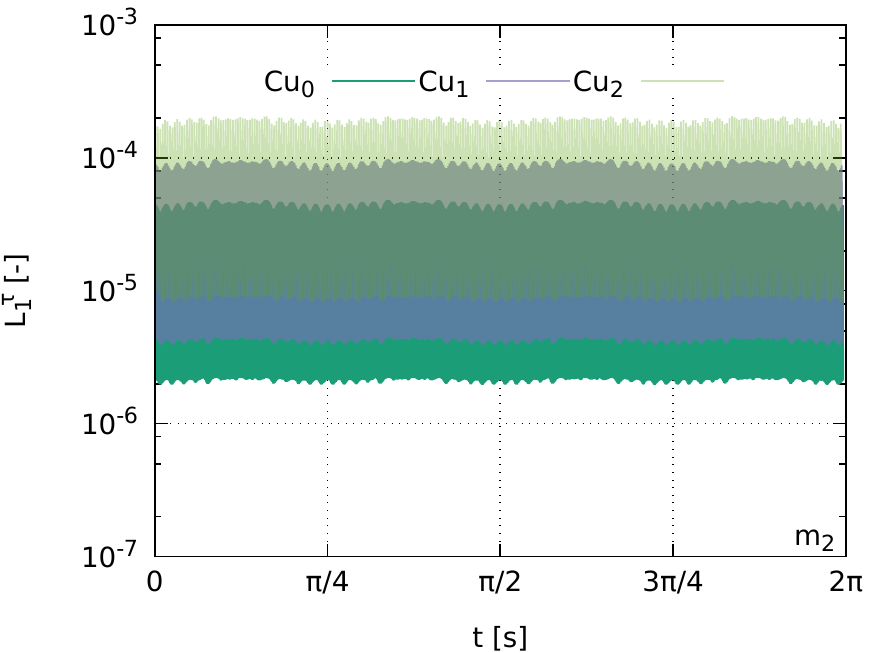}
 \end{minipage}%
 \begin{minipage}{.425\textwidth}
  \includegraphics[width=1.0\textwidth,height=.75\textwidth,angle=0]{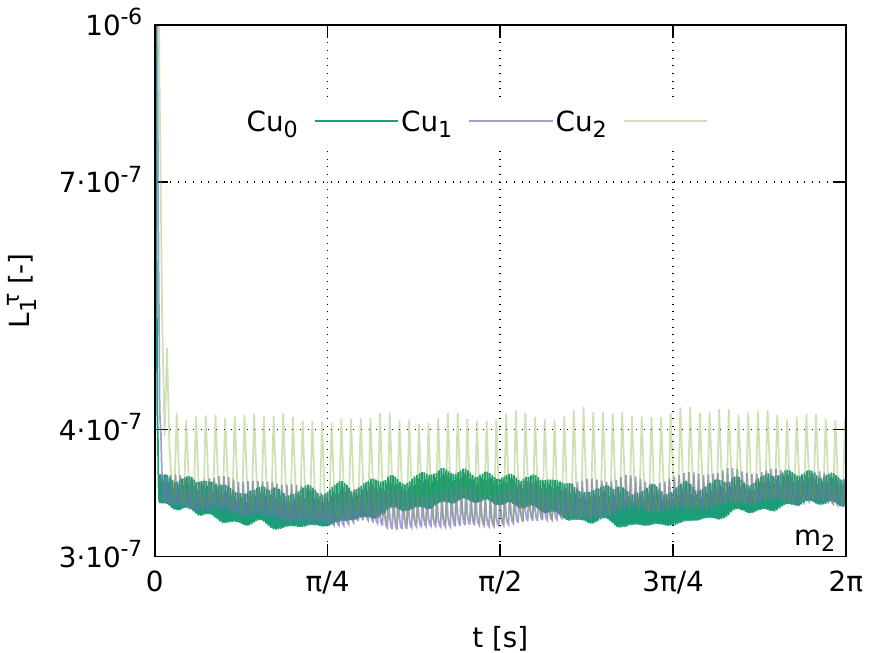}
 \end{minipage}
 \begin{minipage}{.425\textwidth}
  \includegraphics[width=1.0\textwidth,height=.75\textwidth,angle=0]{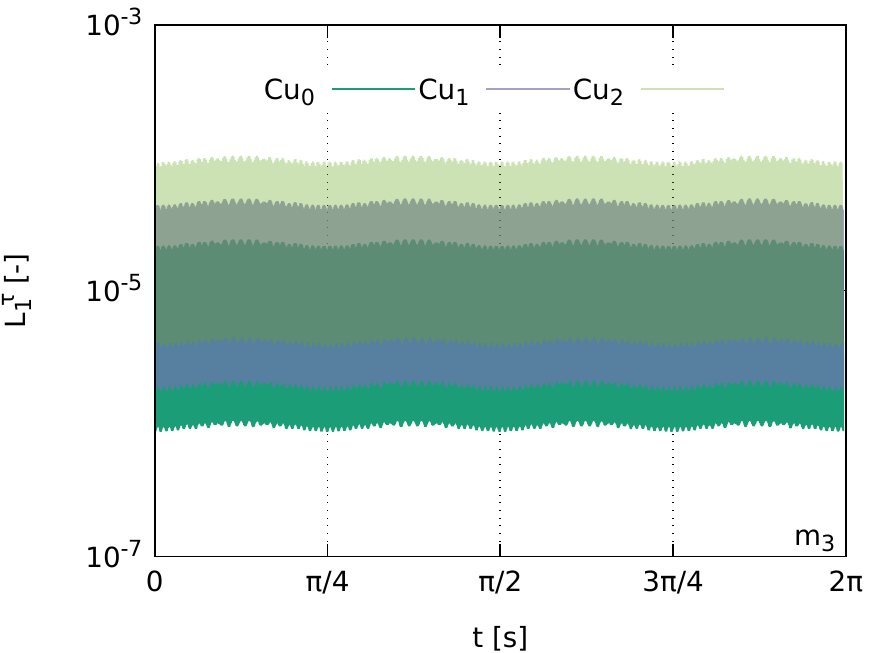}
 \end{minipage}%
 \begin{minipage}{.425\textwidth}
  \includegraphics[width=1.0\textwidth,height=.75\textwidth,angle=0]{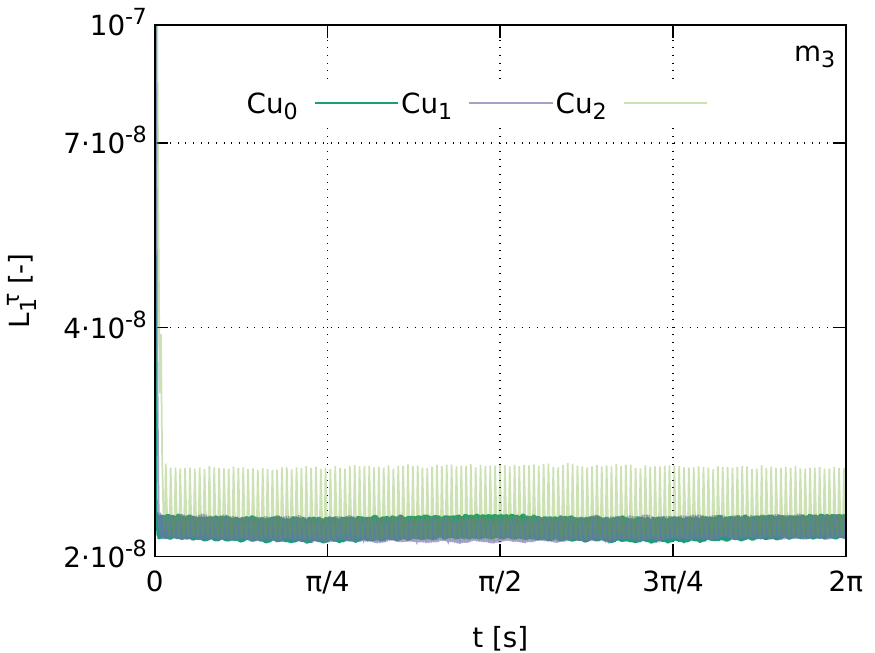}
 \end{minipage}
  \begin{minipage}{.425\textwidth}
  \includegraphics[width=1.0\textwidth,height=.75\textwidth,angle=0]{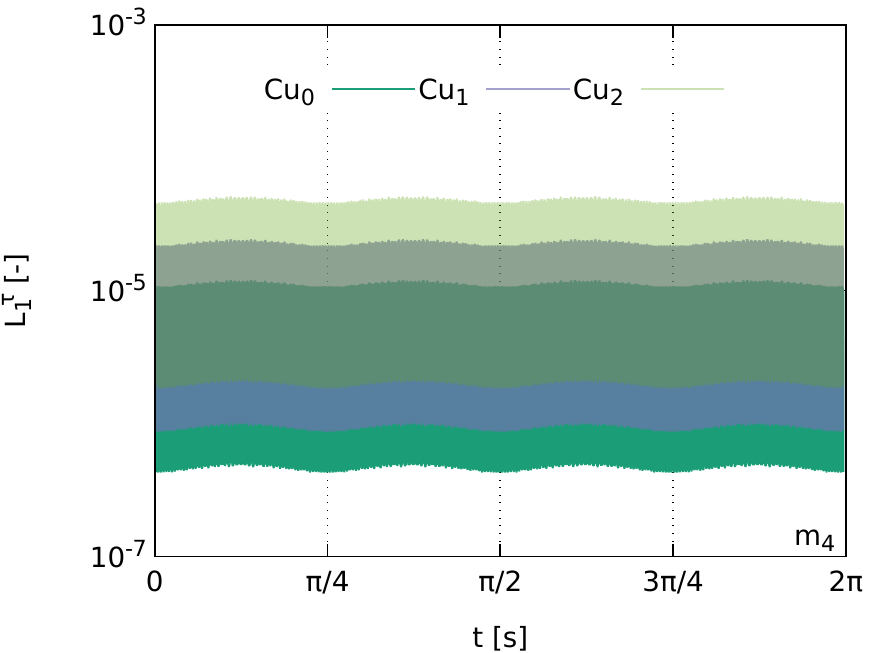}
 \end{minipage}%
 \begin{minipage}{.425\textwidth}
  \includegraphics[width=1.0\textwidth,height=.75\textwidth,angle=0]{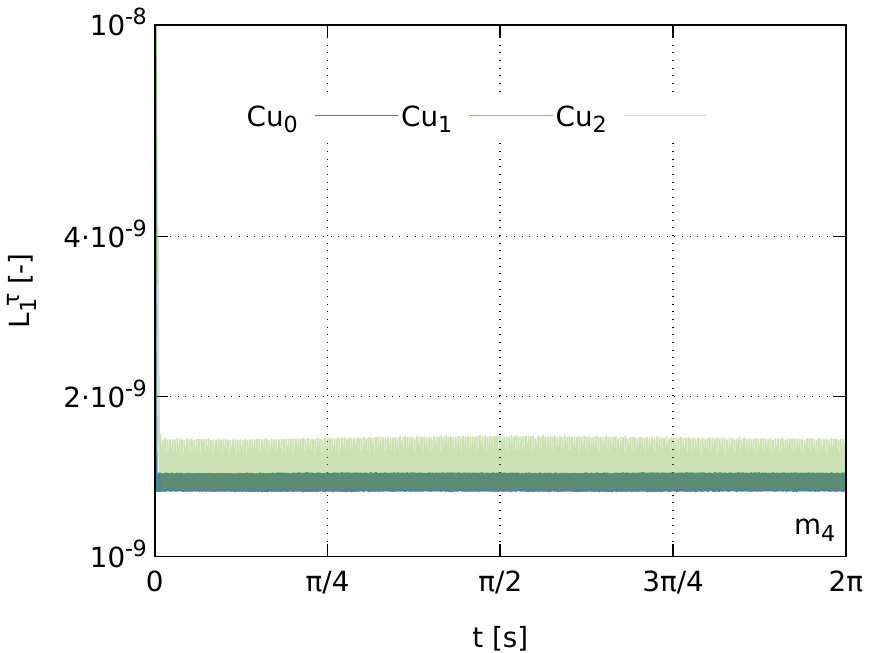}
 \end{minipage}
  \begin{minipage}{.425\textwidth}
  \includegraphics[width=1.0\textwidth,height=.75\textwidth,angle=0]{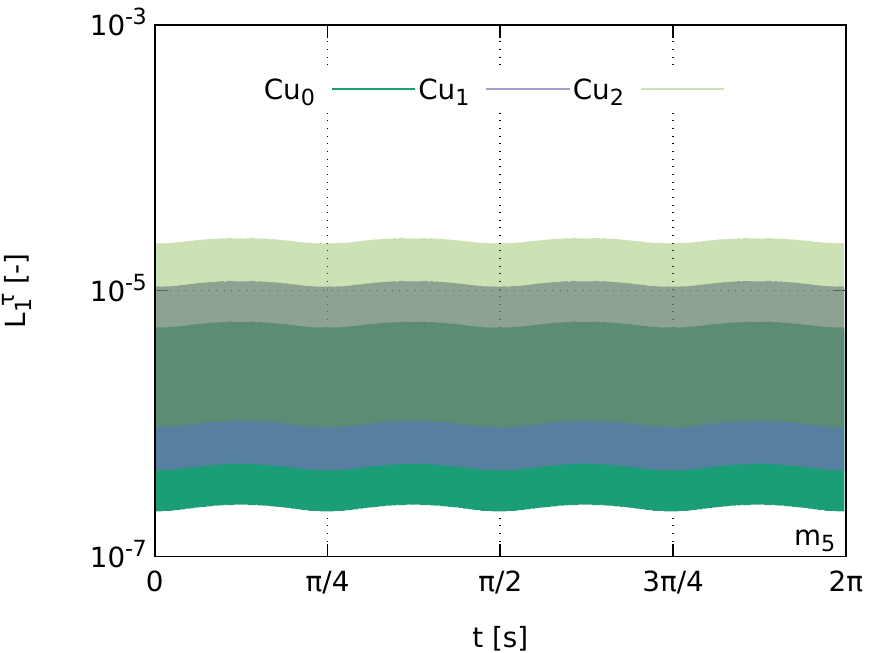}
 \end{minipage}%
 \begin{minipage}{.425\textwidth}
  \includegraphics[width=1.0\textwidth,height=.75\textwidth,angle=0]{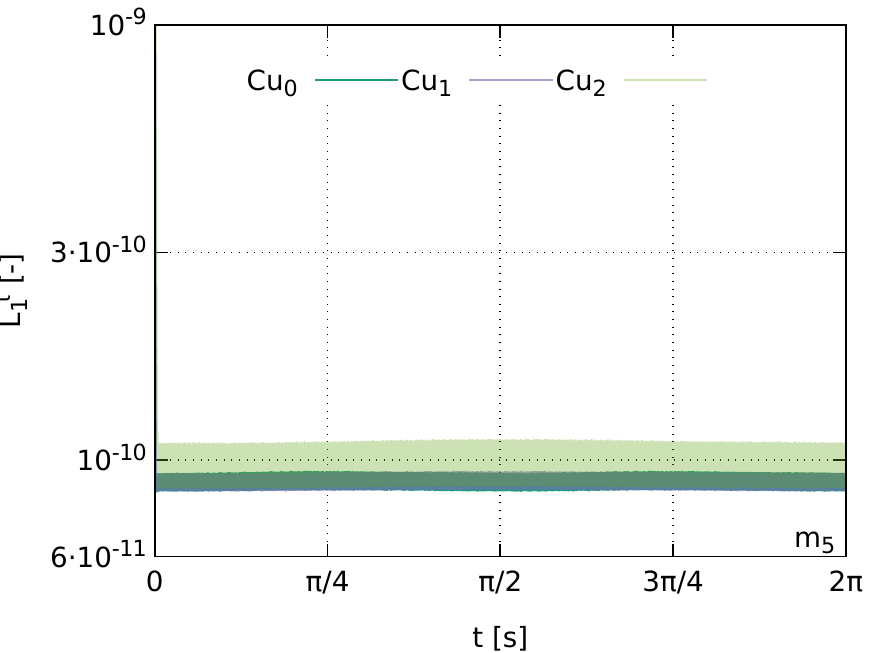}
 \end{minipage} 
 \caption{\small{The convergence
                 of advection and re-initialization \Eqs{eq9}{eq10} 
                 during advection of the circular interface, $L_1^\tau$
                 norm defined by \Eq{Aeq1}
                 where $N_\tau=4$ steps $\Delta \tau$ is plotted
                 after each time step $\Delta t_l$ or $C\!u_l$ number $l\!=\!0,1,2$,
                 on four gradually refined grids
                 $m_k=2^{4+k}\times 2^{4+k},\,k=2,\ldots,5$
                 (from top to bottom).
                 The Eulerian (left) or Lagrangian (right)
                 schemes were used to discretize \Eq{eq9}. }}
 \label{fig6}
 \end{figure}
 \elnm

\Fig{fig6} depicts
histories of joint
convergence
of \Eqs{eq9}{eq10}
during one revolution of
the circular interface.
The joint 
convergence
of the advection
and re-initialization
equations
is illustrated
using the
$L_1^{\tau}$ norm  
plotted
after each time step 
$\Delta t_l$, $l\!=\!0,1,2$
during a single 
re-initialization 
cycle with $N_\tau\!=\!4$ 
steps $\Delta \tau$.
The results
in the left
column of \Fig{fig6}
are obtained
using the implicit
Eulerian scheme
whereas the results
in the right column 
are obtained with
the explicit
Lagrangian scheme
introduced by 
\Eqs{eq29}{eq30}.

One notices,
the diagrams 
in the right
column of \Fig{fig6},
illustrate reduction
of the $L_1^\tau$ norms
by the order of
magnitude on 
each subsequent grid
$m_k$, $k\!=\!2,\ldots,5$,
(figures from top to bottom)
indicating convergence
of the re-initialization
process with the gradual 
grid refinement.
In this case,
the influence of time
step size $\Delta t_l$, $l\!=\!0,1,2$
on the convergence rate
and error level is minor.
The influence
of the time
step size $\Delta t_l$
on the $L_1^\tau$ norms levels
is more
evident 
in the case 
when Eulerian scheme
is used,
compare
convergence
recordings
in the left column
of \Fig{fig6}.
Unlike 
in the case 
of Lagrangian scheme,
convergence of the solution
to \Eqs{eq9}{eq10}
with the mesh refinement
is disputable
when the Eulerian scheme 
is used to
advance \Eq{eq9}
in time $t$.
Some reduction
in the error levels
obtained for different
meshes $m_k$, $k\!=\!2,\ldots,5$
can be observed in \Fig{fig6}(left),
however it does not display
the expected second-order
accuracy.
We note, in the  case 
of Lagrangian scheme
at the beginning
of advection
several iterations
are needed to achieve
constant levels of
convergence, see
for example 
\Fig{fig6}(right)
for grid $m_2$.
Hence,
the net spatial 
and temporal 
discretization 
error introduced by
the Lagrangian scheme
in \Eq{eq9}  
is further
reduced 
by the
third-order
TVD Runge-Kutta 
and CIS
schemes used 
in discretization
of \Eq{eq10}.
This is in 
contrast to
the results
obtained using
the Eulerian scheme,
where the
re-initialization
step does
not reduce 
errors
introduced 
during
advection.
The one 
re-initialization
cycle with 
$N_\tau=4$ steps 
$\Delta \tau$
reduces this error
by one order
of magnitude
but at the beginning
of the new re-initialization
cycle the error returns back
to its previous levels, see 
left column
in \Fig{fig6}.
This 
behavior can
be attributed 
to the errors
introduced by the
TVD MUSCL advection
scheme deforming 
the interface shape
and is the main 
cause
of much slower 
convergence
with the gradual 
mesh refinement
in the case of 
Eulerian scheme.
In the case 
of the Lagrangian
scheme, 
the error variation
during a one re-initialization cycle 
on the single time step $\Delta t$ 
remains almost constant. 
This statement 
is true for all  
$C\!u_l$, $l\!=\!0,1,2$;
we emphasize, 
exactly 
the same 
discretization
of \Eq{eq10} is used
when the Eulerian or
Lagrangian schemes
are used to approximate
\Eq{eq9}.
\blnm
 \begin{figure}[h!] \nonumber
 \begin{minipage}{.5\textwidth}
  \includegraphics[width=1.\textwidth,height=1.\textwidth,angle=0]{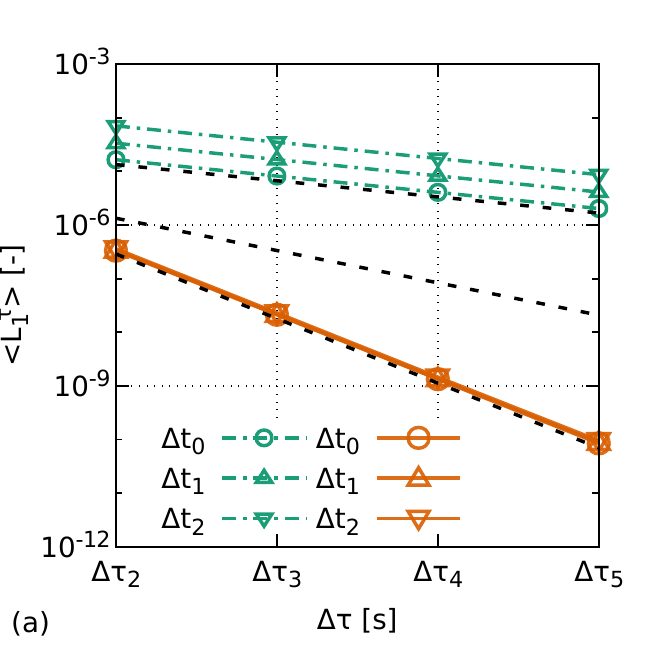}
 \end{minipage}
 \begin{minipage}{.5\textwidth}
  \includegraphics[width=1.\textwidth,height=1.\textwidth,angle=0]{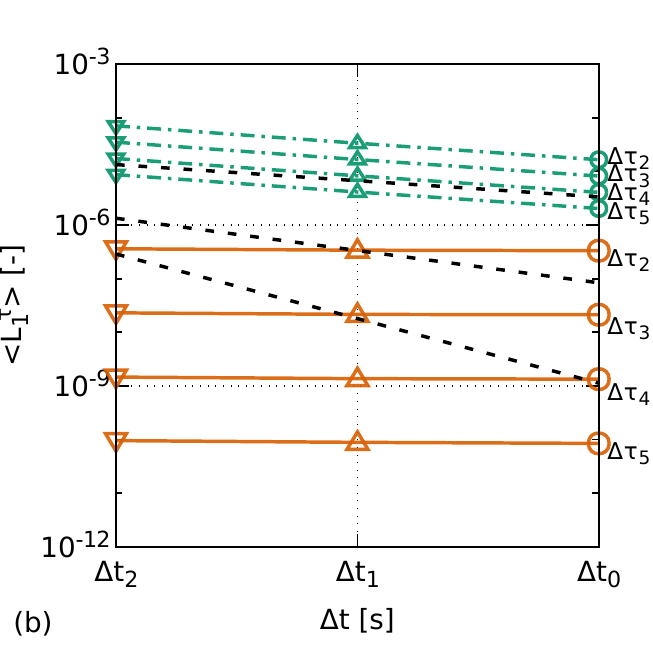}
 \end{minipage}
 \caption{\small{The convergence rate of averaged 
                 $L_1^\tau$ norms from \Fig{fig6}, see \Eq{Aeq4}.
                 The  diagrams (a),(b) 
                 present the same results
                 and illustrate the joint convergence rate 
                 of  advection and re-initialization 
                 equations (\ref{eq9})-(\ref{eq10})
                 during revolution of circular interface 
                 using the Eulerian (green dashed-dotted lines) 
                 and Lagrangian (orange solid lines) schemes
                 with three  $C\!u_l$, i.e.,
                 $\Delta t_l$ time steps $l\!=\!0,1,2$ 
                 on four  grids, i.e., 
                 the four $\Delta \tau_k = \sqrt{2}\Delta x_k/4$,
                 $k=2,\ldots,5$
                 re-initialization time steps.  
                 The dashed-black lines depict
                 slopes of
                 the first, 
                 second and third order
                 convergence rates, respectively.}}
 \label{fig7}
 \end{figure}
\elnm 

To investigate 
in more details 
convergence rates 
illustrated 
in \Fig{fig6}, 
the $L_1^\tau$ norms
depicted in this
figure are averaged 
in times $t$ 
and $\tau$ revealing
information about 
the joint 
convergence rate
of the advection and
re-initialization
equations (\ref{eq9})-(\ref{eq10}),
see \Eq{Aeq4} 
explaining how
$\av{L_1^\tau}$ 
in \Fig{fig7} 
is calculated.
The results
in \Fig{fig7}(a)(b) 
obtained with 
the Eulerian scheme
(green dashed-dotted lines)
show the first order convergence rate
with respect to $\Delta \tau_k$ and $\Delta t_l$
whereas the results obtained with the
new Lagrangian scheme (orange solid lines)
 reveal
the third-order convergence 
rate with regard to  $\Delta \tau_l$ 
and they remain
almost constant
with regard to time 
$\Delta t_l$.
In fact  
$\av{L_1^\tau}$ 
decreases slightly
for different $\Delta t_l$,
$l\!=\!0,1,2$ 
as it can be deduced 
from 
the right 
column
in \Fig{fig6}. 
%
%
%
\Fig{fig7}(b)
shows that
re-initialization
dominates
the convergence rate
of \Eqs{eq9}{eq10}
in the time domain
when the Lagrangian
scheme is used.
Hence,
when 
CIS 
and 
the new  
Lagrangian scheme
are used together
to solve \Eqs{eq9}{eq10}, 
the  convergence rate  
of advection and 
re-initialization 
is the same as 
the theoretical 
order of accuracy 
of the TVD Runge-Kutta  
scheme used 
to integrate \Eq{eq10} 
in time $\tau$,
this result 
is related
to the definition of 
$\Delta \tau_k \!=\! \eph \!=\! \sqrt{2} \Delta x_k/4$,
$k\!=\!2,\ldots,5$.
From this
comparison
it may be deduced 
the Lagrangian
scheme does not
introduce additional
disturbances 
to the shape
of the transported
interface
as its
the case with
its Eulerian
counterpart.
Hence,
re-initialization
governs 
temporal
and spatial
convergence
when \Eq{eq9}
is dicretized
using Eqs.~(\ref{eq24})
and Eqs.~(\ref{eq29})-(\ref{eq30}).

\begin{figure}[!ht] \nonumber
 \begin{minipage}{.5\textwidth}
  \includegraphics[width=1.\textwidth,height=.75\textwidth,angle=0]{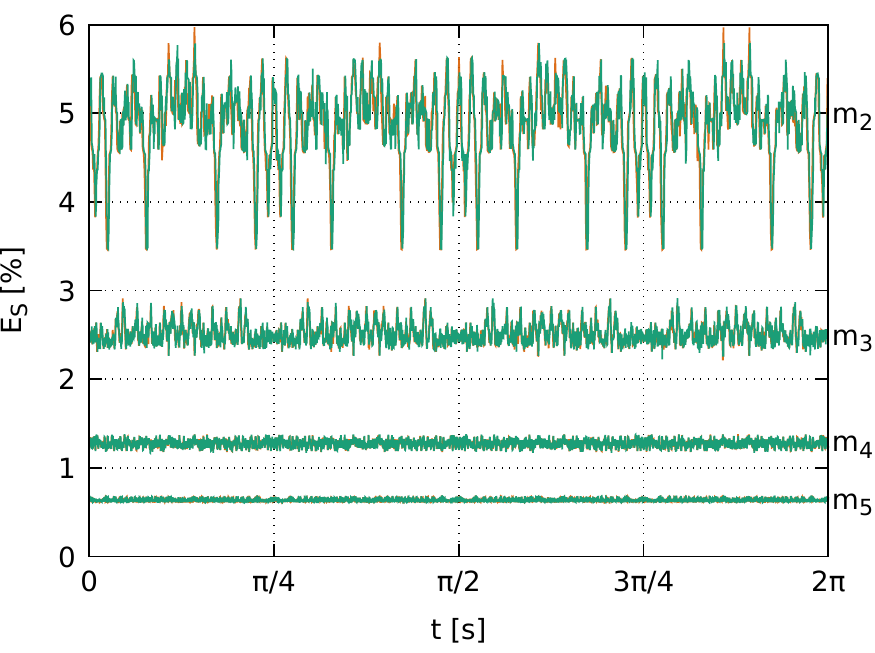}
 \end{minipage}%
 \begin{minipage}{.5\textwidth}
  \includegraphics[width=1.\textwidth,height=.75\textwidth,angle=0]{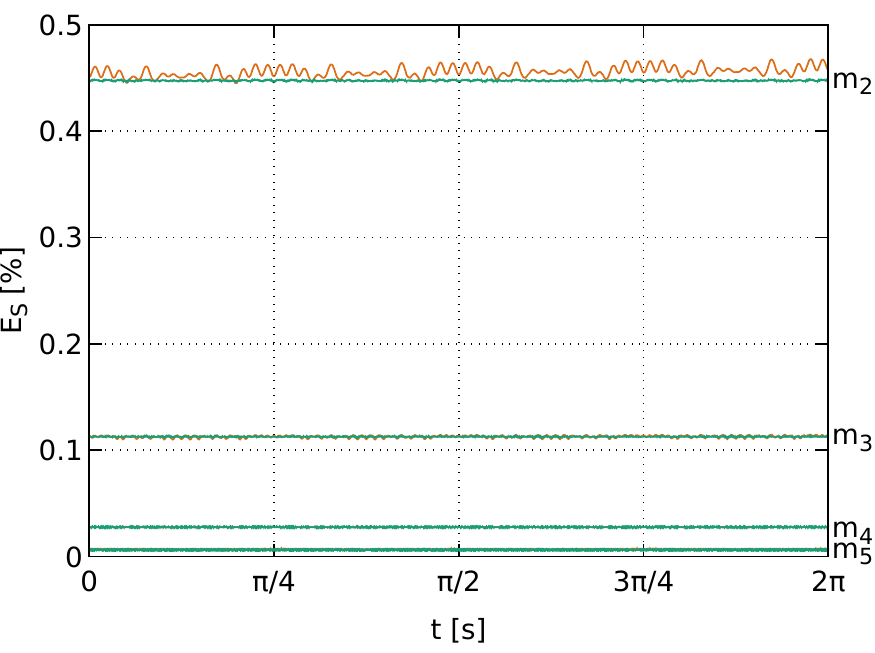}
 \end{minipage}
 \begin{minipage}{.5\textwidth}
  \includegraphics[width=1.\textwidth,height=.75\textwidth,angle=0]{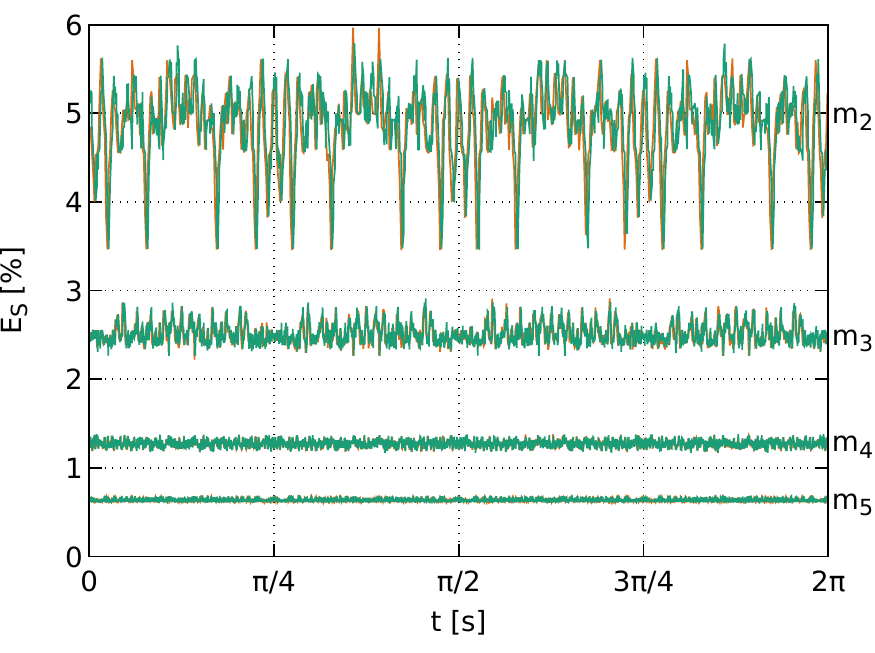}
 \end{minipage}%
 \begin{minipage}{.5\textwidth}
  \includegraphics[width=1.\textwidth,height=.75\textwidth,angle=0]{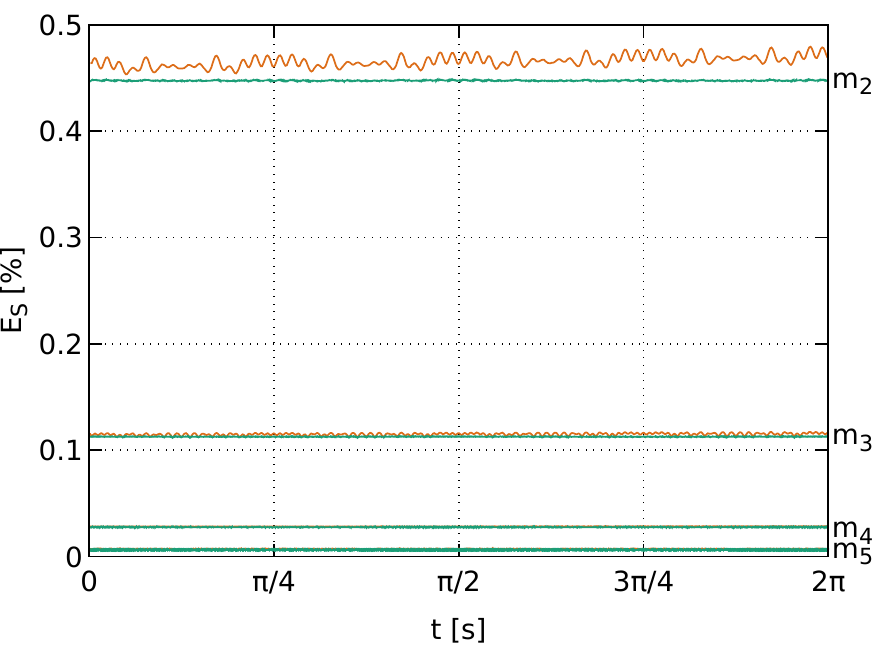}
 \end{minipage}
  \begin{minipage}{.5\textwidth}
  \includegraphics[width=1.\textwidth,height=.75\textwidth,angle=0]{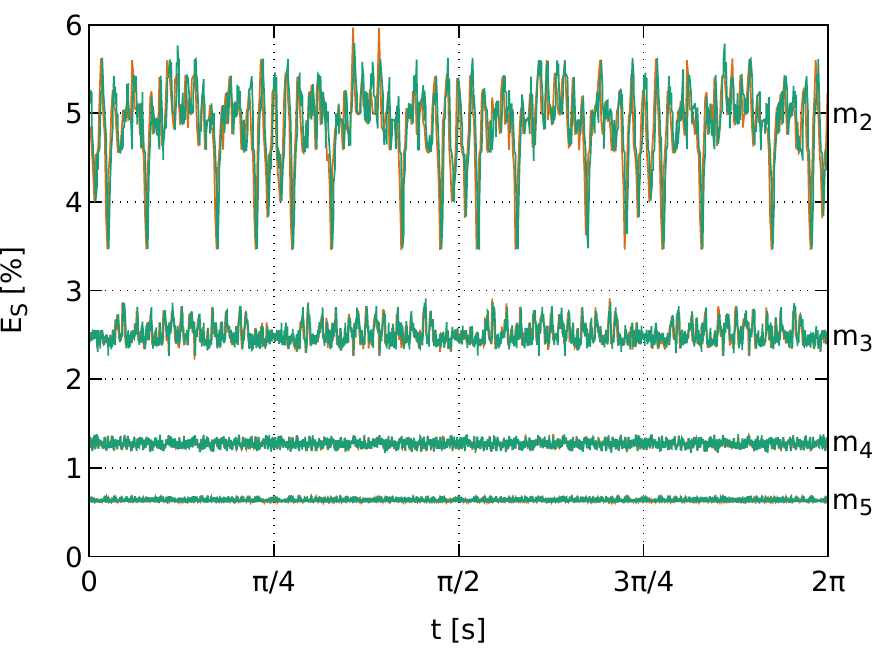}
 \end{minipage}%
 \begin{minipage}{.5\textwidth}
  \includegraphics[width=1.\textwidth,height=.75\textwidth,angle=0]{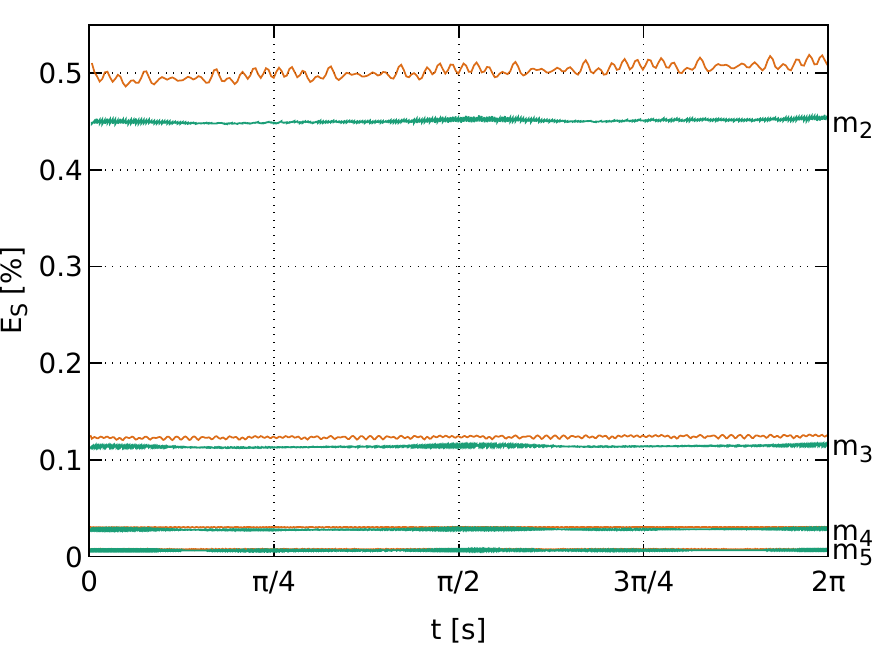}
 \end{minipage}
 \caption{\small{The convergence of mass in regions
                 $R_1 \!=\! \lc x_i | \alpps \ge 0.5 \rc$ (left) and
                 $R_2 \!=\! \lc x_i | \psio \le 8\eph \rc$ (right)
                 during advection of the solid body
                 with the Eulerian (green solid line) and Lagrangian (orange solid line)
                 schemes on four gradually refined grids
                 $m_k,\,k=2,\ldots,5$
                 and three different
                 CFL numbers  $C\!u_0 \approx 0.35$,
                 $C\!u_1 \approx 0.7$,
                 $C\!u_2 \approx 1.4$ 
                 (from top to bottom).
                 The number of
                 re-initialization steps 
                 $N_\tau=4$ per $\Delta t$, $\Delta \tau=\eph$.}}
 \label{fig8}
 \end{figure}

In \Fig{fig8},
the convergence rates 
of mass or volume
of the advected 
circular interface
are presented for three 
$C\!u_l$ numbers $l\!=\!0,1,2$
(top to bottom)
on four gradually 
refined grids 
$m_k$, $k\!=\!2,\ldots,5$.
\blnm
 \elnm
The  
errors $E_S$ 
introduced
to the surface
determined
by the advected interface
are computed 
using \Eq{Aeq8} 
after each
time step $\Delta t$
and at the end 
of each
re-initialization cycle.
As it was 
proposed by
\cite{twacl15},
convergence of
the mass 
is investigated
in the two regions:
$R_1\!=\!\lc x_i | \alpps \ge 0.5 \rc$,
$R_2\!=\!\lc x_i | \psio \le 8\eph \rc$
where $x_i$, $i\!=\!1,\ldots,N_c$ 
denotes the
center of control volume
belonging to one of
the grids $m_k$, $k\!=\!2,\ldots,5$.
The definitions
of $R_1$, $R_2$
regions
exploit 
the two, 
equivalent
representations
of the interface
by  $\alpha \lr \psi\!=\!0\rr=1/2$ 
or $\psi \lr \alpha\!=\!1/2 \rr \!=\! 0$,
respectively.

Surprisingly,
convergence 
of mass
illustrated 
in \Fig{fig8}
does not exhibit
strong dependence
on the advection scheme
or $C\!u_l$ number
used in the simulations,
compare with
the re-initialization
equation norms $L_1^\tau$
and $\av{L_1^\tau}$ 
depicted in \Figs{fig6}{fig7}.
The convergence
histories of mass
recorded during one
revolution of circular
interface are almost 
identical for 
the Eulerian 
and Lagrangian 
schemes.
The largest
differences
are visible
on the coarsest
grids $m_2$ and
$m_3$ in the 
region $R_2$,
see right column
in \Fig{fig8}
for $C\!u_2$.
The oscillations
of the mass are larger
in the case of Lagrangian
scheme, whereas the Eulerian
scheme obtains less
oscillatory 
mass convergence
errors;
on the grids
$m_4, m_5$ the results
obtained using Eulerian
and Lagrangian schemes
are almost identical
in both 
regions $R_1,\,R_2$
and for all $C\!u_l$, $l\!=\!0,1,2$.
In the case of both:
Eulerian and Lagrangian
schemes,  the mass convergence
is achieved independent
from $C\!u_l$, $l\!=\!0,1,2$ used
in the simulation.
\blnm
 \begin{figure}[!ht] \nonumber
 \begin{minipage}{.5\textwidth}
  \includegraphics[width=.9\textwidth,height=.9\textwidth,angle=0]{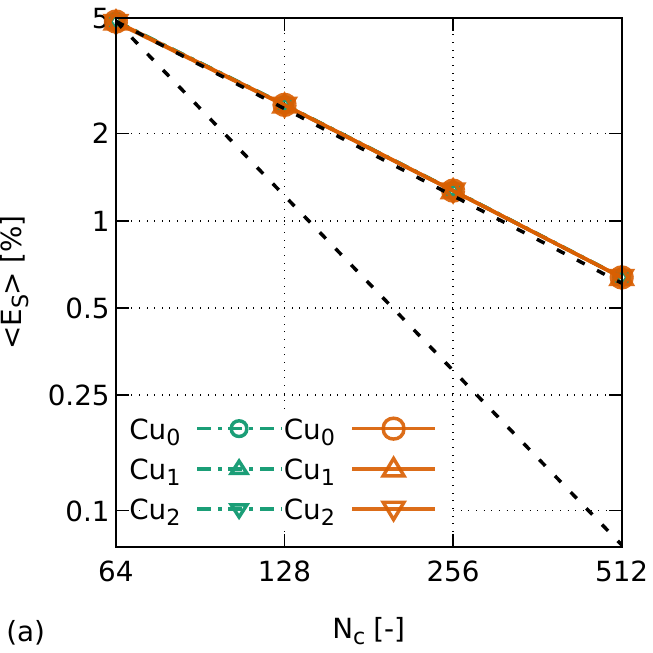}
 \end{minipage}%
 \begin{minipage}{.5\textwidth}
  \includegraphics[width=.9\textwidth,height=.9\textwidth,angle=0]{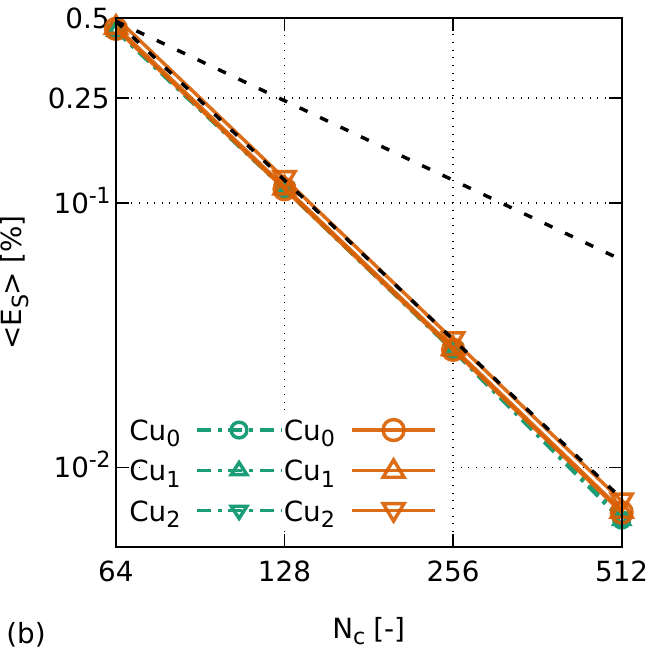}
 \end{minipage}
 \caption{\small{The averaged in time $t$ errors  
                 $E_S$ from \Fig{fig8} computed using \Eq{Aeq9} 
                 in regions
                 $R_1\!=\!\lc x_i | \alpps \ge 0.5 \rc$ (a) and 
                 $R_2\!=\!\lc x_i | \psio < 8\eph \rc$  (b).
                 The results are obtained
                 using 
                 the Eulerian (green dashed-dotted  lines) 
                 and Lagrangian (orange solid lines) advection schemes
                 for the three $C\!u_l$, $l\!=\!0,1,2$ numbers.
                 The black dashed-lines represent, respectively, 
                 the first and second order 
                 convergence slopes.}}
 \label{fig9}
 \end{figure}
 \elnm

The order
of the convergence rate
of mass
depends on the interface
representation
by $\alpha \lr \psi\!=\!0 \rr\!=\!1/2$
or $\psi \lr \alpha\!=\!1/2 \rr\!=\!0$ 
as it is
discussed 
by \cite{twacl15}.
Based on
the results
in \Fig{fig8}
it can be deduced
that in the region $R_1$
the first-order mass
convergence rate
is achieved, whereas in $R_2$
the second order mass
convergence rate is achieved
for all 
$C\!u_l$, $l\!=\!0,1,2$.
This is confirmed 
by the averaged in time $t$
errors $E_S$ from \Fig{fig8}
illustrated in  \Fig{fig9};
the averaged errors
$\av{E_S}$ are  
computed using 
\Eq{Aeq9}.
\blnm
 \begin{figure}[!ht] \nonumber
 \begin{minipage}{.5\textwidth}
  \includegraphics[width=1.\textwidth,height=.75\textwidth,angle=0]{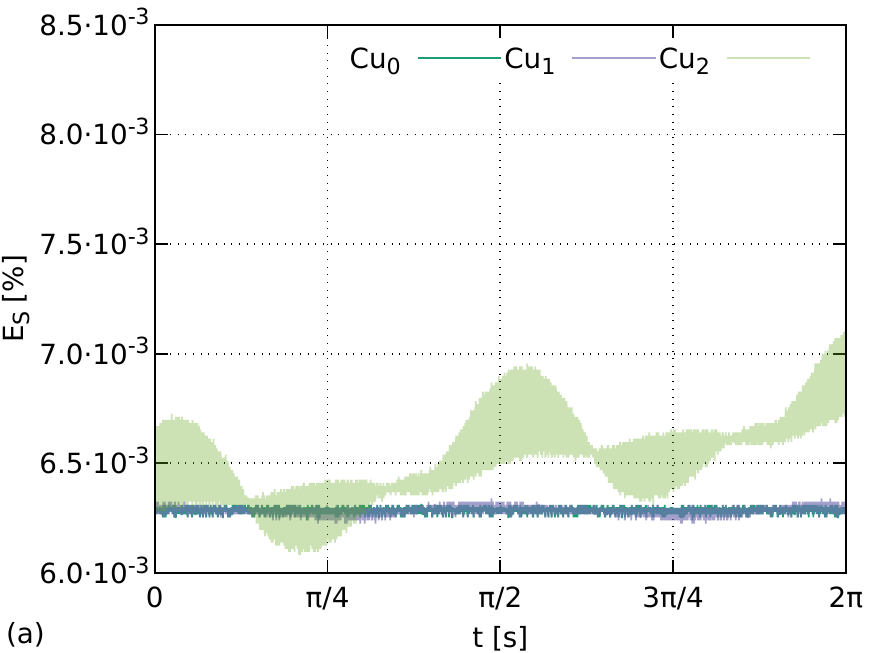}
 \end{minipage}%
 \begin{minipage}{.5\textwidth}
  \includegraphics[width=1.\textwidth,height=.75\textwidth,angle=0]{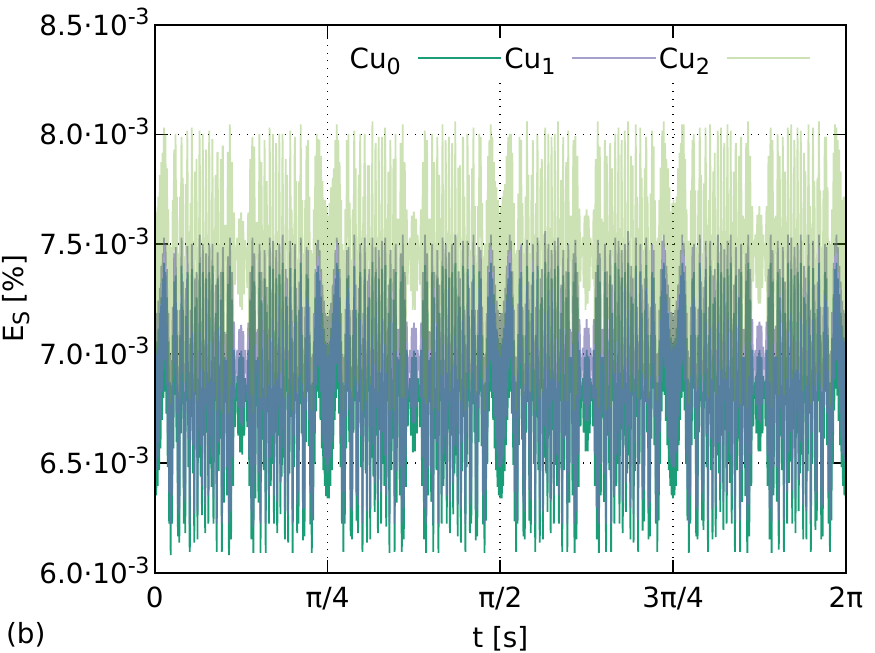}
 \end{minipage}
 \caption{\small{The convergence of mass in the region
                 $R_2 \!=\! \lc x_i | \psio < 8\eph \rc$ on  the grid $m_5$
                 during advection of the solid body
                 with Eulerian (a) and Lagrangian (b)
                 schemes on the grid $m_5$
                 with   $C\!u_0 \approx 0.35$,
                 $C\!u_1 \approx 0.7$,
                 $C\!u_2 \approx 1.4$. 
                 The number of
                 re-initialization steps 
                 $N_\tau \!=\! 4$ per time step $\Delta t$, $\Delta \tau \!=\! \eph$,
                 these results are also depicted in \Fig{fig8}.}}
 \label{fig10}
 \end{figure}
\elnm

The closer
inspection of the
mass convergence results
on the finest grid
$m_5$ in region $R_2$
is illustrated
in \Fig{fig10}.
This comparison 
demonstrates
the mass errors
obtained using
the Eulerian
scheme are  more sensitive
to the selected 
time step
size $\Delta t_l$, $l\!=\!0,1,2$.
The results presented
in \Fig{fig10}
indicate that
for $C\!u_0 \approx 0.35$, $C\!u_1 \approx 0.7$
Eulerian scheme
can achieve 
better mass conservation
(the error level and its oscillations are lower)
than  the Lagrangian scheme.
When $C\!u_2 \approx 1.4$,
slow but  constant divergence
of the mass occurs
in the case of 
advection carried out
with the Eulerian scheme.
In contrast,
the errors in
mass conservation
achieved with
the Lagrangian scheme
seem to be almost
unaffected by
the increment
in the time
step size $\Delta t_l$,
see \Fig{fig10}(b).
The errors obtained with
the Lagrangian scheme indicate
no change in the mass
or volume of the advected
circular shape during the
whole revolution 
for all
tested Currant
numbers.
Hence, in the case of Lagrangian
scheme the mass is conserved during
one revolution of the circular
interface when the conditions
used to derive \Eqs{eq29}{eq30}
are satisfied.
\blnm
 \begin{figure}[!ht] \nonumber
 \begin{minipage}{.5\textwidth}
  \centering\includegraphics[width=.75\textwidth,height=0.75\textwidth,angle=0]{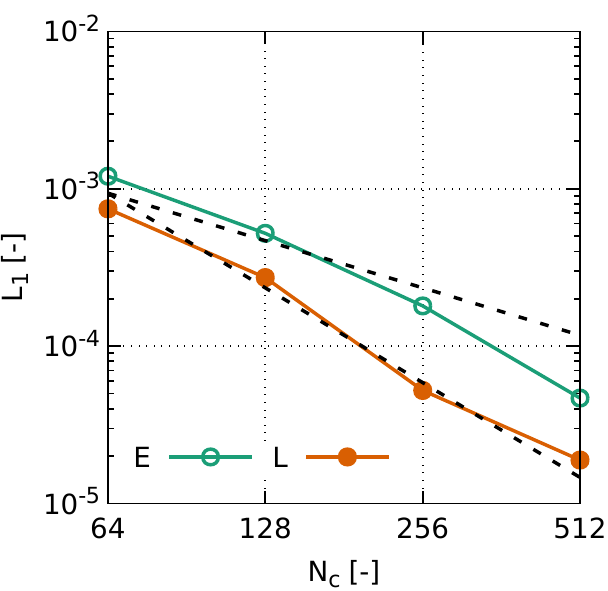}
 \end{minipage}
 \begin{minipage}{.5\textwidth}
  \centering\includegraphics[width=.75\textwidth,height=0.75\textwidth,angle=0]{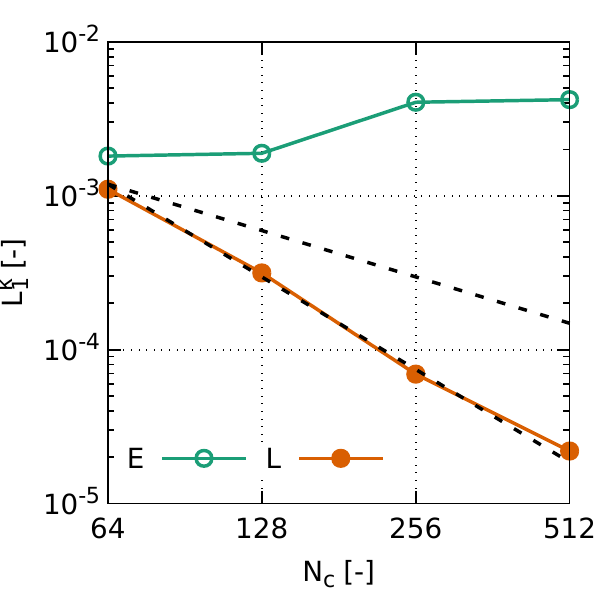}
 \end{minipage}
 \begin{minipage}{.5\textwidth}
  \centering\includegraphics[width=.75\textwidth,height=0.75\textwidth,angle=0]{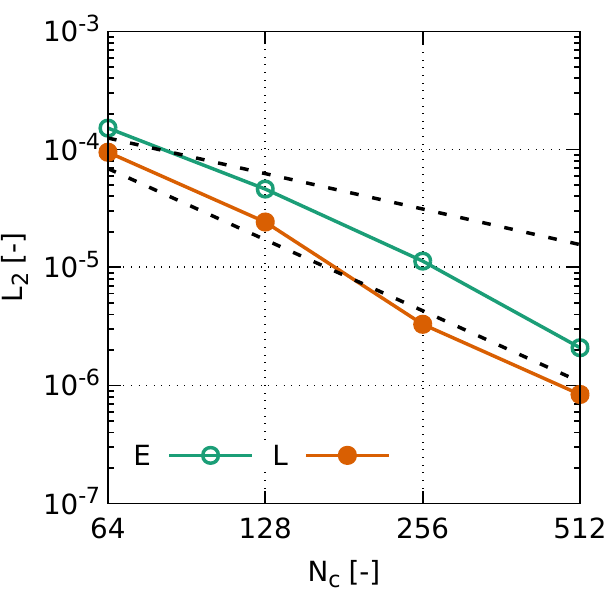}
 \end{minipage}
 \begin{minipage}{.5\textwidth}
  \centering\includegraphics[width=.75\textwidth,height=0.75\textwidth,angle=0]{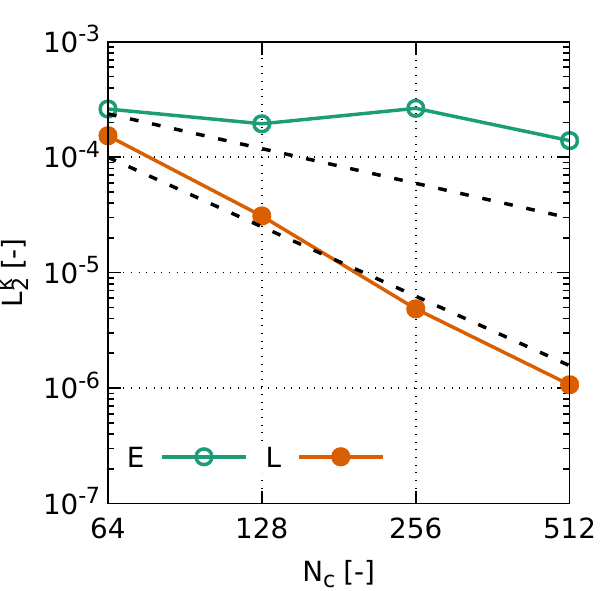}
 \end{minipage}
 \begin{minipage}{.5\textwidth}
  \centering\includegraphics[width=.75\textwidth,height=0.75\textwidth,angle=0]{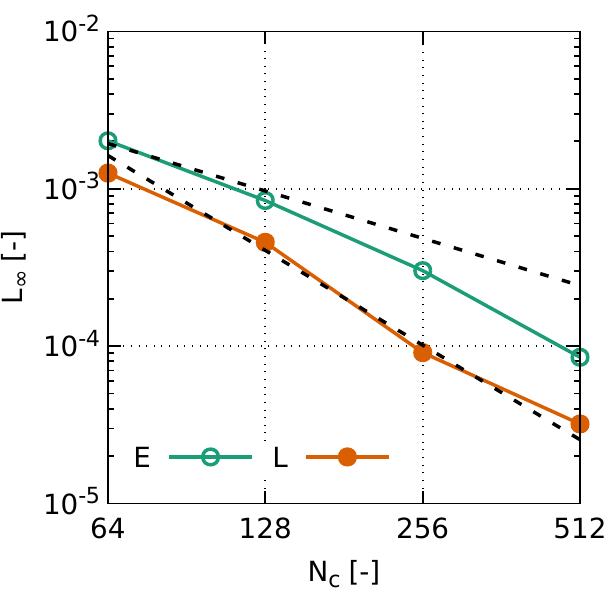}
 \end{minipage}
 \begin{minipage}{.5\textwidth}
  \centering\includegraphics[width=.75\textwidth,height=0.75\textwidth,angle=0]{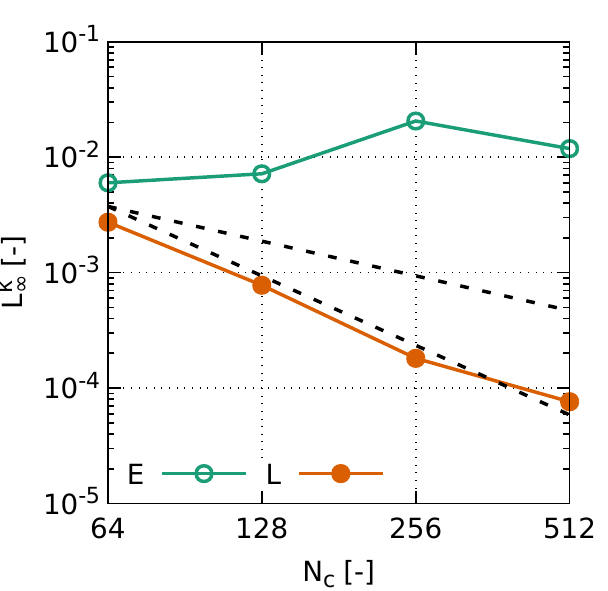}
 \end{minipage} 
 \caption{\small{The convergence of 
                 $L_1,\,L_2,\,L_{\infty}$ 
                 norms defined 
                 by Eqs.~\ref{Aeq5}--\ref{Aeq7} 
                 for $C\!u_0 \approx 0.35$ 
                 computed after one revolution of
                 the circular interface.
                 The convergence
                 of the interface shape (left), the convergence
                 of the interface curvature (right).
                 Symbols $E,\,L$ denote
                 results obtained with 
                 the Eulerian or Lagrangian
                 schemes, the black dashed 
                 lines depict
                 the first and second order 
                 convergence slopes.}}
 \label{fig11}
 \end{figure}
\elnm

\Figs{fig11}{fig13}
illustrate 
the convergence
rates  of the interface
shape and  curvature on
gradually refined grids $m_k$, $k\!=\!2,\ldots,5$.
The interface shape and curvature
are defined, respectively, 
by the level-sets of
the signed-distance function
$\psi \lr \alpha \!=\!1/2 \rr \!=\! 0$
and corresponding
curvature $\kappa \ls \psi \lr \alpha\!=\!1/2 \rr \!=\! 0 \rs \!=\! 1/R$.
The 
convergence 
rates  illustrated
in \Figs{fig11}{fig13},
for three
$C\!u_l$, $l\!=\!0,1,2$ 
numbers
are computed 
at the end of 
one revolution
of the circular
interface.
In the left column 
the convergence rates
of the interface shape,
in the right column 
the convergence rates
of the interface curvature
(denoted using superscript $\kappa$)
are presented.
The 
norms
$L_1$, $L_2$, $L_{\infty}$
in \Figs{fig11}{fig13}
are defined by Eqs.~(\ref{Aeq5})-(\ref{Aeq7})
in  a similar
manner 
for level-sets 
of the interface
shape and curvature,
information
how they are computed
can be found 
in \ref{appA}.

\begin{figure}[h!] \nonumber
 \begin{minipage}{.5\textwidth}
  \centering\includegraphics[width=.75\textwidth,height=0.75\textwidth,angle=0]{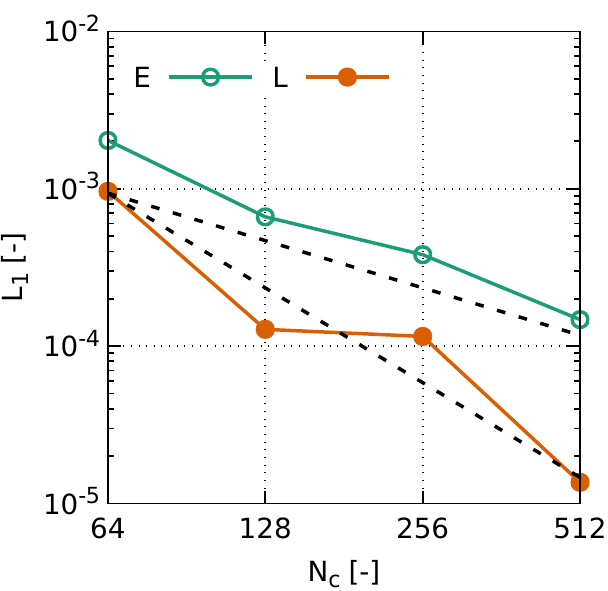}
 \end{minipage}
 \begin{minipage}{.5\textwidth}
  \centering\includegraphics[width=.75\textwidth,height=0.75\textwidth,angle=0]{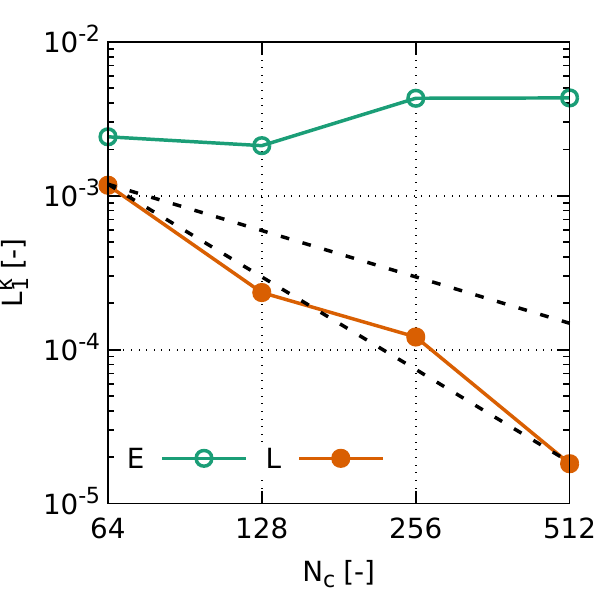}
 \end{minipage}
 \begin{minipage}{.5\textwidth}
  \centering\includegraphics[width=.75\textwidth,height=0.75\textwidth,angle=0]{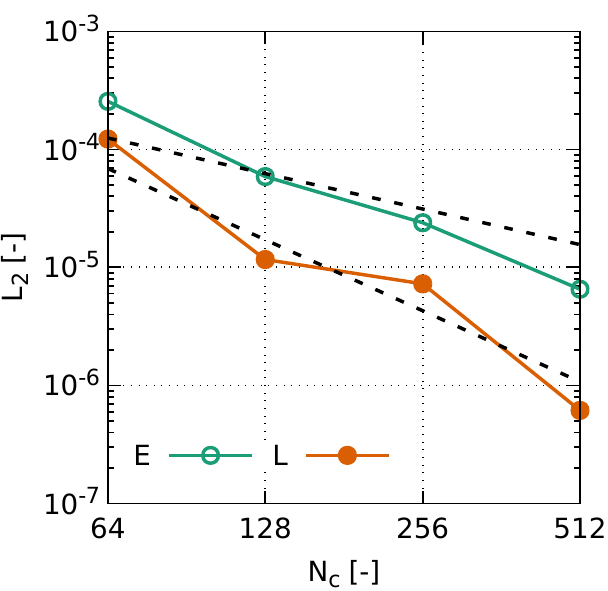}
 \end{minipage}
 \begin{minipage}{.5\textwidth}
  \centering\includegraphics[width=.75\textwidth,height=0.75\textwidth,angle=0]{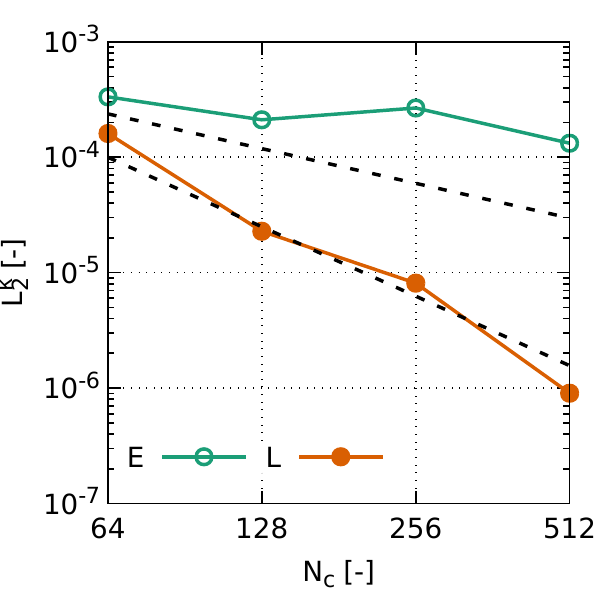}
 \end{minipage}
 \begin{minipage}{.5\textwidth}
  \centering\includegraphics[width=.75\textwidth,height=0.75\textwidth,angle=0]{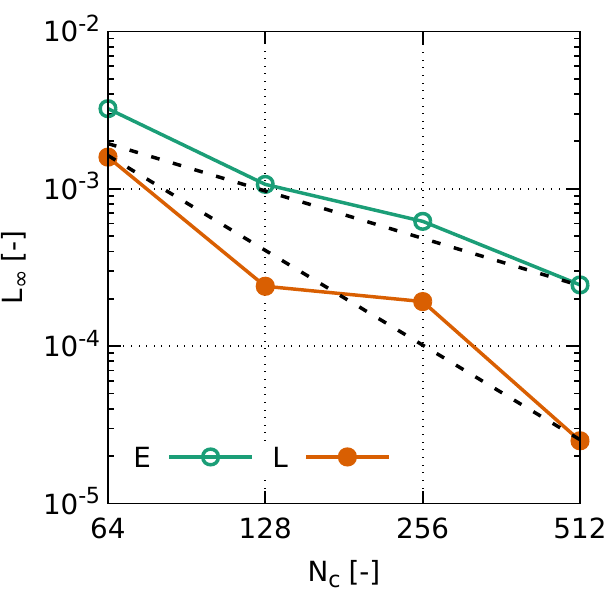}
 \end{minipage}
 \begin{minipage}{.5\textwidth}
  \centering\includegraphics[width=.75\textwidth,height=0.75\textwidth,angle=0]{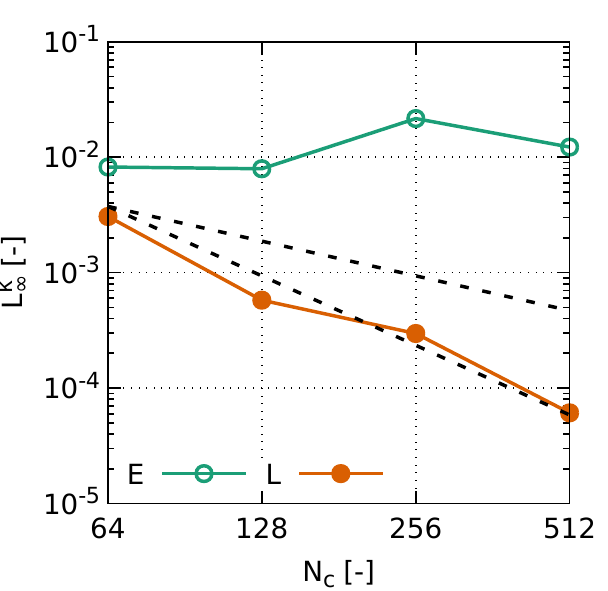}
 \end{minipage} 
 \caption{\small{The convergence of 
                 $L_1,\,L_2,\,L_{\infty}$ 
                 norms defined 
                 by Eqs.~\ref{Aeq5}--\ref{Aeq7} 
                 for $C\!u_1 \approx 0.7$ 
                 computed after one revolution of
                 the circular interface.
                 The convergence
                 of the interface shape (left), the convergence
                 of the interface curvature (right).
                 Symbols $E,\,L$ denote
                 results obtained with 
                 the Eulerian or Lagrangian
                 schemes, the black dashed 
                 lines depict
                 the first and second order 
                 convergence slopes.}}
 
 \label{fig12}
 \end{figure}

\Fig{fig11} 
reveals that only 
the Lagrangian scheme 
achieves
the second-order
convergence
rate of both:
the interface 
shape and
curvature,
compare the
diagrams in
left and right
columns.
This convergence
is however 
affected by
the selected time steps
size $\Delta t_l$, $l\!=\!0,1,2$.
The larger
the Courant number is,
the less obvious the order of
the convergence rate,
although, the second-order
trend can still be deduced
from the results presented
in \Figs{fig12}{fig13}.
\blnm
\elnm
\blnm
 \begin{figure}[!ht] \nonumber
 \begin{minipage}{.5\textwidth}
  \centering\includegraphics[width=.75\textwidth,height=0.75\textwidth,angle=0]{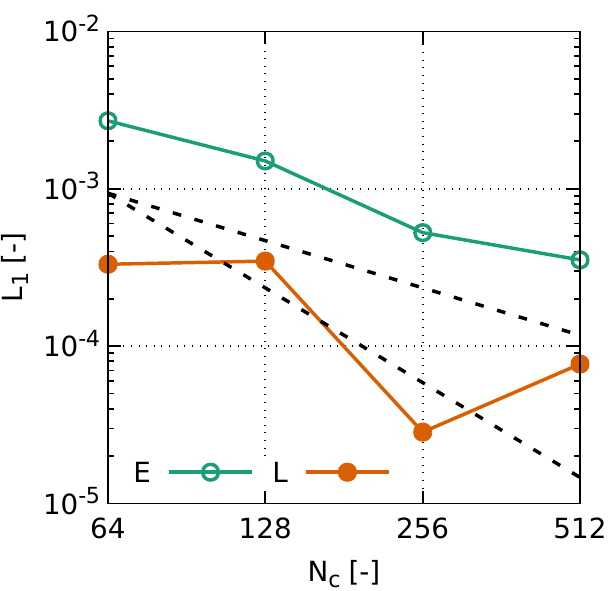}
 \end{minipage}%
 \begin{minipage}{.5\textwidth}
  \centering\includegraphics[width=.75\textwidth,height=0.75\textwidth,angle=0]{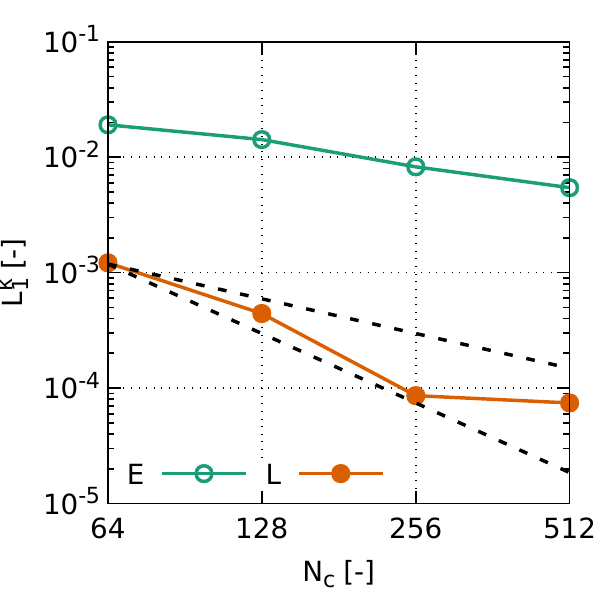}
 \end{minipage}
 \begin{minipage}{.5\textwidth}
  \centering\includegraphics[width=.75\textwidth,height=0.75\textwidth,angle=0]{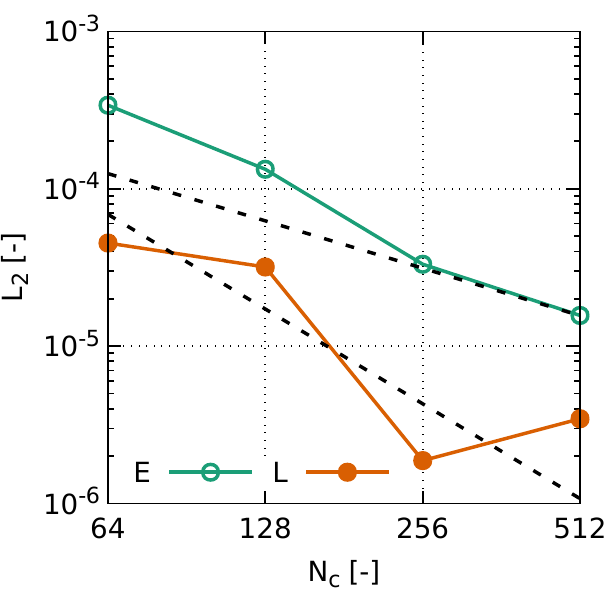}
 \end{minipage}%
 \begin{minipage}{.5\textwidth}
  \centering\includegraphics[width=.75\textwidth,height=0.75\textwidth,angle=0]{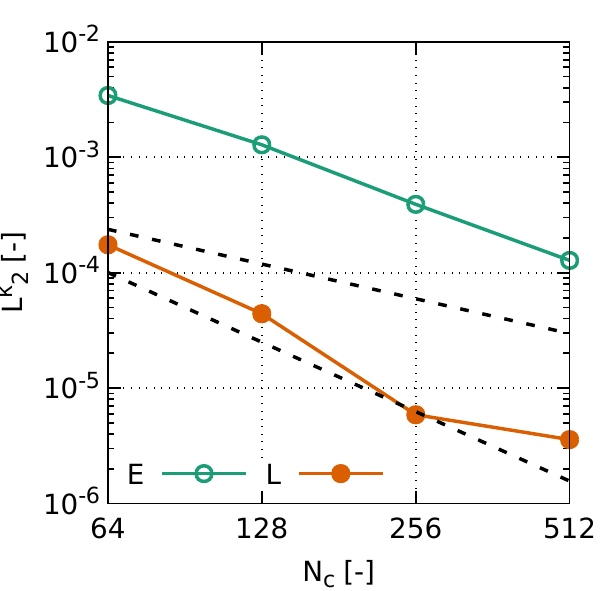}
 \end{minipage}
 \begin{minipage}{.5\textwidth}
  \centering\includegraphics[width=.75\textwidth,height=0.75\textwidth,angle=0]{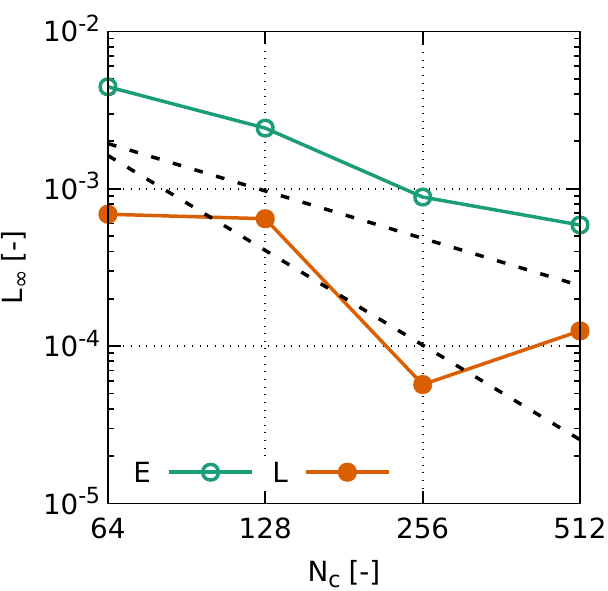}
 \end{minipage}%
 \begin{minipage}{.5\textwidth}
  \centering\includegraphics[width=.75\textwidth,height=0.75\textwidth,angle=0]{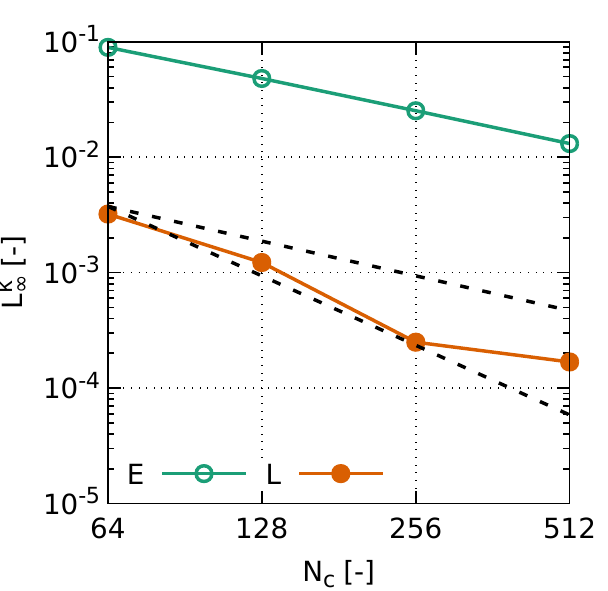}
 \end{minipage}
 \caption{\small{The convergence of 
                 $L_1,\,L_2,\,L_{\infty}$ 
                 norms defined 
                 by Eqs.~\ref{Aeq5}--\ref{Aeq7} 
                 for $C\!u_2 \approx 1.4$ 
                 computed after one revolution of
                 the circular interface.
                 The convergence
                 of the interface shape (left), the convergence
                 of the interface curvature (right).
                 Symbols $E,\,L$ denote
                 results obtained with 
                 the Eulerian or Lagrangian
                 schemes, the black dashed 
                 lines depict
                 the first and second order 
                 convergence slopes.}}
 \label{fig13}
 \end{figure}
\elnm
In the case
of the results
obtained with $C\!u_0 \approx 0.35$
in \Fig{fig11}
there is no doubt
that 
the complete second-order
convergence rate is 
achieved with 
the Lagrangian
scheme.
The growing
uncertainty
in the convergence rate
of the interface shape
and curvature observed
in \Figs{fig12}{fig13} 
may be related to the
explicit formulation of
the Lagrangian scheme
proposed in the
present paper.
\blnm
 \begin{figure}[h!] \nonumber
 \begin{minipage}{.33\textwidth}
  \includegraphics[width=.95\textwidth,height=.95\textwidth,angle=-90]{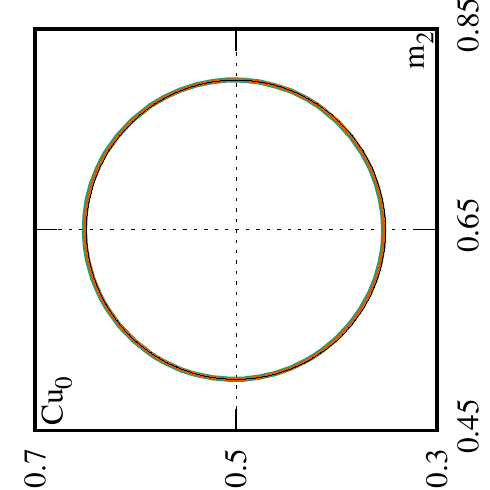}
 \end{minipage}%
 \begin{minipage}{.33\textwidth}
  \includegraphics[width=.95\textwidth,height=.95\textwidth,angle=-90]{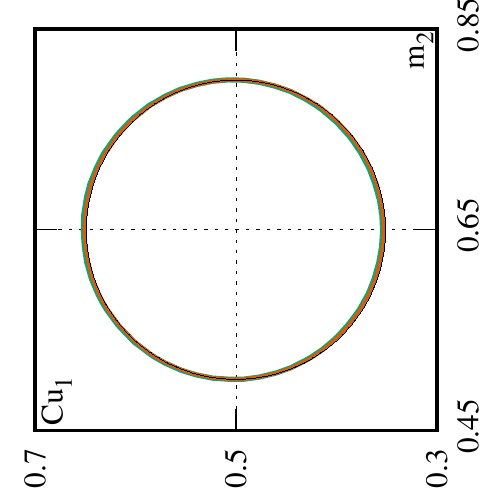}
 \end{minipage}%
 \begin{minipage}{.33\textwidth}
  \includegraphics[width=.95\textwidth,height=.95\textwidth,angle=-90]{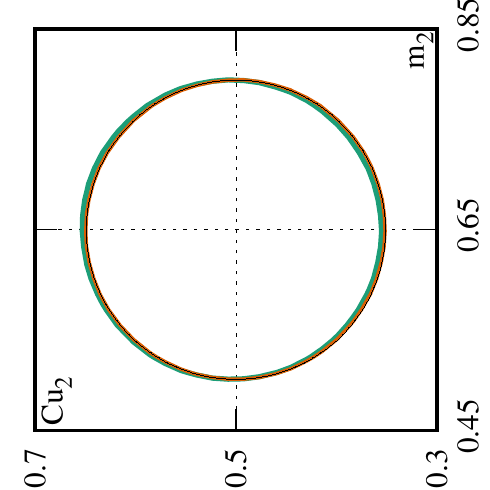}
 \end{minipage}
 \begin{minipage}{.33\textwidth}
  \includegraphics[width=0.95\textwidth,height=0.95\textwidth,angle=-90]{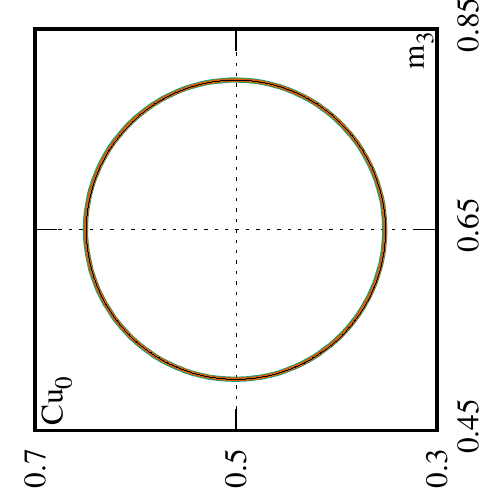}
 \end{minipage}%
 \begin{minipage}{.33\textwidth}
  \includegraphics[width=0.95\textwidth,height=0.95\textwidth,angle=-90]{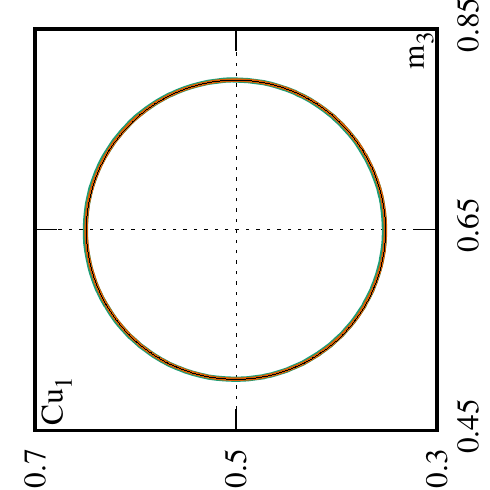}
 \end{minipage}%
 \begin{minipage}{.33\textwidth}
  \includegraphics[width=0.95\textwidth,height=0.95\textwidth,angle=-90]{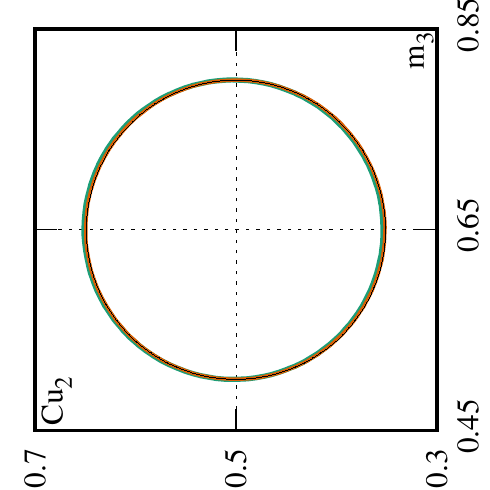}
 \end{minipage}
 \begin{minipage}{.33\textwidth}
  \includegraphics[width=0.95\textwidth,height=0.95\textwidth,angle=-90]{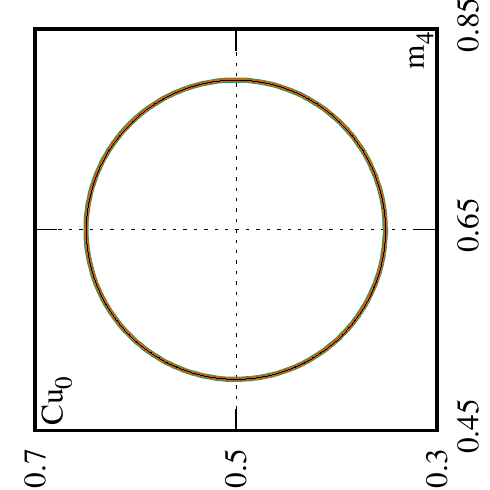}
 \end{minipage}%
 \begin{minipage}{.33\textwidth}
  \includegraphics[width=0.95\textwidth,height=0.95\textwidth,angle=-90]{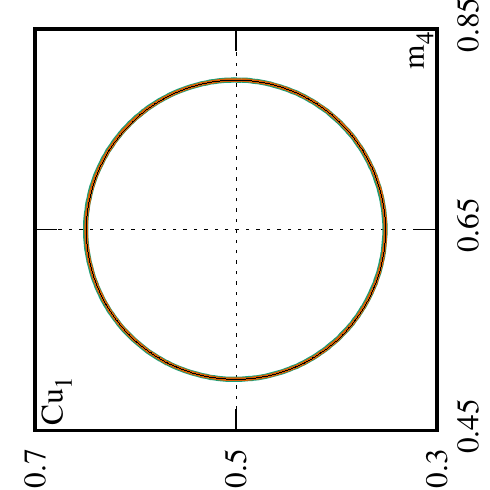}
 \end{minipage}%
 \begin{minipage}{.33\textwidth}
  \includegraphics[width=0.95\textwidth,height=0.95\textwidth,angle=-90]{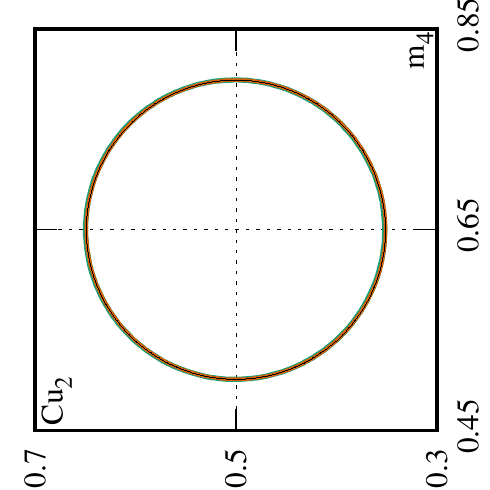}
 \end{minipage}
  \begin{minipage}{.33\textwidth}
  \includegraphics[width=0.95\textwidth,height=0.95\textwidth,angle=-90]{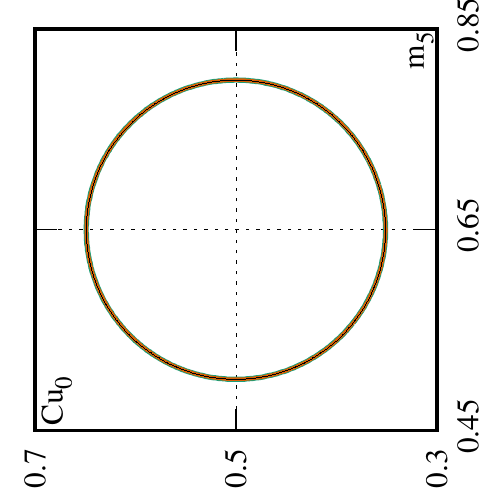}
 \end{minipage}%
 \begin{minipage}{.33\textwidth}
  \includegraphics[width=0.95\textwidth,height=0.95\textwidth,angle=-90]{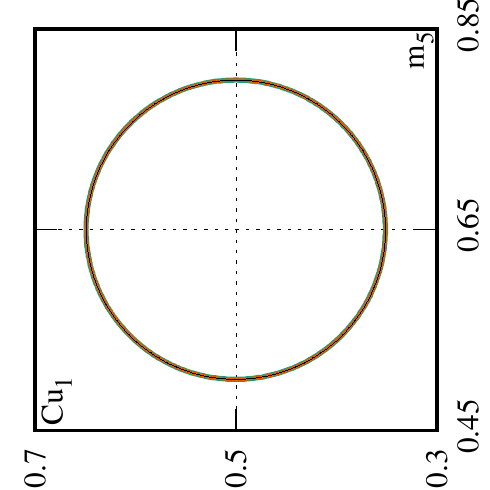}
 \end{minipage}%
 \begin{minipage}{.33\textwidth}
  \includegraphics[width=0.95\textwidth,height=0.95\textwidth,angle=-90]{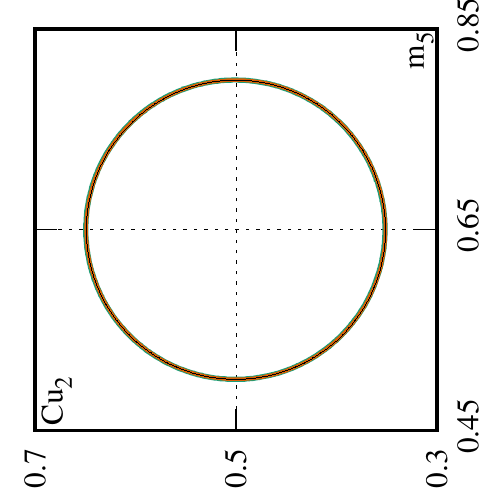}
 \end{minipage} 
 \caption{\small{The level-sets $\psi \lr \alpha \!=\!1/2 \rr \!=\! 0$
                 of the circular interface shape
                 after one revolution with
                 $C\!u_l$, $l\!=\!0,1,2$
                 (left to right)
                 on four grids $m_k\!=\!2^{4+k} \times 2^{4+k}$, $k=2,\ldots,5$
                 (top to bottom).
                 The results are obtained
                 with the Lagrangian (orange solid line) and
                 Eulerian  (green solid line) advection schemes,
                 the exact position of the interface is depicted
                 with the  black solid line.}}
 \label{fig14}
 \end{figure}
\elnm
\blnm
 \begin{figure}[h!] \nonumber
 \begin{minipage}{.33\textwidth}
  \includegraphics[width=0.95\textwidth,height=0.95\textwidth,angle=-90]{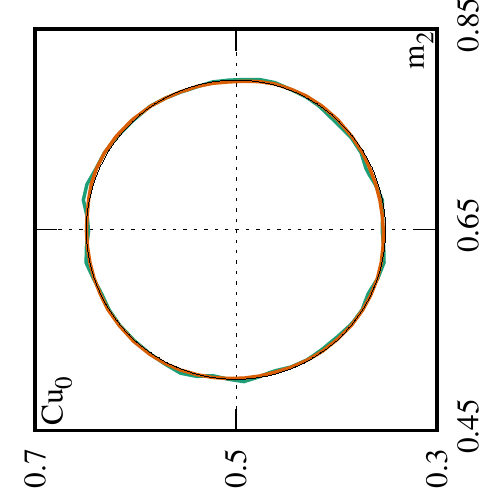}
 \end{minipage}%
 \begin{minipage}{.33\textwidth}
  \includegraphics[width=0.95\textwidth,height=0.95\textwidth,angle=-90]{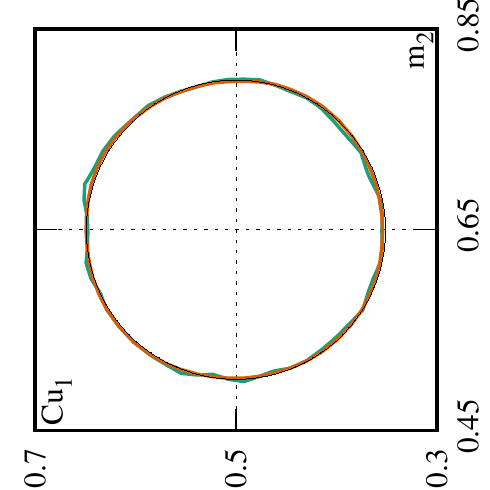}
 \end{minipage}%
 \begin{minipage}{.33\textwidth}
  \includegraphics[width=0.95\textwidth,height=0.95\textwidth,angle=-90]{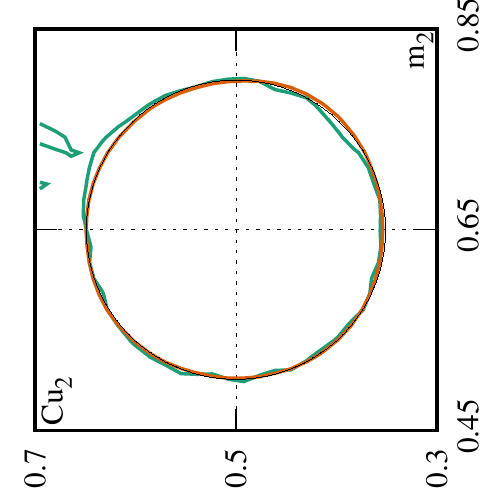}
 \end{minipage}
 \begin{minipage}{.33\textwidth}
  \includegraphics[width=0.95\textwidth,height=0.95\textwidth,angle=-90]{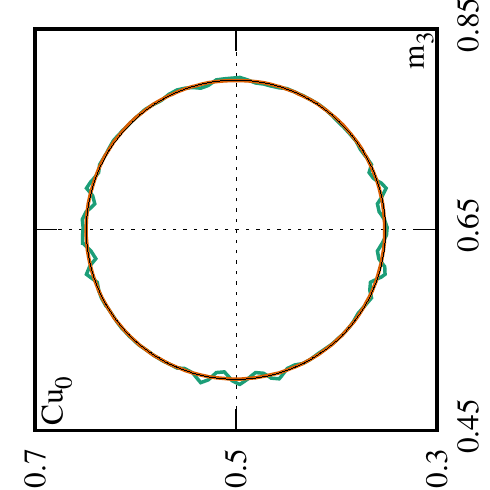}
 \end{minipage}%
 \begin{minipage}{.33\textwidth}
  \includegraphics[width=0.95\textwidth,height=0.95\textwidth,angle=-90]{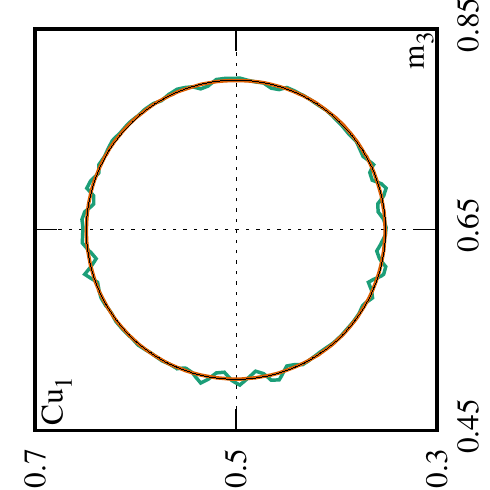}
 \end{minipage}%
 \begin{minipage}{.33\textwidth}
  \includegraphics[width=0.95\textwidth,height=0.95\textwidth,angle=-90]{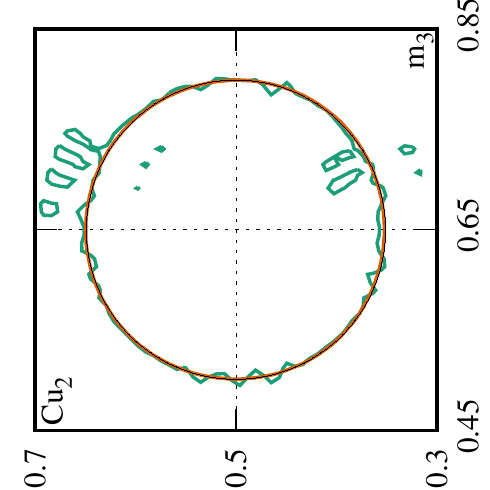}
 \end{minipage}
 \begin{minipage}{.33\textwidth}
  \includegraphics[width=0.95\textwidth,height=0.95\textwidth,angle=-90]{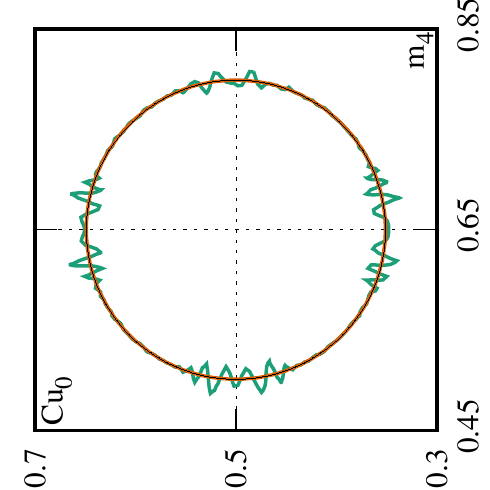}
 \end{minipage}%
 \begin{minipage}{.33\textwidth}
  \includegraphics[width=0.95\textwidth,height=0.95\textwidth,angle=-90]{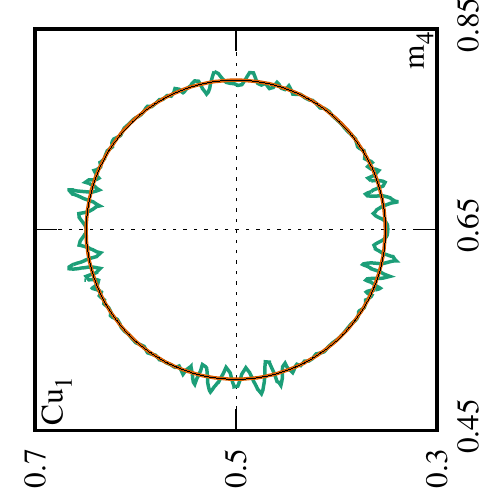}
 \end{minipage}%
 \begin{minipage}{.33\textwidth}
  \includegraphics[width=0.95\textwidth,height=0.95\textwidth,angle=-90]{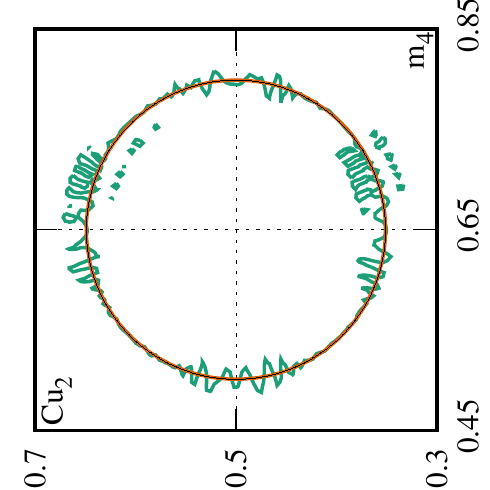}
 \end{minipage}
  \begin{minipage}{.33\textwidth}
  \includegraphics[width=0.95\textwidth,height=0.95\textwidth,angle=-90]{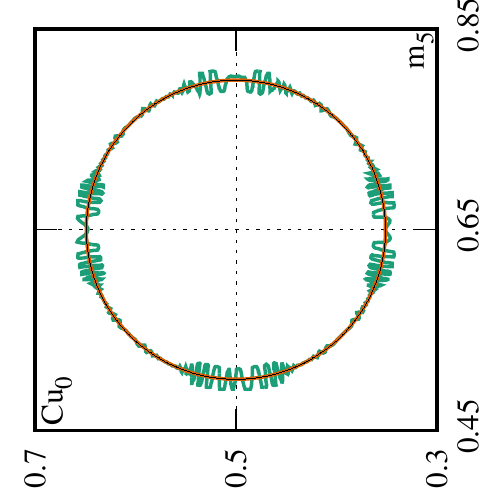}
 \end{minipage}%
 \begin{minipage}{.33\textwidth}
  \includegraphics[width=0.95\textwidth,height=0.95\textwidth,angle=-90]{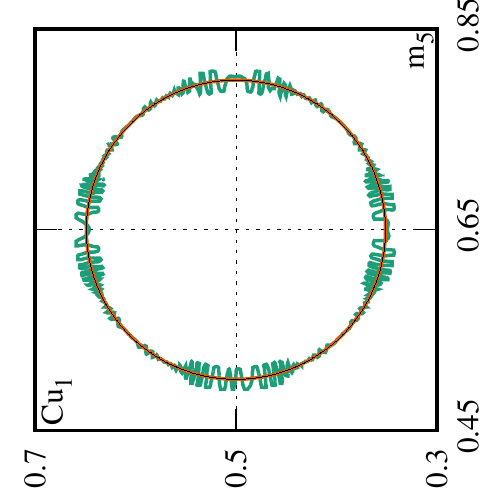}
 \end{minipage}%
 \begin{minipage}{.33\textwidth}
  \includegraphics[width=0.95\textwidth,height=0.95\textwidth,angle=-90]{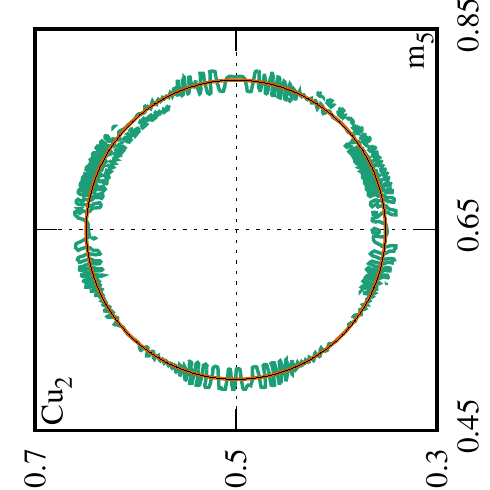}
 \end{minipage}
 \caption{\small{The level-sets of curvature
                 $\kappa \ls \psi \lr \alpha\!=\!1/2 \rr \!=\! 0 \rs \!=\! 1/R$
                 of the circular interfaces from \Fig{fig14}
                 after one revolution with
                 $C\!u_l$, $l\!=\!0,1,2$
                 (left to right)
                 on four grids $m_k\!=\!2^{4+k} \! \times \! 2^{4+k}$, $k\!=\!2,\ldots,5$
                 (top to bottom).
                 The results are obtained
                 with the Lagrangian (orange solid line) and
                 Eulerian  (green solid line) advection schemes,
                 the exact position of 
                 the interface curvature $\kappa = 1/R$
                 is depicted with the black solid line.}}
 \label{fig15}
 \end{figure}
\elnm

Results 
in \Figs{fig11}{fig13}
show  
the implicit
Eulerian scheme
allows for second-order
convergence rate
of the interface
shape when $C\!u_0\approx 0.35$.
Interestingly,  
the interface shape
convergence rates
of the implicit Eulerian scheme,
does not show oscillatory
character unlike
the solution
obtained with
the explicit
Lagrangian scheme,
even for $C\!u_2 \approx 1.4$,
see \Fig{fig13}.
%
%
In \Figs{fig11}{fig13},
it is observed that
for the Eulerian scheme
the order 
of convergence rate
of the interface
becomes lower, 
changing
from second 
for $C\!u_0$
towards the first
order for $C\!u_1,\,C\!u_2$.
This results  
does not correspond 
with the mass convergence
presented in \Fig{fig9}(b)
where second order convergence
was predicted for all $C\!u_l$,
$l\!=\!0,1,2$.
The reason 
of this discrepancy
is related to  
integration 
and averaging 
used 
to compute, 
respectively,
$E_S$ and $\av{E_S}$
in \Figs{fig8}{fig9}
whereas
\Figs{fig11}{fig13}
present
the instantaneous
errors.

The results 
in the right
column in \Figs{fig11}{fig13}
demonstrate 
the interface curvature
does not converge
with the gradual mesh
refinement when 
the Eulerian
scheme is used.
In \Figs{fig14}{fig15}
this result may be 
investigated in detail.
\Fig{fig14} illustrates
shapes of the circular
interface defined by
the level-set $\psi \lr \alpha\!=\!1/2\rr\!=\!0$
after one revolution
on four  grids $m_k$, $k=2,\ldots,5$ 
and for the three
$C\!u_l$, $l\!=\!0,1,2$ numbers, 
obtained with the
Eulerian (green solid lines)
and Lagrangian (orange solid line)
schemes.
The reconstructed 
shapes of the
circular interface
are compared with
the exact analytic
solution (black solid line)
confirming
convergence
towards
the analytic
solution
of both
the Eulerian
and Lagrangian 
schemes
for all
$C\!u_l$, $l\!=\!0,1,2$.
This is in agreement
with the results presented
in the left columns
of \Figs{fig11}{fig13}.

In  \Fig{fig15},
the level-sets
of exact curvature 
(black solid line)
and reconstructed 
curvatures
$\kappa \ls \psi \lr \alpha \!=\!1/2 \rr \!=\! 0 \rs \!=\! 1/R$
computed for 
the corresponding
circular interfaces
from \Fig{fig14}
are compared. 
These results
explain
the apparent 
convergence of 
curvature
observed for
the Eulerian scheme
and the Courant
number $C\!u_2$,
see \Fig{fig13}(right).
The corresponding
level-sets of curvatures
(green solid line)
illustrate
lack of
convergence  
when the Eulerian 
scheme
is used.
All 
level-sets
of curvature obtained
with the Eulerian scheme
show nonphysical
oscillations,
which 
do not 
vanish with
the mesh refinement.
The results in
\Fig{fig15} indicate
the frequency of these
oscillations is amplified
when $\Delta x_k \to 0$.
The origin of these
errors is unclear,
it is supposed
they are artifacts
introduced by 
the second order 
flux limiter controlling
only the slope of $\alpps$.

In the case of
the new Lagrangian
scheme the agreement
between the exact 
(black solid line)
and reconstructed 
(orange solid line)
curvatures is excellent.
On the finest grid $m_5$
it is hard to find any
differences between the
analytic contour
and its numerical 
approximation 
for all 
$C\!u_l$, $l\!=\!0,1,2$ 
used in the present
study.
We recall 
here, these results
are obtained with 
the second-order
accurate finite
volume method resulting
in the second-order accurate
spatial discretization of \Eqs{eq9}{eq10}.

In \Figs{fig11}{fig13} 
and \Figs{fig14}{fig15}
it can be observed 
that in spite
of the first/second-order
accurate  convergence rate
of the interface shape 
with the Eulerian scheme,
the level-set
of curvature
of the same
interfaces
do not show
convergence
towards the
exact solution.
Hence, 
numerical 
convergence of 
the interface shape 
is not a sufficient
condition for
convergence of
the interface curvature.
We conclude,
second-order accurate
TVD MUSCL
used in the present
work is not  able
to reconstruct
the shape
of the circular
interface and its
curvature during 
advection in the
divergence free
velocity field.
The complete second-order
convergence rate
is obtained only
with the Lagrangian
scheme.
The  accuracy
of  reconstruction
of the interface
curvature during advection
does not  affect substantially
the  conservation
of mass in
the conservative
level-set
method, see  \Figs{fig8}{fig10}.

Finally, 
let discuss 
validity of 
the assumption
made after derivation
of \Eq{eq23}
about the existence
of $F_k \! \ls \alpha \rs$
functional minimum.
Based on
the convergence
studies of \Eq{eq10}
presented 
in \Fig{fig2}
and Figs.~\ref{fig6}-\ref{fig15}
we argue
\Eq{eq23} 
provides
the condition 
for existence 
of the functional 
$F_k \! \ls \alpha \rs$
minimum;
$F_k \! \ls \alpha \rs$
defined by \Eq{eq18}
is minimized by $\alpps$ 
given by \Eq{eq6} and hence
$|\nabla \psio| \!=\! 1$. 
%

\section{Conclusions}
\label{sec4}

In this paper
the relation between
the volume of fluid,
level-set and phase-field
interface models
has been introduced.
As a consequence,
the statistical model
of the non-flat interface
in the state of phase
equilibrium 
is postulated.
The statistical
view on the interfaces
agitated by the stochastic
velocity fields has been
already developed in
other works 
(e.g. \cite{hong00,brocchini01a,brocchini01b,freeze03,smolentsev05,mwaclawczyk11,waclawczyk2015}),
the statistical model
introduced herein
is based on
the ensemble
averaged picture
of the sharp interface
disturbed 
by the
stochastic 
velocity
field.
It is derived
by the ensemble 
averaging
of the phase 
indicator function
transport equation (\ref{eq1})
and the conservative
closure of
the correlation
between the sharp interface
velocity fluctuation
$\bW' \! \cdot \! \bng$ and
the exact Dirac's delta
function $\deltap$
indicating its 
instantaneous
position,
see \Eq{eq13} 
and \Eq{eq10}.

Subsequently,
the relation
of this new model
with the modified 
Allen-Cahn 
equation (\ref{eq4})
is established
showing  
the statistical
model of the interface 
describes
the non-flat interface
in the state of phase
equilibrium.
This result 
introduces
the physical
interpretation of
the re-initialization
equation (\ref{eq10}),
determination 
of its stationary
solution is equivalent 
to finding the minimum
of the modified
Ginzburg-Landau 
functional
given by \Eq{eq18}.
The new 
term in \Eq{eq18}
can be interpreted 
as the contribution 
to the interfacial
energy 
density in result
of a local 
change of 
the regularized
interface
shape 
and/or 
size.
The functional 
derivative of
this new term given
by \Eq{eq17}, 
resembles
the model
of capillary
forces used
in the one-fluid 
sharp interface
formulation.

The relation between
the statistical interface
model and 
modified Allen-Cahn equation 
shows the CLS method
is equivalent
of the phase field
interface model.
The order parameter
$\alpps$ of this new phase
field interface model
is a conserved
quantity,
it may be interpreted as
the probability of finding one
of the two phases sharing
the regularized interface.
The probability 
$\alpps$ is defined 
in terms of 
the logistic
distribution 
where $\eph \!>\! 0$
is measure of
deviation 
of the instantaneous
sharp interface position
from its expected position, 
see Eqs.~(\ref{eq6})-(\ref{eq7})
and \Eq{eq12}.
This latter
result
along with
the results 
presented 
by the author 
(see \cite{twacl15})
introduces
the relation
between
the sharp and diffusive
interface models, 
see \Eqs{eq9}{eq10}
in the limit $\eph \!\to\! 0$.

In the second 
part of 
the present paper,
two numerical
techniques are
introduced to reduce
numerical errors
during solution 
of \Eqs{eq9}{eq10}
and guarantee 
the balance 
of interfacial
energies 
predicted by
\Eq{eq23}.
At first,
dependence of
the known 
$\alpps$ profile
on the signed-distance
function $\psio$
is exploited
to approximate
the RHS fluxes
in  \Eq{eq10}
leading to 
the constrained
interpolation
scheme (CIS), 
see Eqs.~(\ref{eq25}).
It is demonstrated,
CIS 
improves 
stability
of the numerical 
solution of \Eq{eq10}
with regard to 
the selected time
step size $\Delta \tau$ 
and avoids
oscillatory 
errors, 
see \Fig{fig2} 
and \Figs{fig3}{fig5},
respectively.
Furthermore,
result in \Figs{fig6}{fig7} 
show  CIS guarantees
the theoretical 
convergence rate
of the numerical solution 
of \Eqs{eq9}{eq10}
on gradually
refined grids 
during advection
of the regularized
interface.

Next,
the new semi-analytical
Lagrangian scheme
for advection 
of $\alpps$ and $\psio$
functions has 
been derived, 
see \Eqs{eq29}{eq30}.
In \Sec{sec32} 
it is demonstrated
the new advection scheme
avoids introduction of 
the high-frequency
oscillatory errors
in curvature of 
the regularized
interface,
see \Fig{fig15}.
For this reason,
the Lagrangian 
scheme
reaches
the theoretical
second-order
convergence
of the interface 
shape 
and curvature, see
results in \Figs{fig11}{fig13}
and \Figs{fig14}{fig15}, 
respectively.

\section*{Acknowledgments}
This work is supported by the grant
of National Science Center, Poland
(Narodowe Centrum Nauki, Polska)
in the project 
\emph{``Statistical modeling of turbulent two-fluid flows with interfaces''},
ref. nr. 2016/21/B/ST8/01010, ID:334165.


\appendix

\section{Error norms}
\label{appA}

To compute 
errors
during 
the numerical solution 
of \Eqs{eq9}{eq10}
different error 
norms are used
in the present work,
this appendix 
provides their
definitions.
In \Fig{fig2},
the distance between solutions
on two different time $\tau$ levels
is measured by 
the first-order norm 
\blnm
\be
 L_1^\tau= \frac{1}{N_c} \sum_{i=1}^{N_{c}} |\alpha_i^{n+1}-\alpha_i^{n}|,
\label{Aeq1}
\ee
\elnm
where $N_c$ is the number of
control volumes and $n+1$ denotes
a new time level $\tau$, summation is
performed over the control volumes
centers in the entire computational
domain $\Omega$.
In \Fig{fig6}, 
$L_1^\tau$ norm 
is plotted 
after each
time step $\Delta t$.

In  Figs.~\ref{fig3}-\ref{fig5}, 
normalized first-order norms
are used to visualize numerical
errors introduced by LIS or CIS 
interpolation during re-initialization
\blnm
\be
 L_{1,an} \lr \phi \rr 
 = \frac{| \phi_{an} - \phi_{num}|}{|\phi_{an} |+\epsilon}, 
\label{Aeq2}
\ee
\elnm
\blnm
\be
 L_{1,max} \lr \phi \rr 
 = \frac{| \phi_{an} - \phi_{num}|}{\phi_{an,max}}, 
\label{Aeq3}
\ee
\elnm
where 
values $\phi_{an}$, $\phi_{num}$ 
are calculated,
respectively,
analytically and numerically 
in each control volume,
$\epsilon \!=\! 5\cdot10^{-16}$ and
$\phi \!=\! \alpha$,
or is the first 
component of  $\nabla \alpha$
or $\nabla^2 \alpha$.
$L_{1,max} \lr \phi \rr$
norm is obtained
using
$\phi_{an,max} \!=\! max \ls \phi_{an,i} \rs$
where $i \!=\! 1,\ldots,N_c$.

In \Fig{fig7} 
the averaged in
times $t$ and $\tau$ norms 
$L_1^\tau$
given by \Eq{Aeq1} 
are summarized,
the averaged  norms
are calculated
according 
to the formula
\blnm
\be
\av{L_1^\tau}  
 = \frac{1}{N_\tau N_t} \sum_{m=1}^{N_t} \sum_{n=1}^{N_\tau} L_{1,m,n}^\tau,
\label{Aeq4}
\ee
\elnm
where  $N_t,\,N_\tau$ denote the number
of time steps $\Delta t,\,\Delta \tau$,
respectively.

In \Figs{fig11}{fig13}
$L_1$, 
$L_2$ and $L_\infty$
norms are used to investigate
convergence of 
the interface shape
and curvature, their
definitions read
\blnm
\be
 L_1 = \frac{1}{N_p} \sum_{l=1}^{N_{p}} |\phi_e^l-\phi_n^l |,
\label{Aeq5}
\ee
\be
 L_2 = \frac{1}{N_p} \ls \sum_{l=1}^{N_{p}} \lr \phi_e^l-\phi_n^l \rr^2 \rs^{1/2},
\label{Aeq6}
\ee
\elnm
\blnm
\be
 L_{\infty}= max \ls |\phi_e^l-\phi_n^l | \rs,\,\, \text{where}\,\, l\!=\!1,\ldots,N_p
\label{Aeq7}
\ee
\elnm
and $N_p$ denotes
the number of probes 
on the contour $\phi$ 
representing 
the level-sets of 
the interface
$\psi \lr \alpha \!=\!1/2 \rr \!=\! 0$
or curvature $\kappa \ls \psi \lr \alpha\!=\!1/2 \rr\!=\! 0 \rs \!=\! 1/R$
computed on the $k\!-\!th$ grid $m_k$
(for brevity, 
 the grid index is 
 omitted in Eqs.~(\ref{Aeq5})-(\ref{Aeq7})),
$\phi_e$ denotes 
the point on the exact level-set
and $\phi_n$ denotes
its numerical approximation.
Moreover, 
we assume that norms
$L_1^\kappa$,
$L_2^\kappa$ and
$L_\infty^\kappa$ 
computed using
Eqs.~(\ref{Aeq5})-(\ref{Aeq7})
when $\phi$ 
is the level-set
of $\kappa$
are dimensionless
as they are divided
by $\kappa_1=1\,[1/m]$.

Another class of 
the numerical error 
indicator is obtained by 
calculation of the difference
between analytical and reconstructed 
surface/volume of 
the advected circular 
interface in \Sec{sec32}. 
In order to measure
departure of the numerical solution
$S_n$ from the exact value $S_e = \pi R^2$
following formula is used to compute
\blnm
\be
 E_S = 100 \cdot | 1-S_n/S_e |
\label{Aeq8}
\ee
\elnm
after each time step
$\Delta t$ and 
at the end of 
the re-initialization
cycle, i.e., after $N_\tau \!=\! 4$
time steps $\Delta \tau$.
This error is averaged
in time $t$ to closely
inspect convergence
of mass during
one revolution of the
circular interface, see \Fig{fig9};
the averaging is carried 
out using 
the equation
\blnm
\be
 \av{E_{S}} = \frac{1}{N_t} \sum_{l=1}^{N_t} E_{S}^l.
\label{Aeq9}
\ee
\elnm
%
%
\section{Discretization of the re-initialization equation}
\label{appB}
%
In this appendix, 
discretization
of \Eq{eq10} 
in the framework 
of the second-order 
accurate finite volume 
method is presented.
After integration 
of \Eq{eq10}
in the control 
volume $V_P$, 
employment of the Gauss 
theorem and mid-point
rule in centers
of the faces $f$ 
and in  
center of
the given control
volume $P$,
one obtains
\blnm
\be
  \pd{\alpha}{\tau} \Bigr|_P = \frac{1}{V_P} 
   \sum_{f=1}^{n_b} \ls \deltaa \lr |\nabla \psi | -  1 \rr \bng \cdot \bn \rs_f S_f,
\label{Beq1}
\ee
\elnm
where $n_b$ 
is  the number 
of neighbors 
of the  control volume $P$,
$\ls \bng \cdot \bn \rs_f$ is
dot product of the normal 
$\bng \!=\! \nabla \psi/|\nabla \psi|$ 
interpolated
at the face $f$ 
and normal 
$\bn_{f} \!=\! \bS_f/|\bS_f|$
where $\bS_f$ denotes surface
vector of the face $f$;
$\tilde{\delta} \lr \alpha_f \rr \!=\! \alpha_f \lr 1-\alpha_f \rr$
is approximated using the
linear interpolation (LIS) on the face $f\!=\!e$ leading to
$\alpha_e \!=\! \lr \alpha_P \!+\! \alpha_E \rr/2$,
or constrained interpolation (CIS) defined by Eqs.~(\ref{eq25}).

$|\nabla \psi|$
in \Eq{Beq1}
is computed using
the second-order 
central-difference 
approximation of
$\nabla \psi$ 
components; 
at the
face $f \!=\! e$ 
this approximation
reads
\blnm
\begin{align}
  \bs
  \pd{\psi}{x_1} \Big|_e &\approx  
   \frac{\lr \psi_{E}-\psi_{P} \rr}{ \Delta x}, \\
  \pd{\psi}{x_2} \Big|_e &\approx
   \frac{\lr \psi_{N}+\psi_{NE}-\psi_{S}-\psi_{SE} \rr}{4 \Delta y}, \\
  \es
\label{Beq2}
\end{align}
\elnm
where subscript $E,N,T,\ldots$ 
represent the centers 
of the neighbor 
control volumes
on uniform, orthogonal
structured grid.
%

\bibliography{mybibfile}

\begin{thebibliography}{41}
\expandafter\ifx\csname natexlab\endcsname\relax\def\natexlab#1{#1}\fi
\providecommand{\url}[1]{\texttt{#1}}
\providecommand{\href}[2]{#2}
\providecommand{\path}[1]{#1}
\providecommand{\DOIprefix}{doi:}
\providecommand{\ArXivprefix}{arXiv:}
\providecommand{\URLprefix}{URL: }
\providecommand{\Pubmedprefix}{pmid:}
\providecommand{\doi}[1]{\href{http://dx.doi.org/#1}{\path{#1}}}
\providecommand{\Pubmed}[1]{\href{pmid:#1}{\path{#1}}}
\providecommand{\bibinfo}[2]{#2}
\ifx\xfnm\relax \def\xfnm[#1]{\unskip,\space#1}\fi
\bibitem[{Aarts et~al.(2004)Aarts, Schmidt and Lekkerkerker}]{aarts2004}
\bibinfo{author}{Aarts, D. G. A.~L.}, \bibinfo{author}{Schmidt, M.},
  \bibinfo{author}{Lekkerkerker, H. N.~W.}, \bibinfo{year}{2004}.
\newblock \bibinfo{title}{Direct visual observation of thermal capillary
  waves}.
\newblock {\it \bibinfo{journal}{Science}\/},  {\it \bibinfo{volume}{304}\/},
  \bibinfo{pages}{847--850}. \DOIprefix\doi{10.1126/science.1097116}.
\bibitem[{Allen and Cahn(1979)}]{allen1979}
\bibinfo{author}{Allen, S.}, \bibinfo{author}{Cahn, J.}, \bibinfo{year}{1979}.
\newblock \bibinfo{title}{A microscopic theory for antiphase domain boundary
  motion and its application to antiphase domain coarsening}.
\newblock {\it \bibinfo{journal}{Acta Metall.}\/},  {\it
  \bibinfo{volume}{27}\/}, \bibinfo{pages}{1085--1095}.
\bibitem[{Anderson et~al.(1998)Anderson, McFadden and Wheeler}]{anderson1998}
\bibinfo{author}{Anderson, D.~M.}, \bibinfo{author}{McFadden, G.~B.},
  \bibinfo{author}{Wheeler, A.~A.}, \bibinfo{year}{1998}.
\newblock \bibinfo{title}{{D}iffuse-{I}nterface {M}ethods in {F}luid
  {M}echanics}.
\newblock {\it \bibinfo{journal}{Annu. Rev. Fluid Mech.}\/},  {\it
  \bibinfo{volume}{30}\/}, \bibinfo{pages}{139--165}.
  \DOIprefix\doi{10.1146/annurev.fluid.30.1.139}.
\bibitem[{Balakrishnan(1992)}]{bala1992}
\bibinfo{author}{Balakrishnan, N.}, \bibinfo{year}{1992}.
\newblock {\it \bibinfo{title}{Handbook of the logistic distribution}\/}.
\newblock \bibinfo{publisher}{Marcel Deker {INC}}.
\bibitem[{Balcazar et~al.(2014)Balcazar, Jofre, Lehmkuhl, Castro and
  Rigola}]{balcazar2014}
\bibinfo{author}{Balcazar, N.}, \bibinfo{author}{Jofre, L.},
  \bibinfo{author}{Lehmkuhl, O.}, \bibinfo{author}{Castro, J.},
  \bibinfo{author}{Rigola, J.}, \bibinfo{year}{2014}.
\newblock \bibinfo{title}{A finite-volume/level-set method for simulating
  two-phase flows on unstructured grids}.
\newblock {\it \bibinfo{journal}{Int. J. Multiphase Flow}\/},  {\it
  \bibinfo{volume}{64}\/}, \bibinfo{pages}{55 -- 72}.
  \DOIprefix\doi{http://dx.doi.org/10.1016/j.ijmultiphaseflow.2014.04.008}.
\bibitem[{Bao et~al.(2012)Bao, Shi, Sun and Wang}]{bao2012}
\bibinfo{author}{Bao, K.}, \bibinfo{author}{Shi, Y.}, \bibinfo{author}{Sun,
  S.}, \bibinfo{author}{Wang, X.-P.}, \bibinfo{year}{2012}.
\newblock \bibinfo{title}{A finite element method for the numerical solution of
  the coupled {C}ahn-{H}illiard and {N}avier-{S}tokes system for moving contact
  line problems}.
\newblock {\it \bibinfo{journal}{J. Comp. Phys.}\/},  {\it
  \bibinfo{volume}{231}\/}, \bibinfo{pages}{8083 -- 8099}.
  \DOIprefix\doi{http://dx.doi.org/10.1016/j.jcp.2012.07.027}.
\bibitem[{Brassel and Bretin(2011)}]{brassel11}
\bibinfo{author}{Brassel, M.}, \bibinfo{author}{Bretin, E.},
  \bibinfo{year}{2011}.
\newblock \bibinfo{title}{A modified phase field approximation for mean
  curvature flow with conservation of the volume}.
\newblock {\it \bibinfo{journal}{Math. Method. Appl. Sci.}\/},  {\it
  \bibinfo{volume}{34}\/}, \bibinfo{pages}{1157--1180}. \URLprefix
  \url{http://dx.doi.org/10.1002/mma.1426}. \DOIprefix\doi{10.1002/mma.1426}.
\bibitem[{Brocchini and Peregrine(2001{\natexlab{a}})}]{brocchini01a}
\bibinfo{author}{Brocchini, M.}, \bibinfo{author}{Peregrine, D.~H.},
  \bibinfo{year}{2001}{\natexlab{a}}.
\newblock \bibinfo{title}{The dynamics of strong turbulence at free surfaces.
  {P}art 1. {D}escription}.
\newblock {\it \bibinfo{journal}{J. Fluid Mech.}\/},  {\it
  \bibinfo{volume}{449}\/}, \bibinfo{pages}{225--254}.
\bibitem[{Brocchini and Peregrine(2001{\natexlab{b}})}]{brocchini01b}
\bibinfo{author}{Brocchini, M.}, \bibinfo{author}{Peregrine, D.~H.},
  \bibinfo{year}{2001}{\natexlab{b}}.
\newblock \bibinfo{title}{The dynamics of strong turbulence at free surfaces.
  {P}art 2. {F}ree-surface boundary conditions}.
\newblock {\it \bibinfo{journal}{J. Fluid Mech.}\/},  {\it
  \bibinfo{volume}{449}\/}, \bibinfo{pages}{255--290}.
\bibitem[{Cahn and Hilliard(1958)}]{cahn1958}
\bibinfo{author}{Cahn, J.~W.}, \bibinfo{author}{Hilliard, J.~E.},
  \bibinfo{year}{1958}.
\newblock \bibinfo{title}{{F}ree {E}nergy of a {N}onuniform {S}ystem. {I.}
  {I}nterfacial {F}ree {E}nergy}.
\newblock {\it \bibinfo{journal}{J. Chem. Phys.}\/},  {\it
  \bibinfo{volume}{28}\/}, \bibinfo{pages}{258--267}.
  \DOIprefix\doi{http://dx.doi.org/10.1063/1.1744102}.
\bibitem[{Chiu and Lin(2011)}]{chiu11}
\bibinfo{author}{Chiu, P.-H.}, \bibinfo{author}{Lin, Y.-T.},
  \bibinfo{year}{2011}.
\newblock \bibinfo{title}{A conservative phase field method for solving
  incompressible two-phase flows}.
\newblock {\it \bibinfo{journal}{J. Comp. Phys.}\/},  {\it
  \bibinfo{volume}{230}\/}, \bibinfo{pages}{185--204}.
  \DOIprefix\doi{http://dx.doi.org/10.1016/j.jcp.2010.09.021}.
\bibitem[{Fedeli(2017)}]{fedeli2017}
\bibinfo{author}{Fedeli, L.}, \bibinfo{year}{2017}.
\newblock \bibinfo{title}{Computer simulations of phase field drops on
  super-hydrophobic surfaces}.
\newblock {\it \bibinfo{journal}{J. Comp. Phys.}\/},  {\it
  \bibinfo{volume}{344}\/}, \bibinfo{pages}{247--259}.
  \DOIprefix\doi{http://dx.doi.org/10.1016/j.jcp.2017.04.068}.
\bibitem[{Ferziger and Peri{\'c}(2002)}]{peric02}
\bibinfo{author}{Ferziger, J.~H.}, \bibinfo{author}{Peri{\'c}, M.},
  \bibinfo{year}{2002}.
\newblock {\it \bibinfo{title}{Computational {M}ethods for {F}luid
  {D}ynamics}\/}.
\newblock \bibinfo{publisher}{Springer Verlag, Berlin Heidelberg New York}.
\bibitem[{Freeze et~al.(2003)Freeze, Smolentsev, Morley and M.}]{freeze03}
\bibinfo{author}{Freeze, B.}, \bibinfo{author}{Smolentsev, S.},
  \bibinfo{author}{Morley, N.}, \bibinfo{author}{M., A.}, \bibinfo{year}{2003}.
\newblock \bibinfo{title}{Characterization of the effect of {F}roude number on
  surface waves and heat transfer in inclined turbulent open channel flows}.
\newblock {\it \bibinfo{journal}{Heat Mass Transfer}\/},  {\it
  \bibinfo{volume}{46}\/}, \bibinfo{pages}{3765--3775}.
\bibitem[{Gottlieb and Shu(1998)}]{gottlieb98}
\bibinfo{author}{Gottlieb, S.}, \bibinfo{author}{Shu, C.-W.},
  \bibinfo{year}{1998}.
\newblock \bibinfo{title}{Total variation diminishing {R}unge-{K}utta schemes}.
\newblock {\it \bibinfo{journal}{Math. Comp.}\/},  {\it
  \bibinfo{volume}{67}\/}, \bibinfo{pages}{73--85}.
\bibitem[{Herrmans(2005)}]{herrmans05}
\bibinfo{author}{Herrmans, M.}, \bibinfo{year}{2005}.
\newblock \bibinfo{title}{{R}efined {L}evel-{S}et {G}rids method for tracking
  interfaces}.
\newblock In {\it \bibinfo{booktitle}{Annual Research Briefs}\/} (pp.
  \bibinfo{pages}{3--18}).
\newblock \bibinfo{address}{Center of Turbulence Research, University of
  Stanford}.
\bibitem[{Hong and Walker(2000)}]{hong00}
\bibinfo{author}{Hong, W.-L.}, \bibinfo{author}{Walker, D.},
  \bibinfo{year}{2000}.
\newblock \bibinfo{title}{Reynolds-averaged equations for free surface flows
  with application to high-{F}roude-number jet spreding}.
\newblock {\it \bibinfo{journal}{J. Fluid Mech.}\/},  {\it
  \bibinfo{volume}{417}\/}, \bibinfo{pages}{183--209}.
\bibitem[{Kim et~al.(2014)Kim, Lee and Choi}]{kim14}
\bibinfo{author}{Kim, J.}, \bibinfo{author}{Lee, S.}, \bibinfo{author}{Choi,
  Y.}, \bibinfo{year}{2014}.
\newblock \bibinfo{title}{A conservative {A}llen-{C}ahn equation with a
  spacetime dependent {L}agrange multiplier}.
\newblock {\it \bibinfo{journal}{Int. J. Eng. Sci.}\/},  {\it
  \bibinfo{volume}{84}\/}, \bibinfo{pages}{11 -- 17}.
  \DOIprefix\doi{http://dx.doi.org/10.1016/j.ijengsci.2014.06.004}.
\bibitem[{McCaslin and Desjardins(2014)}]{mccaslin2014}
\bibinfo{author}{McCaslin, J.~O.}, \bibinfo{author}{Desjardins, O.},
  \bibinfo{year}{2014}.
\newblock \bibinfo{title}{A localized re-initialization equation for the
  conservative level set method}.
\newblock {\it \bibinfo{journal}{J. Comp. Phys.}\/},  {\it
  \bibinfo{volume}{262}\/}, \bibinfo{pages}{408 -- 426}.
  \DOIprefix\doi{http://dx.doi.org/10.1016/j.jcp.2014.01.017}.
\bibitem[{Moelans et~al.(2008)Moelans, Blanpain and Wollants}]{moelans08}
\bibinfo{author}{Moelans, N.}, \bibinfo{author}{Blanpain, B.},
  \bibinfo{author}{Wollants, P.}, \bibinfo{year}{2008}.
\newblock \bibinfo{title}{An introduction to phase-field modeling of
  microstructure evolution}.
\newblock {\it \bibinfo{journal}{Calphad}\/},  {\it \bibinfo{volume}{32}\/},
  \bibinfo{pages}{268 -- 294}.
  \DOIprefix\doi{http://dx.doi.org/10.1016/j.calphad.2007.11.003}.
\bibitem[{Olsson and Kreiss(2005)}]{olsson05}
\bibinfo{author}{Olsson, E.}, \bibinfo{author}{Kreiss, G.},
  \bibinfo{year}{2005}.
\newblock \bibinfo{title}{A conservative level-set method for two phase flow}.
\newblock {\it \bibinfo{journal}{J. Comp. Phys.}\/},  {\it
  \bibinfo{volume}{210}\/}, \bibinfo{pages}{225--246}.
\bibitem[{Osher and Fedkiw(2003)}]{osher03}
\bibinfo{author}{Osher, S.}, \bibinfo{author}{Fedkiw, R.},
  \bibinfo{year}{2003}.
\newblock {\it \bibinfo{title}{{L}evel {S}et {M}ethods and {D}ynamic {I}mplicit
  {S}urfaces}\/}.
\newblock \bibinfo{publisher}{Springer Verlag, INC. New-York}.
\bibitem[{Osher and Sethian(1988)}]{osher1988}
\bibinfo{author}{Osher, S.}, \bibinfo{author}{Sethian, J.~A.},
  \bibinfo{year}{1988}.
\newblock \bibinfo{title}{Fronts propagating with curvature-dependent speed:
  {A}lgorithms based on {H}amilton-{J}acobi formulations}.
\newblock {\it \bibinfo{journal}{J. Comp. Phys.}\/},  {\it
  \bibinfo{volume}{79}\/}, \bibinfo{pages}{12 -- 49}.
  \DOIprefix\doi{http://dx.doi.org/10.1016/0021-9991(88)90002-2}.
\bibitem[{Pashos et~al.(2015)Pashos, Kokkoris and Boudouvis}]{pashos2015}
\bibinfo{author}{Pashos, G.}, \bibinfo{author}{Kokkoris, G.},
  \bibinfo{author}{Boudouvis, A.}, \bibinfo{year}{2015}.
\newblock \bibinfo{title}{A modified phase-field method for the investigation
  of wetting transitions of droplets on patterned surfaces}.
\newblock {\it \bibinfo{journal}{J. Comp. Phys.}\/},  {\it
  \bibinfo{volume}{283}\/}, \bibinfo{pages}{258 -- 270}.
  \DOIprefix\doi{http://dx.doi.org/10.1016/j.jcp.2014.11.045}.
\bibitem[{Pope(1998)}]{pope88}
\bibinfo{author}{Pope, S.}, \bibinfo{year}{1998}.
\newblock \bibinfo{title}{The evolution of surfaces in turbulence}.
\newblock {\it \bibinfo{journal}{Int. J. Eng. Sciences}\/},  {\it
  \bibinfo{volume}{26}\/}, \bibinfo{pages}{445--469}.
\bibitem[{Sch{\"a}fer(2006)}]{schaefer06}
\bibinfo{author}{Sch{\"a}fer, M.}, \bibinfo{year}{2006}.
\newblock {\it \bibinfo{title}{{C}omputational {E}ngineering, {I}ntroduction to
  {N}umerical {M}ethods}\/}.
\newblock \bibinfo{publisher}{Springer-Verlag Berlin Heidelberg New York}.
\bibitem[{Smolentsev and Miraghaie(2005)}]{smolentsev05}
\bibinfo{author}{Smolentsev, S.}, \bibinfo{author}{Miraghaie, R.},
  \bibinfo{year}{2005}.
\newblock \bibinfo{title}{Study of a free surface in open-channel water flows
  in the regime from "weak" to "strong" turbulence}.
\newblock {\it \bibinfo{journal}{Int. J. Multiphase Flows}\/},  {\it
  \bibinfo{volume}{31}\/}, \bibinfo{pages}{921--939}.
\bibitem[{Smoluchowski(1908)}]{smol1908}
\bibinfo{author}{Smoluchowski, M.}, \bibinfo{year}{1908}.
\newblock \bibinfo{title}{{M}olekular-kinetische {T}heorie der {O}paleszenz von
  {G}asen im kritischen zustande, sowie einiger verwandter erscheinungen}.
\newblock {\it \bibinfo{journal}{Ann. Phys.}\/},  {\it
  \bibinfo{volume}{330}\/}, \bibinfo{pages}{205--226}.
  \DOIprefix\doi{10.1002/andp.19083300203}.
\bibitem[{Sussman et~al.(1998)Sussman, Fatemi, Smereka and Osher}]{sussman98}
\bibinfo{author}{Sussman, M.}, \bibinfo{author}{Fatemi, E.},
  \bibinfo{author}{Smereka, P.}, \bibinfo{author}{Osher, S.},
  \bibinfo{year}{1998}.
\newblock \bibinfo{title}{An improved level set method for incompressible
  two-phase flows}.
\newblock {\it \bibinfo{journal}{Comput. Fluids}\/},  {\it
  \bibinfo{volume}{27}\/}, \bibinfo{pages}{663 -- 680}.
  \DOIprefix\doi{http://dx.doi.org/10.1016/S0045-7930(97)00053-4}.
\bibitem[{Sussman et~al.(1994)Sussman, Smereka and Osher}]{sussman94}
\bibinfo{author}{Sussman, M.}, \bibinfo{author}{Smereka, P.},
  \bibinfo{author}{Osher, S.~J.}, \bibinfo{year}{1994}.
\newblock \bibinfo{title}{A level set approach for computing solutions to
  incompressible two-phase flows}.
\newblock {\it \bibinfo{journal}{J. Comp. Phys.}\/},  {\it
  \bibinfo{volume}{114}\/}, \bibinfo{pages}{146--159}.
\bibitem[{Tryggvason et~al.(2011)Tryggvason, Scardovelli and Zaleski}]{trygg11}
\bibinfo{author}{Tryggvason, G.}, \bibinfo{author}{Scardovelli, R.},
  \bibinfo{author}{Zaleski, S.}, \bibinfo{year}{2011}.
\newblock {\it \bibinfo{title}{Direct Numerical Simulations of Gas-Liquid
  Multiphase Flows}\/}.
\newblock \bibinfo{publisher}{Cambridge University Press}.
\bibitem[{Vrij(1973)}]{vrij1973}
\bibinfo{author}{Vrij, A.}, \bibinfo{year}{1973}.
\newblock \bibinfo{title}{Light scattering from liquid interfaces}.
\newblock {\it \bibinfo{journal}{Chemie Ingenieur Technik}\/},  {\it
  \bibinfo{volume}{45}\/}, \bibinfo{pages}{1113--1114}.
  \DOIprefix\doi{10.1002/cite.330451807}.
\bibitem[{van~der Waals(1979)}]{waals1979}
\bibinfo{author}{van~der Waals, J.}, \bibinfo{year}{1979}.
\newblock \bibinfo{title}{The thermodynamic theory of capillarity under the
  hypothesis of a continuous variation of density}.
\newblock {\it \bibinfo{journal}{J. Statist. Phys.}\/},  {\it
  \bibinfo{volume}{20}\/}, \bibinfo{pages}{200--244}.
\bibitem[{Wac{\l}awczyk and Oberlack(2011)}]{mwaclawczyk11}
\bibinfo{author}{Wac{\l}awczyk, M.}, \bibinfo{author}{Oberlack, M.},
  \bibinfo{year}{2011}.
\newblock \bibinfo{title}{Closure proposals for the tracking of
  turbulence-agitated gas-liquid interfaces in stratified flows}.
\newblock {\it \bibinfo{journal}{Int. J. Multiphase Flow}\/},  {\it
  \bibinfo{volume}{37}\/}, \bibinfo{pages}{967--976}.
\bibitem[{Wac{\l}awczyk and Wac{\l}awczyk(2015)}]{waclawczyk2015}
\bibinfo{author}{Wac{\l}awczyk, M.}, \bibinfo{author}{Wac{\l}awczyk, T.},
  \bibinfo{year}{2015}.
\newblock \bibinfo{title}{A priori study for the modelling of
  velocity-interface correlations in the stratified air-water flows}.
\newblock {\it \bibinfo{journal}{Int. J. Heat Fluid Flow}\/},  {\it
  \bibinfo{volume}{52}\/}, \bibinfo{pages}{40 -- 49}.
  \DOIprefix\doi{http://dx.doi.org/10.1016/j.ijheatfluidflow.2014.11.004}.
\bibitem[{Wac{\l}awczyk(2015)}]{twacl15}
\bibinfo{author}{Wac{\l}awczyk, T.}, \bibinfo{year}{2015}.
\newblock \bibinfo{title}{A consistent solution of the re-initialization
  equation in the conservative level-set method}.
\newblock {\it \bibinfo{journal}{J. Comp. Phys.}\/},  {\it
  \bibinfo{volume}{299}\/}, \bibinfo{pages}{487 -- 525}.
  \DOIprefix\doi{http://dx.doi.org/10.1016/j.jcp.2015.06.029}.
\bibitem[{Wac{\l}awczyk and Koronowicz(2006)}]{waclawczyk06}
\bibinfo{author}{Wac{\l}awczyk, T.}, \bibinfo{author}{Koronowicz, T.},
  \bibinfo{year}{2006}.
\newblock \bibinfo{title}{Modelling of the free surface flows with
  high-resolution schemes}.
\newblock {\it \bibinfo{journal}{Chemical and Process Engineering}\/},  {\it
  \bibinfo{volume}{27}\/}, \bibinfo{pages}{783--802}.
\bibitem[{Wac{\l}awczyk and Koronowicz(2008{\natexlab{a}})}]{waclawczyk08_2}
\bibinfo{author}{Wac{\l}awczyk, T.}, \bibinfo{author}{Koronowicz, T.},
  \bibinfo{year}{2008}{\natexlab{a}}.
\newblock \bibinfo{title}{Comparison of {CICSAM} and {HRIC} high resolution
  schemes for interface capturing}.
\newblock {\it \bibinfo{journal}{J. Theoretical and Applied Mechanics}\/},
  {\it \bibinfo{volume}{46}\/}, \bibinfo{pages}{325--345}.
\bibitem[{Wac{\l}awczyk and Koronowicz(2008{\natexlab{b}})}]{waclawczyk08_3}
\bibinfo{author}{Wac{\l}awczyk, T.}, \bibinfo{author}{Koronowicz, T.},
  \bibinfo{year}{2008}{\natexlab{b}}.
\newblock \bibinfo{title}{Remarks on prediction of wave drag using {VOF} method
  with interface capturing approach}.
\newblock {\it \bibinfo{journal}{Archives of Civil and Mechanical
  Engineering}\/},  {\it \bibinfo{volume}{8}\/}, \bibinfo{pages}{5 -- 14}.
  \DOIprefix\doi{http://dx.doi.org/10.1016/S1644-9665(12)60262-3}.
\bibitem[{Wac{\l}awczyk et~al.(2014)Wac{\l}awczyk, Wac{\l}awczyk and
  Kraheberger}]{twaclawczyketal14}
\bibinfo{author}{Wac{\l}awczyk, T.}, \bibinfo{author}{Wac{\l}awczyk, M.},
  \bibinfo{author}{Kraheberger, S.~V.}, \bibinfo{year}{2014}.
\newblock \bibinfo{title}{Modeling of turbulence-interface interactions in
  stratified two-phase flows}.
\newblock {\it \bibinfo{journal}{Journal of Physics: Conference Series}\/},
  {\it \bibinfo{volume}{530}\/}.
\bibitem[{Yue et~al.(2007)Yue, Zhou and Feng}]{yue2007}
\bibinfo{author}{Yue, P.}, \bibinfo{author}{Zhou, C.}, \bibinfo{author}{Feng,
  J.~J.}, \bibinfo{year}{2007}.
\newblock \bibinfo{title}{Spontaneous shrinkage of drops and mass conservation
  in phase-field simulations}.
\newblock {\it \bibinfo{journal}{J. Comp. Phys.}\/},  {\it
  \bibinfo{volume}{223}\/}, \bibinfo{pages}{1 -- 9}.
  \DOIprefix\doi{http://dx.doi.org/10.1016/j.jcp.2006.11.020}.

\end{thebibliography}

\end{document}